\documentclass{raa}
\usepackage{graphicx,times}             
\input{epsf.sty}                        

\usepackage{graphicx}             
\usepackage{epstopdf}
\begin{document}
   \title{Chemical abundance analysis of 19 barium stars
\footnotetext{\small $*$ Based on spectroscopic observations in
           National Astronomical Observatory, Chinese Academy
           of Sciences (NAOC) (Xinglong, China) .}
}
   \volnopage{Vol.0 (200x) No.0, 000--000}      
   \setcounter{page}{1}          

   \author{G. C. Yang \inst{1,2,3}, Y. C. Liang\inst{2},  M.
   Spite\inst{4}, Y. Q. Chen\inst{2}, G. Zhao\inst{2}, B. Zhang\inst{1,2}, G. Q.
   Liu\inst{5,2},  Y. J. Liu\inst{2}, N. Liu\inst{6}, L. C. Deng\inst{2}, F. Spite\inst{4},
       V. Hill\inst{4}, C. X. Zhang\inst{7}}
   \offprints{gcyang@bao.ac.cn,ycliang@bao.ac.cn}
   \institute{
  $^1$College of Physics and Information Engineering, Hebei Normal University, Shijiazhuang 050024, China\\
  $^2$Key Laboratory of Optical Astronomy, National Astronomical
Observatories, Chinese Academy of Sciences, Beijing 100012, China; {\it ycliang@bao.ac.cn}\\
 $^3$Graduate School of the Chinese Academy of Sciences, Beijing 100049, China \\
 $^4$GEPI, Observatoire de Paris-Meudon, 92195 Meudon, France \\
 $^5$Physics Department and Tsinghua Center for Astrophysics, Tsinghua University, Beijing 100084, China \\
 $^6$Astronomy Department, Beijing Normal University, Beijing 100875, China \\
 $^7$Department of Physics, Shijiazhuang University, Shijiazhuang 050035, China
   }
    \authorrunning{G. C. Yang, Y. C. Liang, M. Spite et al.}
 \titlerunning{Chemical abundance analysis of 19 barium stars}
   \date{Received, accepted,}
 \abstract
{
We aim at deriving accurate atmospheric parameters and chemical abundances
of 19 barium (Ba) stars, including both strong and mild Ba stars, based on
the high signal-to-noise ratio and high resolution Echelle spectra obtained
from the 2.16\,m telescope at Xinglong station of National Astronomical Observatories,
Chinese Academy of Sciences. The chemical abundances of the sample stars were obtained
from an LTE, plane-parallel and line-blanketed atmospheric model by inputting the
atmospheric parameters (effective temperatures $T_{\rm eff}$, surface gravities log\,$g$,
metallicity [Fe/H] and microturbulent velocity $\xi_{\rm t}$) and equivalent widths
of stellar absorption lines. These samples of Ba stars are giants indicated by atmospheric parameters,
metallicities and kinematic analysis about UVW velocity. Chemical abundances of 17 elements
were obtained for these Ba stars. Their Na, Al, $\alpha$- and iron-peak elements
(O, Na, Mg, Al, Si, Ca, Sc, Ti, V, Cr, Mn, Ni) are similar to the solar abundances.
Our samples of Ba stars show obvious overabundances of
neutron-capture ($n$-capture) process elements relative to the Sun.
Their median abundances of [Ba/Fe], [La/Fe] and [Eu/Fe] are 0.54, 0.65 and
0.40, respectively. The \ion{Y}{i} and \ion{Zr}{i} abundances are lower than Ba, La and Eu,
but higher than the $\alpha$- and iron-peak elements for the strong Ba stars
and similar to the iron-peak elements for the mild stars.
There exists a positive correlation between Ba intensity
and [Ba/Fe].  For the $n$-capture elements (Y, Zr, Ba, La), there is an anti-correlation between
their [X/Fe] and [Fe/H]. We identify nine of our sample stars as strong Ba stars with [Ba/Fe]$>$0.6
where seven of them have Ba intensity Ba=2-5, one has Ba=1.5 and another one has Ba=1.0. The remaining
ten stars are classified as mild Ba stars with 0.17$<$[Ba/Fe]$<$0.54.
  \keywords{  Methods: observational --
              stars: abundances --
              stars: atmospheres --
              Stars: chemically peculiar --
              Stars: kinematics and dynamics --
              Stars: late-type}
}
  \maketitle
\section{Introduction}
Barium (Ba) stars belong to a distinct class of peculiar giants
which were first identified by Bidelman \& Keenan (1951). The stars
of this group are G to K giants, and in their spectra there are
abnormally strong lines of $s$-process elements, especially
\ion{Ba}{ii} at 4554\,\AA, as well as enhanced CH, CN and C$_{2}$
molecular bands. Besides Ba and Sr, other
heavy elements such as Y, Zr, La, Ce, Pr, Nd and Sm are also
enhanced. The general properties of Ba stars were reviewed in detail in McClure (1984),
and the chemical composition of Ba stars have been reviewed by Lambert (1985, 1988).

Burbidge et al. (1957) identified the mode of synthesis with slow $n$-capture
process ($s$-process) being responsible for the
production of the majority of the isotopes heavier than iron.
The later works on $s$-process enrichment usually
parameterize the amount of $^{13}$C burnt during the $s$-process
operation in the interior of asymptotic giant branch
(AGB) stars (Straniero et al. 1995, 1997; Busso et al. 1995, 1999,
2001; Gallino et al. 1998). They suggested that $^{13}$C was
completely burnt in the radiative condition, and the resulting
$s$-process nucleosynthesis occurred during the quiescent interpulse
period, instead of the convective thermal pulse.
$^{22}$Ne$(\alpha,n)^{25}$Mg was still active for a very short
period during the convective pulse and had a minor influence on the whole
process of nucleosynthesis. In thermally pulsating AGB stars (TP-AGBs), the deep
dredge-up phenomenon that follows the thermal pulses, the so-called third dredge-up,
mixes some of the processed material to the atmosphere, where it becomes accessible to
observations. The mechanism of $s$-process is relatively clear, whereas the
astrophysical site of the rapid $n$-capture process ($r$-process) is debated,
although it is believed to take place in
supernova explosions of massive stars (Cowan et al. 1991; Sneden et al. 2008).

Because of their low
luminosity, low mass and the absence of the unstable nucleus
$^{99}$Tc ($\tau_{1\over 2}=2\times 10^{5}$\,yr), Ba stars are not TP-AGB stars.
It is generally believed that they belong to binary systems and
that the overabundance of their heavy elements would be caused by the accretion of the matter
ejected by their more massive companions (now a white dwarf) in their TP-AGB phase about
1$\times$ 10$^{6}$ years ago (McClure
et al. 1980; Han et al. 1995; Jorissen et al. 1998; Liang et al.
2000). They should not have evolved to the TP-AGB stage to synthesize these heavy elements yet.

McClure et al. (1980) revealed that most Ba stars, maybe all of
them, show variations in radial-velocity suggesting the presence of
companions. The orbital elements of a number of Ba stars in binary
systems have been measured (Carquillat et al. 1998; Udry et al.
1998a,b; Jorissen et al. 1998). The determination of the chemical
abundances of a large sample of Ba stars must be very
important for understanding their formation scenario and
the abundance patterns in the ejecta of AGB stars.

It is often difficult to study the chemical
abundances of such cold stars with M, S and C spectral types and with
strong TiO absorption lines.
Lu (1991) presented a table of 389 Ba stars with an estimate of their Ba intensities (from 1 to 5)
following Lu et al. (1983) on the scale defined
by Warner (1965) and based on the strength of \ion{Ba}
{ii} 4554\,\AA~ line. We could roughly divide the Ba stars into two classes:  weak (mild)
Ba stars (with Ba$<$2) and strong Ba stars (with Ba=2-5). But we will consider [Ba/Fe] abundance
ratios as a stricter diagnostic for strong and mild Ba stars in this work.

Many studies have tried to analyze in detail the abundances of Ba stars based on high
resolution and high
signal-to-noise ($S/N$) spectra. Za\v{c}s (1994)  analyzed the chemical abundances of a
sample of 31 Ba stars
and normal G-K giants. Allen \& Barbuy (2006a,b) analyzed 26 Ba stars. Smiljanic et al. (2007)
analyzed 13 sample
stars. In addition, the chemical composition of 5, 16 and 1 Ba stars was studied by
Boyarchuk et al. (2002),
Antipova et al. (2004) and Yushchenko et al. (2004), respectively. Other available analyses
are mostly based on
data with lower $S/N$ ratio (Pilachowski et al. 1977; Smith 1984; Kovacs 1985; Luck \& Bond 1991).

As a consequence, the number of well studied Ba stars is still small compared to
the 389 stars listed in Lu (1991).
A detailed analysis of the abundances of a large sample of Ba stars is necessary to
reveal their properties and
formation scenario. Moreover, most of the previous studies have focussed on strong Ba stars.
In this work, we aim to extend the sample to many mild Ba stars with a relatively low Ba intensity.

In 2000, we started a project  of analyzing the chemical composition of a large sample of Ba stars.
In 2000, 2001,
2004 and 2005, high resolution Echelle spectra with high $S/N$ ratio were obtained with
the 2.16\,m telescope located
at Xinglong station, China. Liang et al. (2003) and Liu et al. (2009) analyzed four and eight
typical Ba stars, respectively.
In our present work, we analyze in detail the chemical abundances of the remaining 19 Ba stars.

This paper is organized as follows. In Section 2, we describe the
observations and analysis methods. In Section 3, we present the
atmospheric parameters of the stars. In Section 4, we define the
adopted atmospheric model and discuss the choice of spectral lines. We analyze
the element abundances and the uncertainty, and the relation
between the Ba intensity and [Ba/Fe] in Section 5. In Section 6, we
analyze their UVW kinematics and orbital periods. The results are discussed
in Section 7.

\section{Observations and data reduction}
The sample of star was selected from Lu (1991) and Jorissen et
al. (1998), and the corresponding Ba intensities were taken from there.
One normal giant in a Ba-like binary system, HD\,66216,
was further taken from Jorissen et al. (1998) (originally from the
catalogue compiled by Boffin et al. 1993) to check whether
this star also has an abnormal Ba abundance
based on the high quality spectral data. The basic
data on the sample of stars are given in Table~\ref{simbad}.

The spectroscopic observations were carried out with the Coud\'{e}
Echelle Spectrograph equipped with a 1024$\times$1024 TeKtronix CCD
attached to the 2.16\,m telescope at National Astronomical
Observatories, Chinese Academy of Sciences (Xinglong station, China). The
red arm of the spectrograph with a 31.6\,grooves/mm grating was used
in combination with a prism as cross-disperser, providing a good
separation between the Echelle orders. With a 0.5\,mm slit (1.1''),
the resolving power is of the order of R$\sim$30000 in the
camera system at the middle focus. The total wavelength coverage was 5500
$\sim$ 9000\,\AA~ over 36 orders, and most of the spectra have an $S/N$ ratio
greater than 60.

\begin{figure}
\centering
\includegraphics[width=6.0cm]{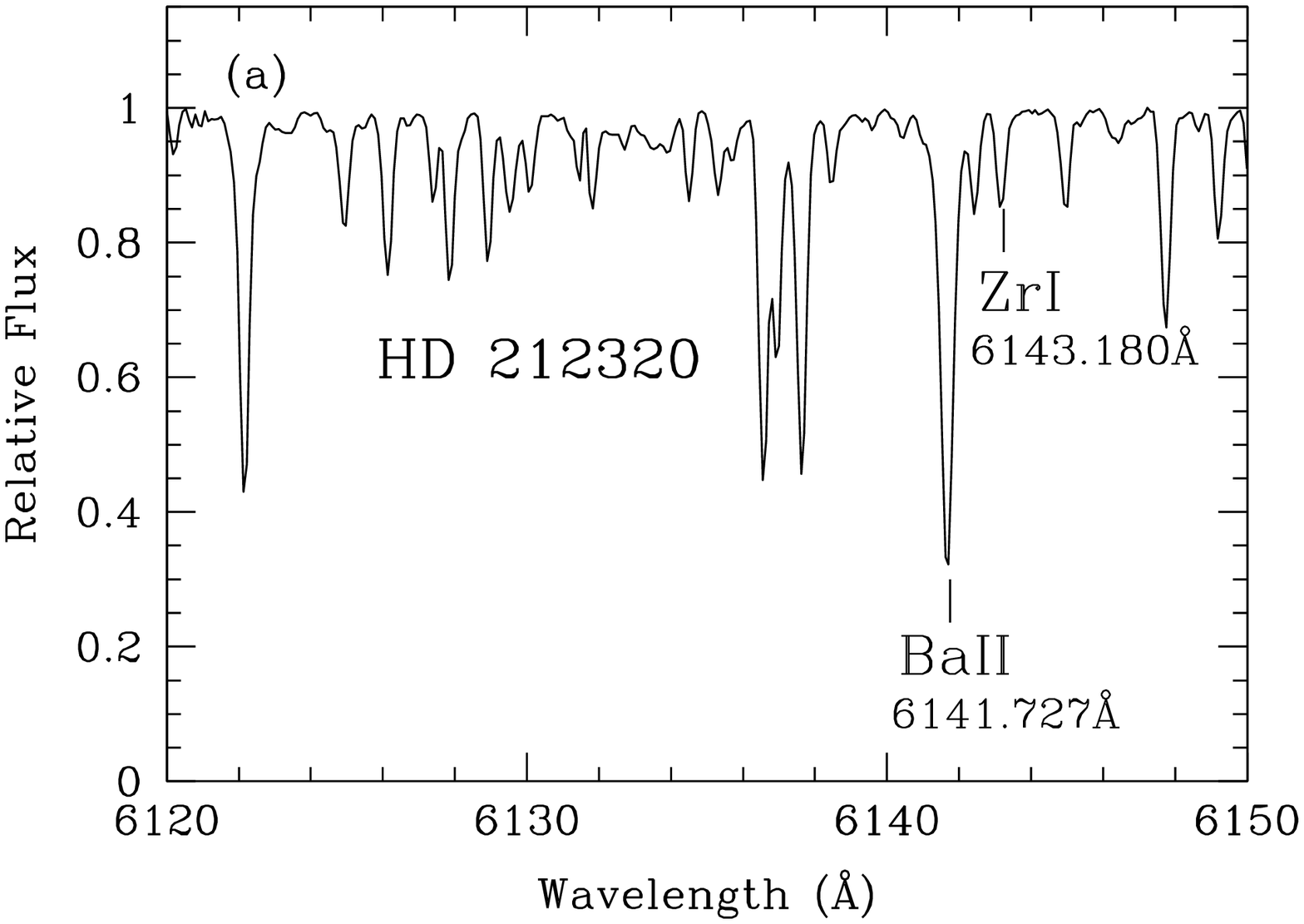}
\includegraphics[width=6.0cm]{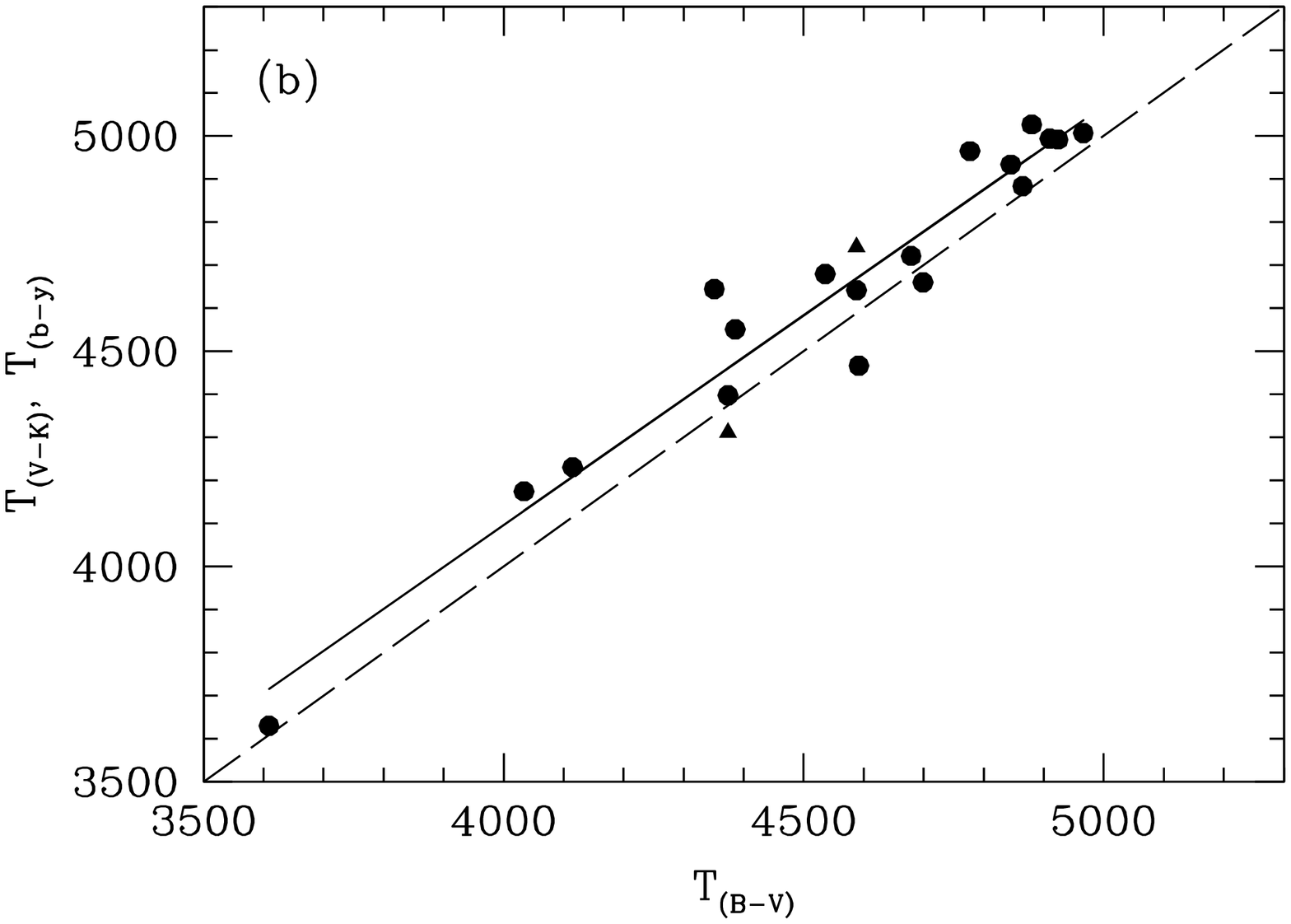}
\vspace{-2.0cm}
\caption {{\bf (a)} A part of spectrum of HD\,212320 showing \ion{Ba}
{ii} 6141.727\,\AA~, \ion{Zr} {i} 6143.180\,\AA~. {\bf (b)} Comparisons among the three sets of $T_{\rm eff}$ values
calculated from $(B-V)$, $(V-K)$ and $(b-y)$ colors for stars in the sample. The filled circles and triangles refer to the temperatures
calculated with $(V-K)$ and $(b-y)$, respectively. The solid
line is the least-square fit to the
points of $T_{\rm eff}(V-K)$ vs. $T_{\rm
eff}(B-V)$, whereas the dashed diagonal line is the
one-to-one relation.} \label{flux.eps}
\end{figure}

We used the standard ECHELLE package in the MIDAS environment for data
reduction and equivalent widths ($EW$s) measurements of lines. The
procedure is: order identification, background subtraction,
flat-field correction, order extraction, wavelength calibration,
radial velocity shift correction, and continuum normalization. Bias,
dark current and scattered-light corrections are included in the
background subtraction. The pixel-to-pixel sensitivity variations
were corrected by using the flat-field. The wavelength calibration
was based on thorium-argon lamp spectra. The EW measurements of
spectral lines were done by applying two different methods: direct
integration of the line profile and Gaussian fitting. The latter is
preferable in the case of weak lines, but unsuitable for strong
lines in which the damping wings contribute significantly to the
EW. The final EWs are weighted averages of these two
measurements, depending on the line intensity (see Zhao et al. 2000
for details). We present part of the spectrum of a star
in the sample, HD\,212320, in Figure \ref{flux.eps}(a).

\section{Atmospheric parameters}
The stellar atmospheric model is specified by four parameters: effective temperature,
surface gravity, metallicity and microturbulent velocity. For more accurate atmospheric
parameters, we perform extinction correction on the observed magnitudes of these stars.

\subsection{Extinction}
The effective temperature of the stars are determined from photometric indices. Since
most of our stars are close to the Galactic plane ($|b|<30$\,deg), careful attention
must be paid to the reddening corrections.
Firstly we calculate the
reddening values $E(B-V)$ based on the dust maps of Schlegel et al. (1998).
Then we revise the $E(B-V)$ by
using the formula presented in Beers et al. (2002, their Eq.(1))
since the Schlegel et al. map may overestimate the reddening values
when their reported color excess exceeds about 0.15 mag (Arce \&
Goodman 1999) or 0.1\,mag (Beers et al. 2002).
Next, we can obtain
the extinction magnitudes in different wavelength bands, $A_{\rm B}$,
$A_{\rm V}$, $A_{\rm K}$, $A_{\rm b}$ and $A_{\rm y}$, by using the
relative extinction table of Schlegel et al. (1998) which is based
on the Galactic extinction law of Cardelli et al. (1989) and
O'Donnell (1994). Finally, the extinction-corrected colors $(B-V)$,
$(V-K)$ and $(b-y)$ are obtained. The distances $D$ (=1/$\pi$) are
estimated from the parallax. All these values are
presented in Table~\ref{E.table}.

\subsection{Effective temperature}

Effective temperatures ($T_{\rm eff}$) are determined from color
indices $(B-V)$, $(V-K)$ or $(b-y)$ by using the empirical
calibrations of Alonso et al. (1999, 2001) for giant stars.
Since the Alonso formula depends on [Fe/H], these calculations are done iteratively.
Figure~\ref{flux.eps}(b) shows the consistency of the
effective temperatures of the sample stars calculated from $(B-V)$,
$(V-K)$ and $(b-y)$. The solid line is the least-square fitting (with
a slope of 0.97) of $T_{\rm eff}$($V-K$) vs. $T_{\rm eff}$($B-V$),
and the dashed diagonal line is the one-to-one relation.

Alonso et al. (1999) showed that the effective temperatures
determined from $(b-y)$ are more accurate than those from $(B-V)$.
Unfortunately, only two stars with $(b-y)$ index are available in our
samples. For calculating abundances, we use the effective
temperature derived from $(B-V)$ rather than $(V-K)$ since the $K$
magnitude of many samples here are brighter than 4\,mag, and these
$K$-band indices often have larger uncertainties. Furthermore, half
of the errors for the $K$ magnitude here are higher than 0.2\,mag, and
a 0.2\,mag error in the $K$ band causes an error of about 250\,K in
effective temperature (for details see Liu et al. 2007). We should
keep in mind that the color $(B-V)$ is affected by the CN and
$C_{2}$, making the stars redder, and as a result, the temperature
derived from this color should be lower (Allen \& Barbuy 2006a).
Considering the uncertainties in spectral data, in [Fe/H], and the
errors in the calibration of color-temperature, we estimate that the
uncertainty in the
effective temperature is about 200\,K.

\subsection{Surface gravity}

There are two methods used to determine the gravity (log\,$g$). One is requiring
the \ion{Fe} {i} and \ion{Fe} {ii} lines to give the same iron abundance.
But it is well known that the derivation of iron abundance from
\ion{Fe} {i} and \ion{Fe} {ii} lines may be affected by many factors
such as unreliable oscillator strengths, non-local thermodynamic equilibrium
(non-LTE) effects and uncertainties
in the temperature structure of the model atmospheres. The other method is using
the parallaxes measured by the Hipparcos satellite.
This method is more precise for bright stars with Hipparcos parallaxes that are
as good as the
stars studied here. In this method the following relations are used:
$$
\log {g \over g_{\odot}}=\log {M\over M_{\odot}}
+4\log {{T_{\rm eff}}\over T_{{\rm eff}\odot}}+0.4(M_{\rm bol}-M_{{\rm bol}\odot})
$$
and
$$
M_{\rm bol}=V+5+5\log \pi +BC,
$$
where $M$ is the stellar mass, $M_{\rm bol}$ the absolute bolometric
magnitude, $V$ the visual magnitude, $BC$ the bolometric correction,
and $\pi$ the parallax. We adopt solar value log\,$g_\odot=4.44$,
$T_{{\rm eff},\odot}=5770$ K, and $M_{{\rm bol},\odot}$=4.75 mag. The
parallax ($\pi$) and its errors are taken from observations made by
the Hipparcos satellite (ESA 1997). Stellar mass was determined from
the position of the star in the $M_{\rm bol}$- log\,$T_{\rm eff}$
diagram. We adopt the stellar evolution tracks given by Girardi et
al. (2000). The bolometric correction, $BC$, is determined using the
empirical calibration of Alonso et al. (1999).

\subsection{Metallicities and microturbulent velocity}

\begin{figure}
\centering
\includegraphics [width=4.0cm]{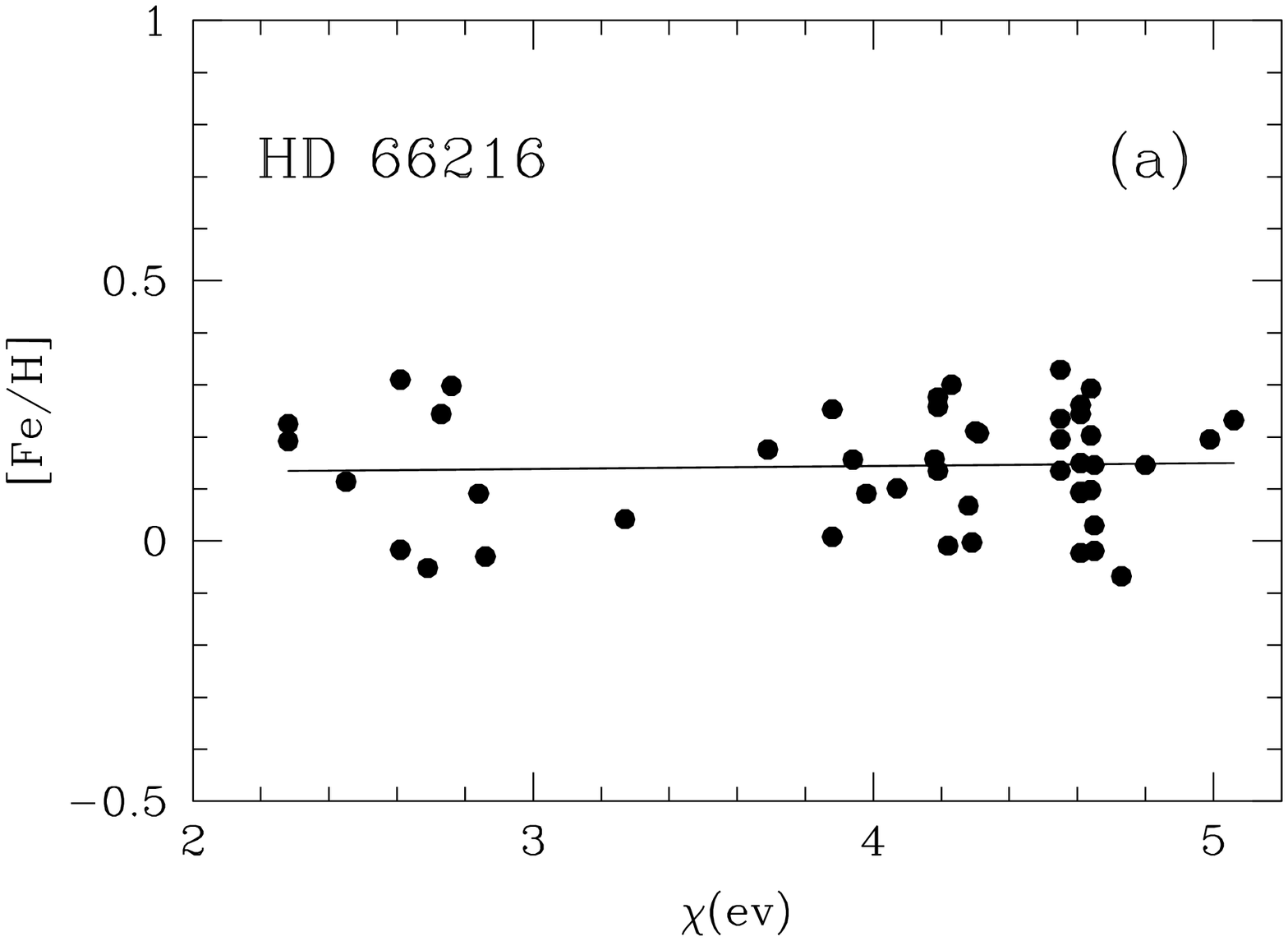}
\includegraphics [width=4.0cm]{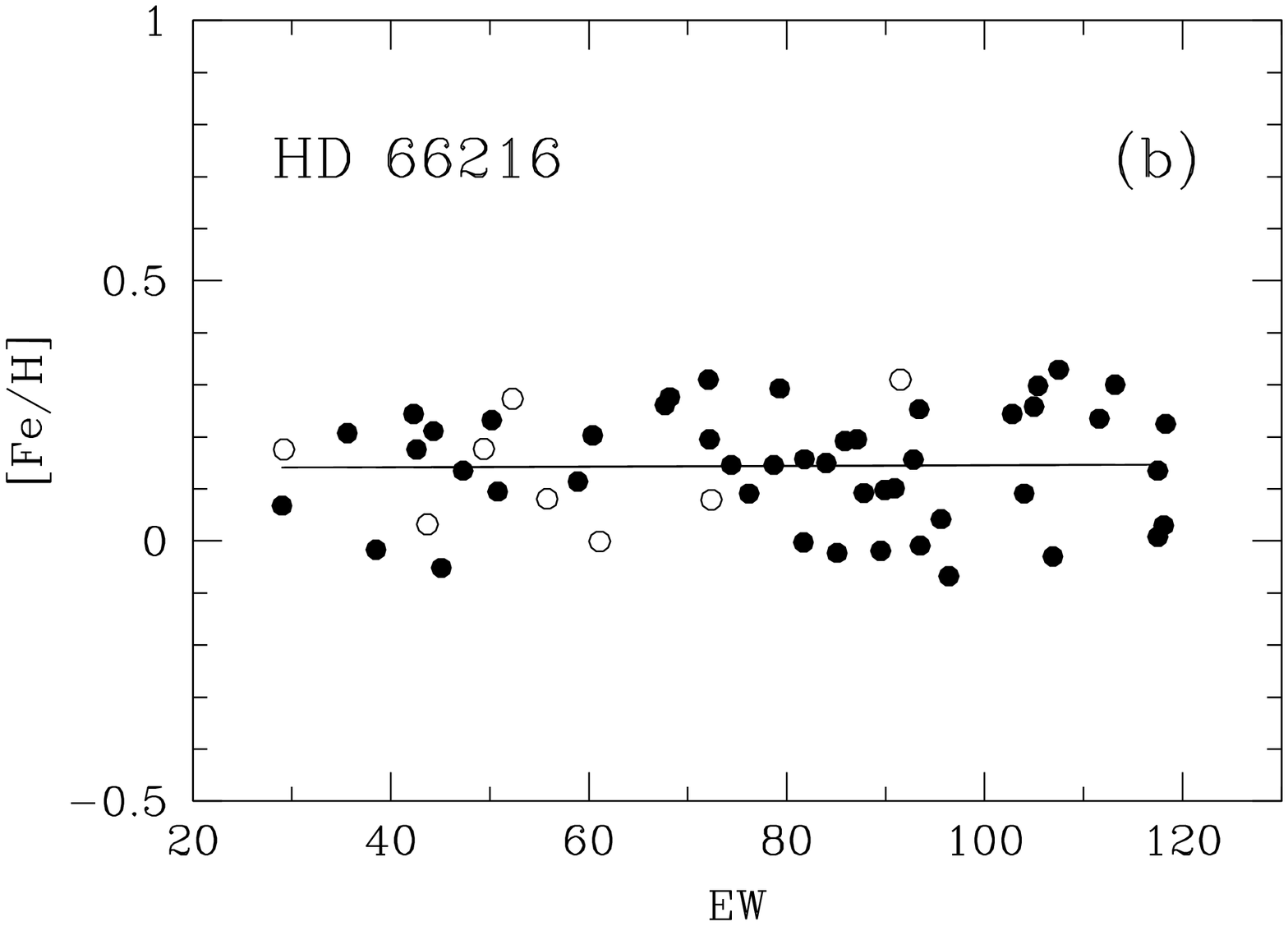}
\includegraphics [width=4.0cm]{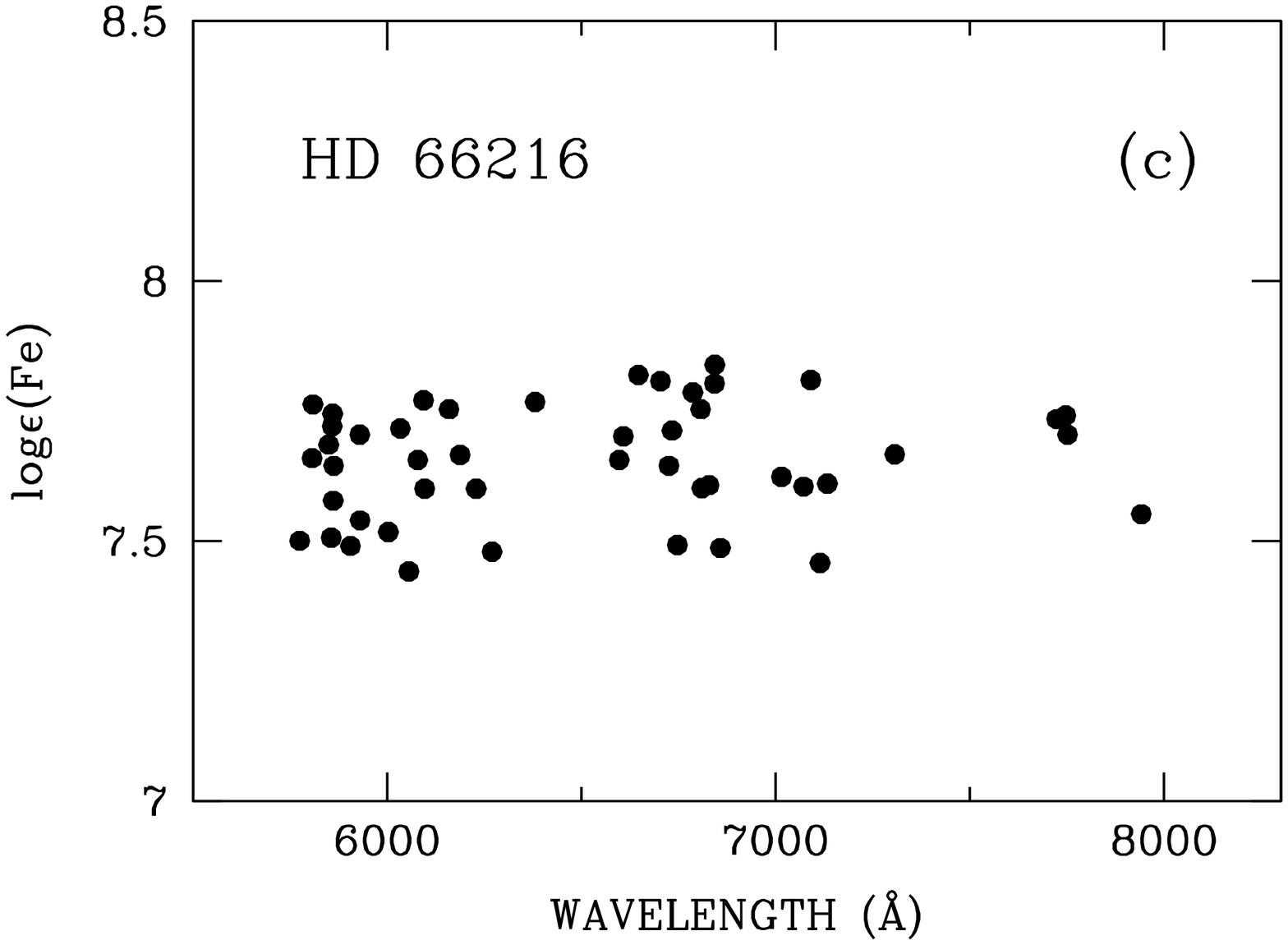}
\vspace{-1.0cm}
\caption {We check the atmospheric parameters $T_{\rm eff}$,
log\,$g$ and $\xi_{\rm t}$ of the stars by taking HD\,66216 as an
example: {\bf (a)} There is no significant trend of [Fe/H] vs.
$\chi$ (\ion{Fe} {i} line) and the excitation equilibrium suggests that the effective temperature we adopt is
reliable. {\bf (b)} The Fe abundances from \ion{Fe} {i} lines and
\ion{Fe} {ii} lines are consistent within 0.2 dex, which illustrates
the ionization equilibrium. It suggests that the log\,$g$ is
reliable. It also shows that there is no trend between Fe
abundances and EWs of the lines. This means that the $\xi _{t}$ is
reliable. The filled and open circles refer to \ion{Fe}{i} and
\ion{Fe}{ii} abundances respectively. The solid lines are the least
square fits of the data points. {\bf (c)} It shows no trend from the correlations
of Fe abundances (\ion {Fe}{i}) with wavelengths, and also indicates
the correctly chosen continuum levels. } \label{check66216.eps}
\end{figure}

The initial metallicities of the sample stars were taken from the literature
if available. Otherwise, we guessed an initial value by judging from the spectra
and the color indices. The final model and chemical composition were derived iteratively.
Microturbulent velocity ($\xi_{\rm t}$) was determined from the abundance analysis
by requiring a zero slope of zero in the relation [Fe/H] vs. EWs.
The derived atmospheric parameters of the sample stars are summarized in
Table~\ref{parameter} including star name, effective temperatures
calculated from $(V-K)$, $(b-y)$ and $(B-V)$, metallicity, surface
gravity and microturbulent velocity.
We checked the atmospheric parameters to make sure that the values we
obtained are reliable. As an example in
Figure~\ref{check66216.eps} we show for HD\,66216 that the abundance deduced from the \ion{Fe} {i}
lines is independent of the excitation potential (test of effective temperature $T_{\rm eff}$,
Fig.~\ref{check66216.eps}a), independent of the EWs of the lines (test of the
microturbulent velocity $\xi_{\rm t}$, Fig.~\ref{check66216.eps}b) and independent of the wavelength
(Fig.~\ref{check66216.eps}c).
In Figure~\ref{check66216.eps}(b) it can also be seen that the iron abundance deduced from \ion{Fe} {i}
lines is the same as
the abundance deduced from the \ion{Fe} {ii} lines (test of gravity log\,$g$).

As a comparison, we have six stars in common with McWilliam (1990). He observed 671 giants but in a narrow
spectral range (6500 - 6850\,\AA) and obtained the atmospheric parameters and elemental abundances.
In Table~\ref{parameter} we also list the atmospheric parameters of the stars
also studied by McWilliam (1990) in the last four columns.
The median differences between ours and theirs are $\Delta$T$_{\rm eff}$=48\,K,
$\Delta$log\,$g$=$-$0.54, $\Delta$[Fe/H]=0.14 and $\Delta$$\xi_{t}$=$-$0.75.

\section{Model atmospheres and spectral lines}

We carried out a classical LTE abundance analysis based on a set of plane parallel, line blanketed,
flux constant model atmospheres (Kurucz 1993). Abundances are determined by using the atmospheric
parameters given in Table~\ref{parameter} ($T_{\rm eff}(B-V)$,
log\,$g$, [Fe/H] and $\xi_{\rm t}$) and the EWs of the lines.
For Fe we selected the lines with $\rm 20< EW <120$\,m\AA, and  $\rm 20< EW <200$\,m\AA~ for the other elements,
because the weaker lines would increase random errors (and possibly some systematic overestimates),
while the stronger lines are very sensitive to the microturbulent velocity and the damping.
Great care was taken to insure consistency of the continuum level.

In Figure~\ref{check66216.eps}(c), as an example,
it can be seen that for HD\,66216 the abundance deduced from the \ion{Fe} {i} lines does not depend on wavelength.
The line list of Liu et al. (2009, their table\,2) and Liang et al. (2003, their table\,3) has been adopted with
their excitation potential $\chi$ and oscillator strengths log$gf$.

As mentioned in Liu et al. (2009), the oscillator strengths
log$gf$ of the spectral lines are taken from the NIST database
($http://physics.nist.gov$), Lambert \& Warner (1968), Weise \& Martin
(1980), Bi$\acute{e}$mont et al. (1981, 1982), Hannaford et al.
(1982), Fuhr et al. (1988), Luck \& Bond (1991), O'Brian et al.
(1991), Bard \& Kock (1994), Lambert et al. (1996), Nissen \&
Schuster (1997), Chen et al. (2000), Liang et al. (2003) and the
references therein. The log$gf$ values of the O~{\sc i} 7771-7774
triplet lines and of the neutron capture ($n$-capture) process elements were taken from
Liang et al. (2003, their table\,3).

\section{Chemical Abundances}
\subsection{The abundance and their uncertainties}

The abundances of 17 elements (O,
Na, Mg, Al, Si, Ca, Sc, Ti, V, Cr, Mn, Ni, Y, Zr, Ba, La, Eu) in our 19 stars are given in
Table~\ref{abundance1}.
The solar abundances are taken from Grevesse \& Sauval (1998).
Uncertainties in the abundances mainly come from
the systematic errors introduced by uncertainties in the atmospheric parameters and the stochastic
error arising from random errors in the oscillator strengths, the damping constants and in the measurements
of EWs. We ignore the uncertainties in atomic data.

In Table~\ref{error.table1} and \ref{error.table2} we computed the variation of [Fe/H] and [X/Fe]
(where X is one of the 17 elements considered here) for a change of 200\,K in the effective temperature,
0.2\,dex in the surface gravity and $\rm0.2\,km/s$ in the micro turbulence velocity, for two typical stars in
our sample, HD\,66216 and HD\,212320.
We expect that the uncertainties in the EW measurements are about $\pm 5$ m\AA. The resulting error in the
abundances depends on the EW of the line and has been computed line by line.
For an element represented by $N$ lines, the final error on the abundances is divided by a factor $\sqrt {N}$ .
The total uncertainty is estimated with the $Least~Squares~Method$ (LSM) and given in Table
\ref{error.table1} and \ref{error.table2}.

We should keep in mind that since we are using the red spectra for abundance analysis, we only have limited
spectral lines of the $n$-capture process elements to use.
For example, only one line of \ion {Y}{i} (6435.000\,\AA) is used, but the \ion {Y} {ii} lines can result
in higher abundances than the \ion {Y}{i} line (Gratton \& Sneden 1994; Yushchenko et al. 2004). There are four lines
of \ion {Zr}{i} used to derive Zr abundances, but \ion {Zr} {ii} line can result in higher abundances than
\ion {Zr}{i} line (Allen \& Barbuy 2006a; Gratton \& Sneden 1994; Yushchenko et al. 2004). There are three lines of
\ion {Ba} {ii} (5853.688, 6141.727 and 6496.908\,\AA) used to derive Ba abundances and two lines of \ion {La} {ii}
(6390.480 and 6774.330\,\AA) used to derive La abundances. The line of \ion {Eu} {ii} at 6645.110\,\AA~ is used to derive
Eu abundances for our sample stars.

\subsubsection{[X/Fe] vs. the Atomic number Z}

\begin{figure*}[b]
\centering
\includegraphics[width=4.0cm]{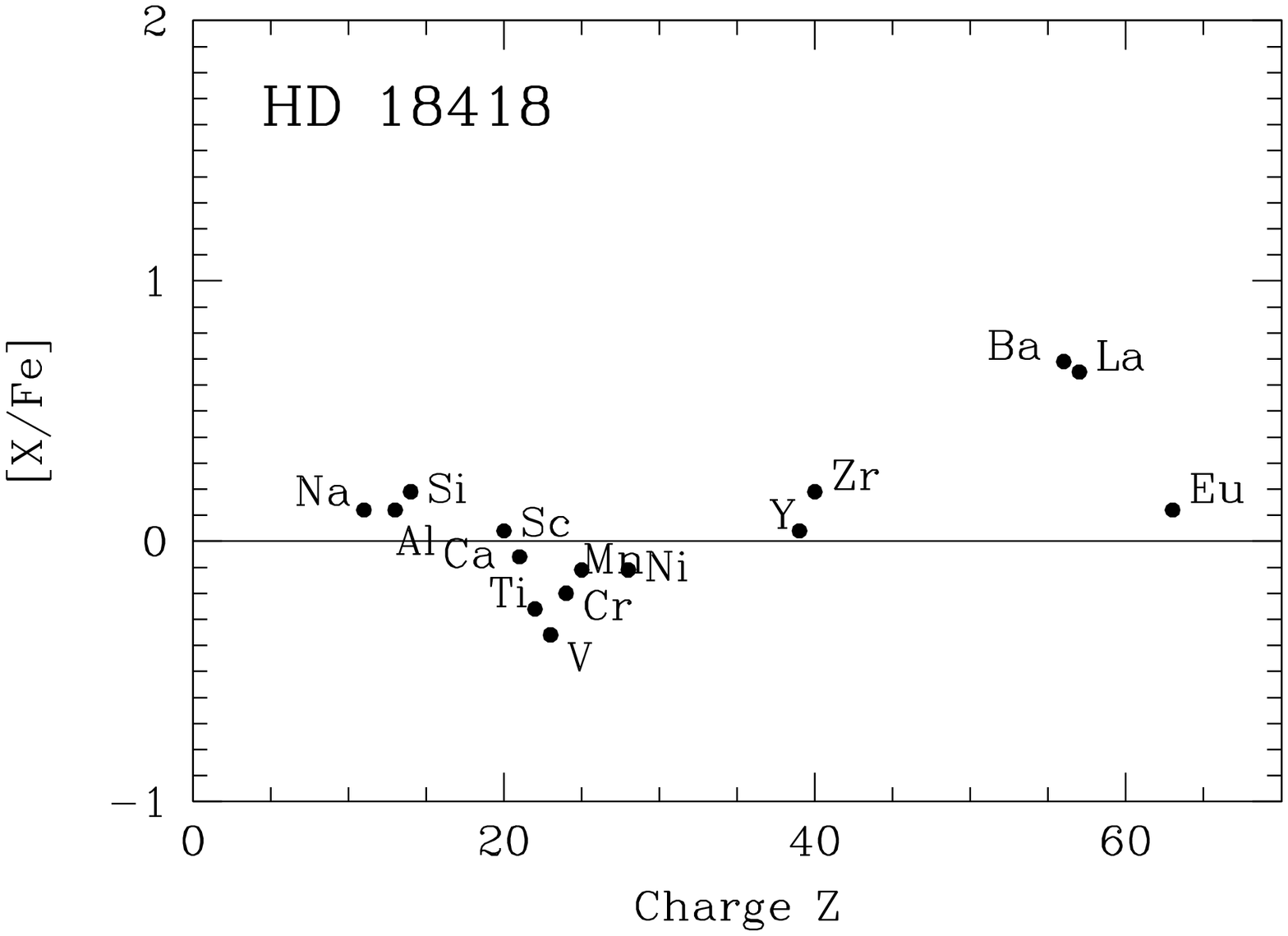}
\includegraphics[width=4.0cm]{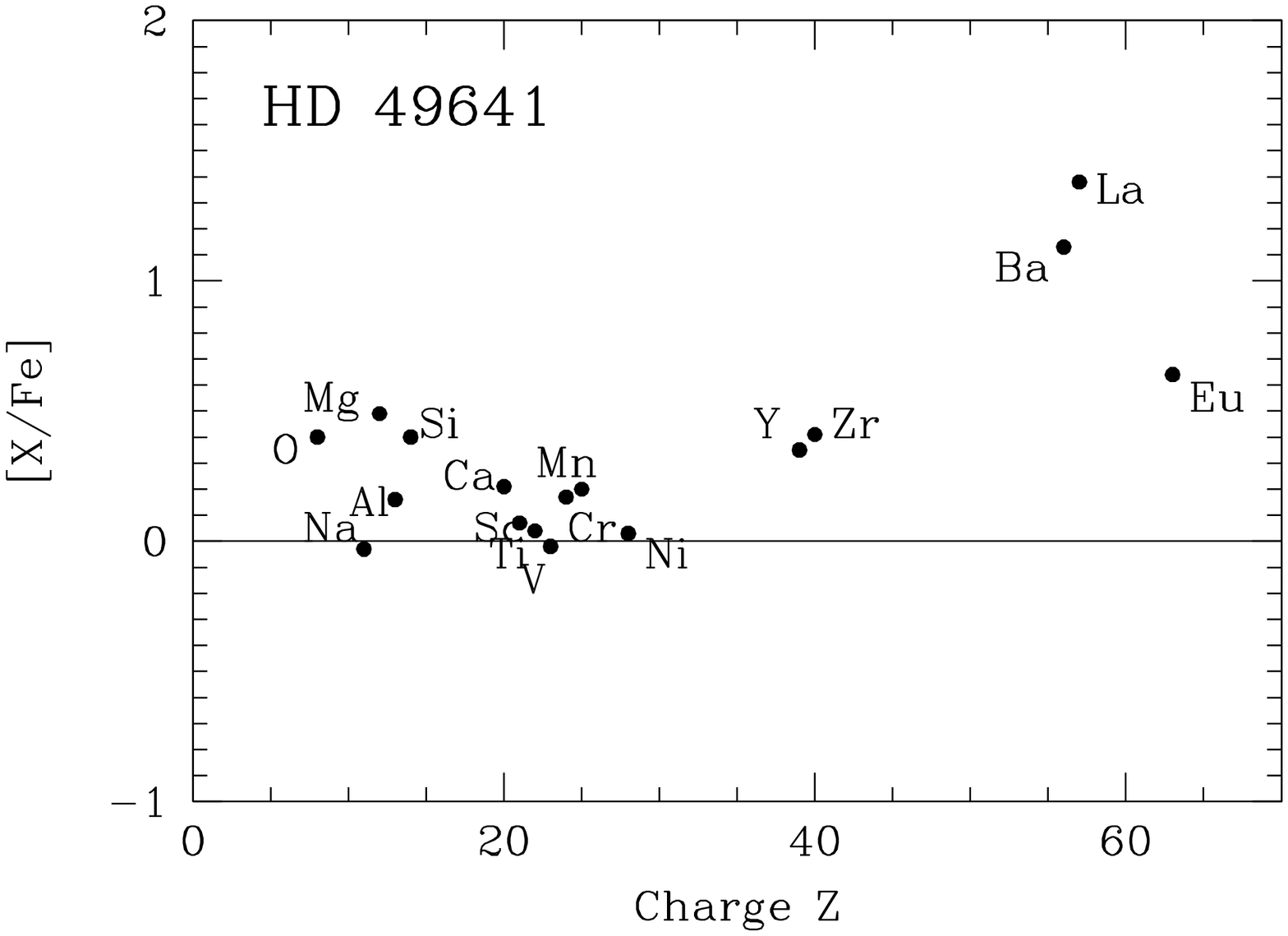}
\includegraphics[width=4.0cm]{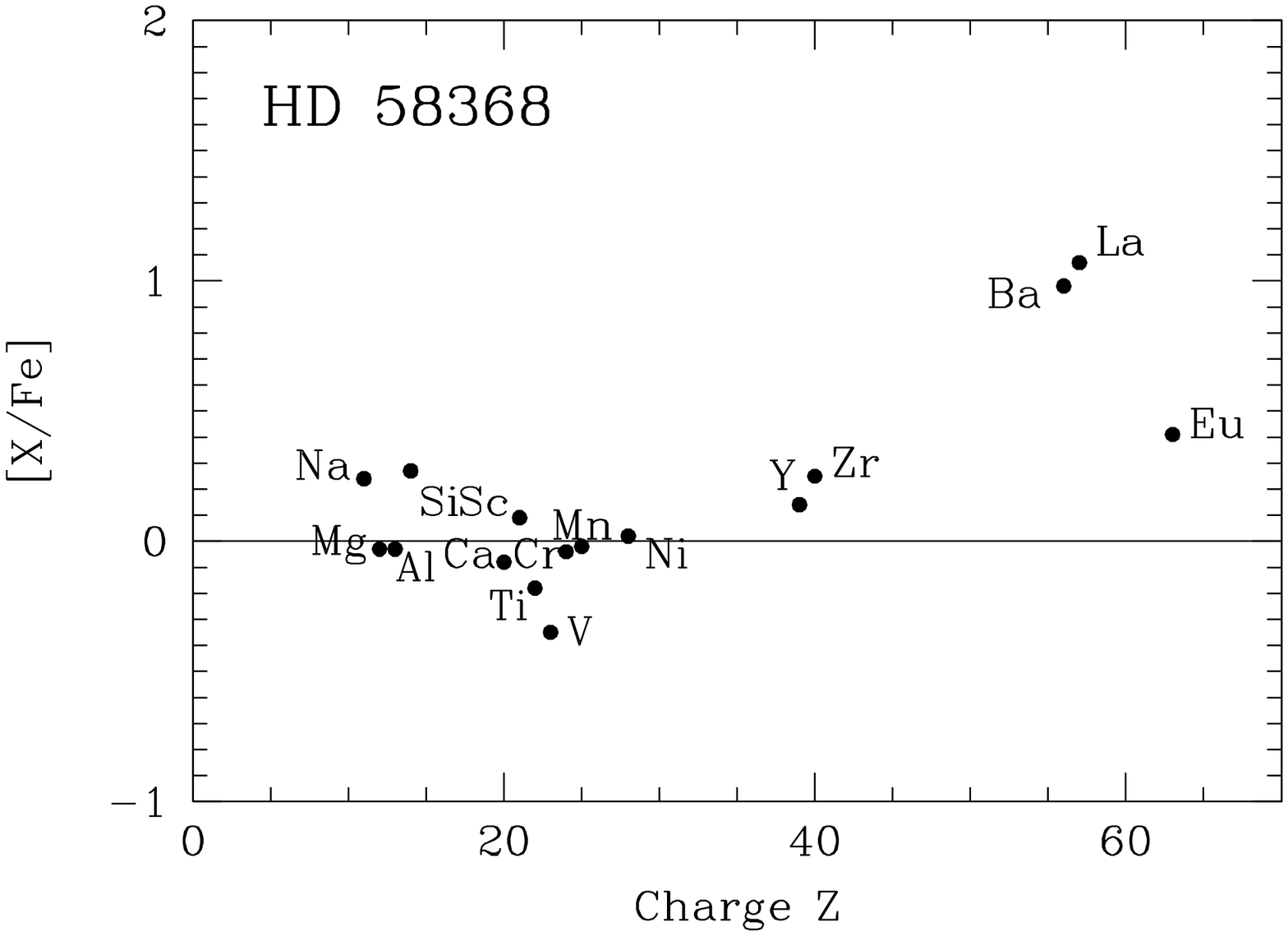} \\  \vspace{-1.cm}
\includegraphics[width=4.0cm]{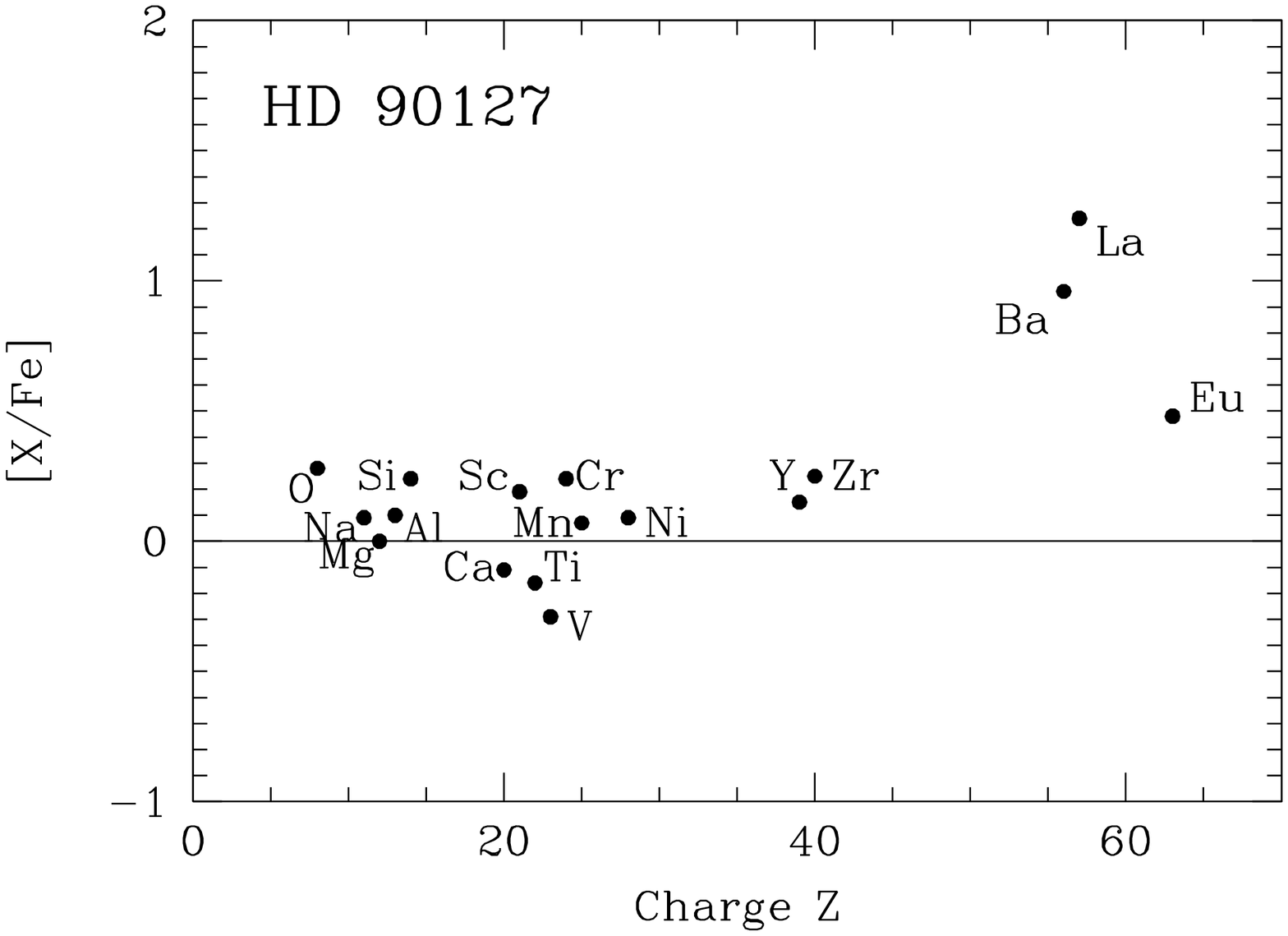}
\includegraphics[width=4.0cm]{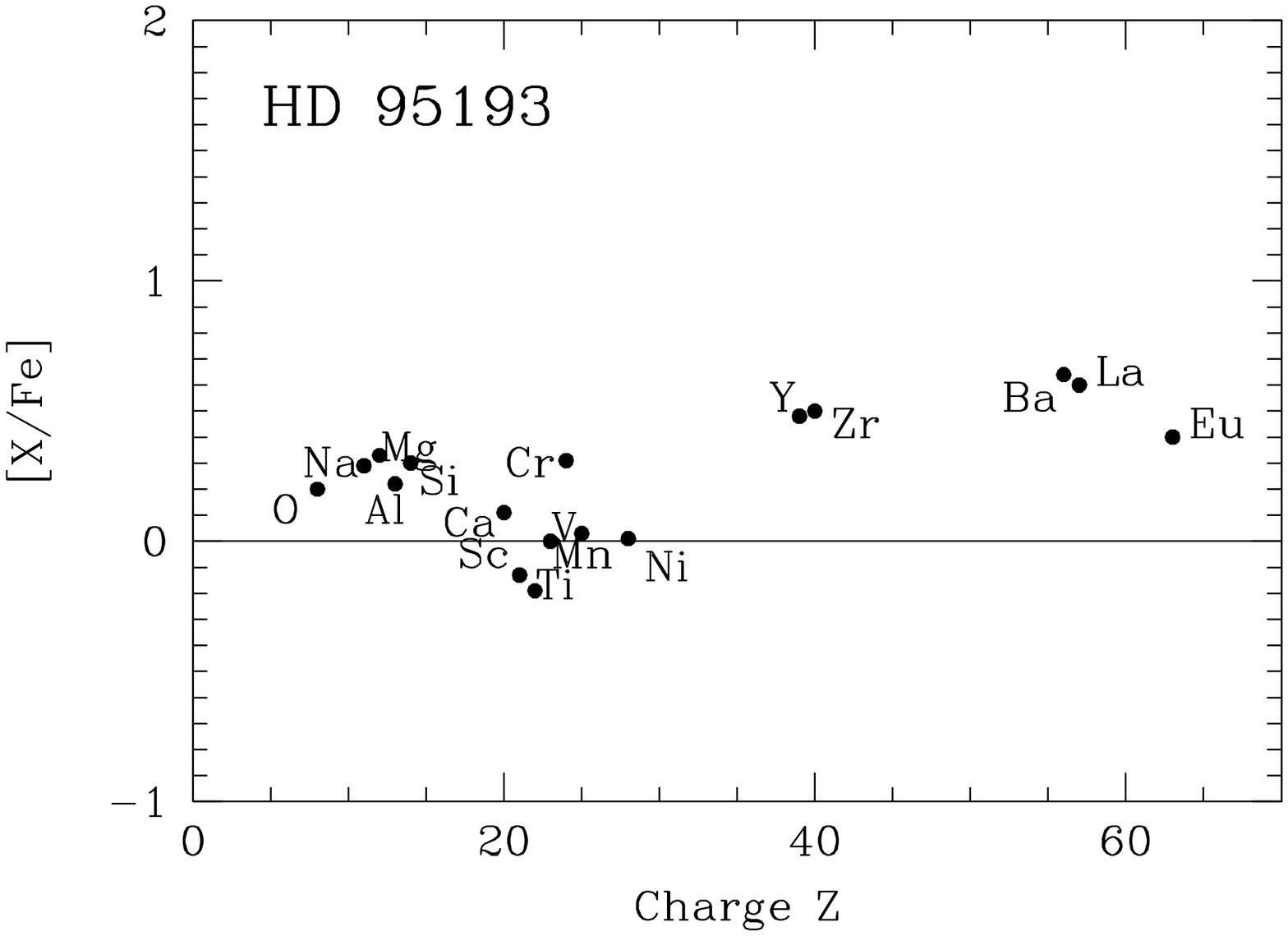}
\includegraphics[width=4.0cm]{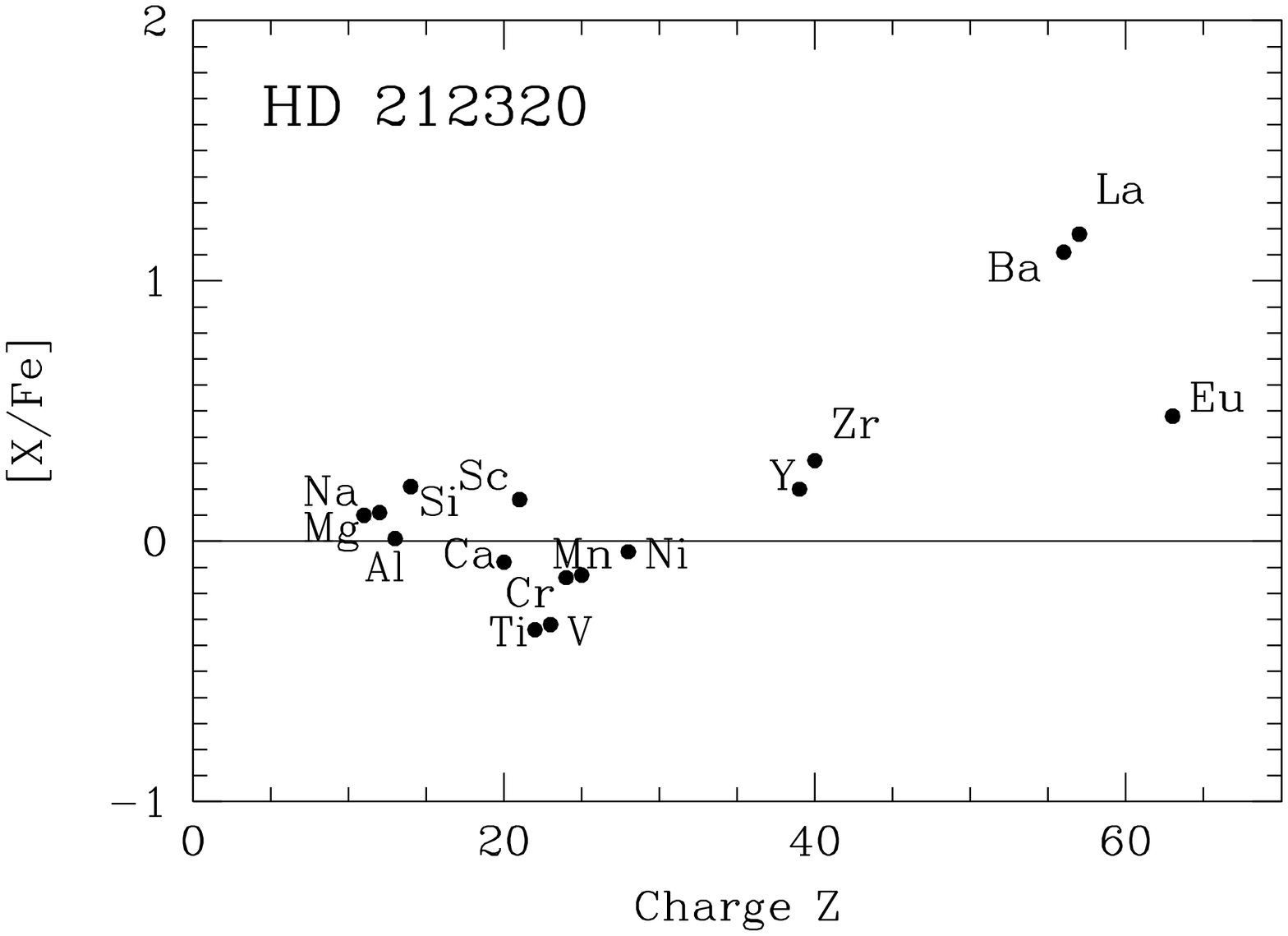}  \\ \vspace{-1.cm}
\includegraphics[width=4.0cm]{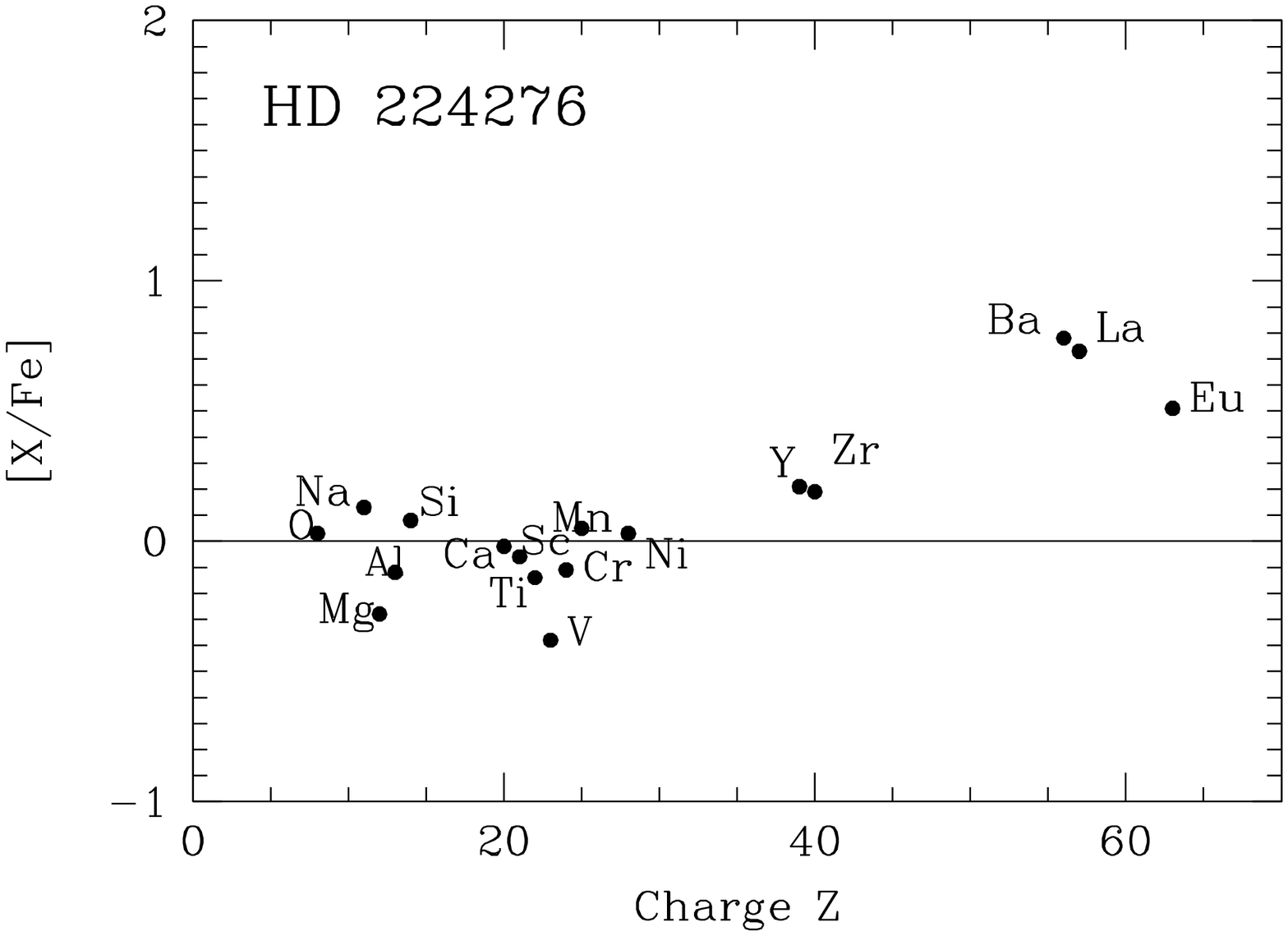}
\includegraphics[width=4.0cm]{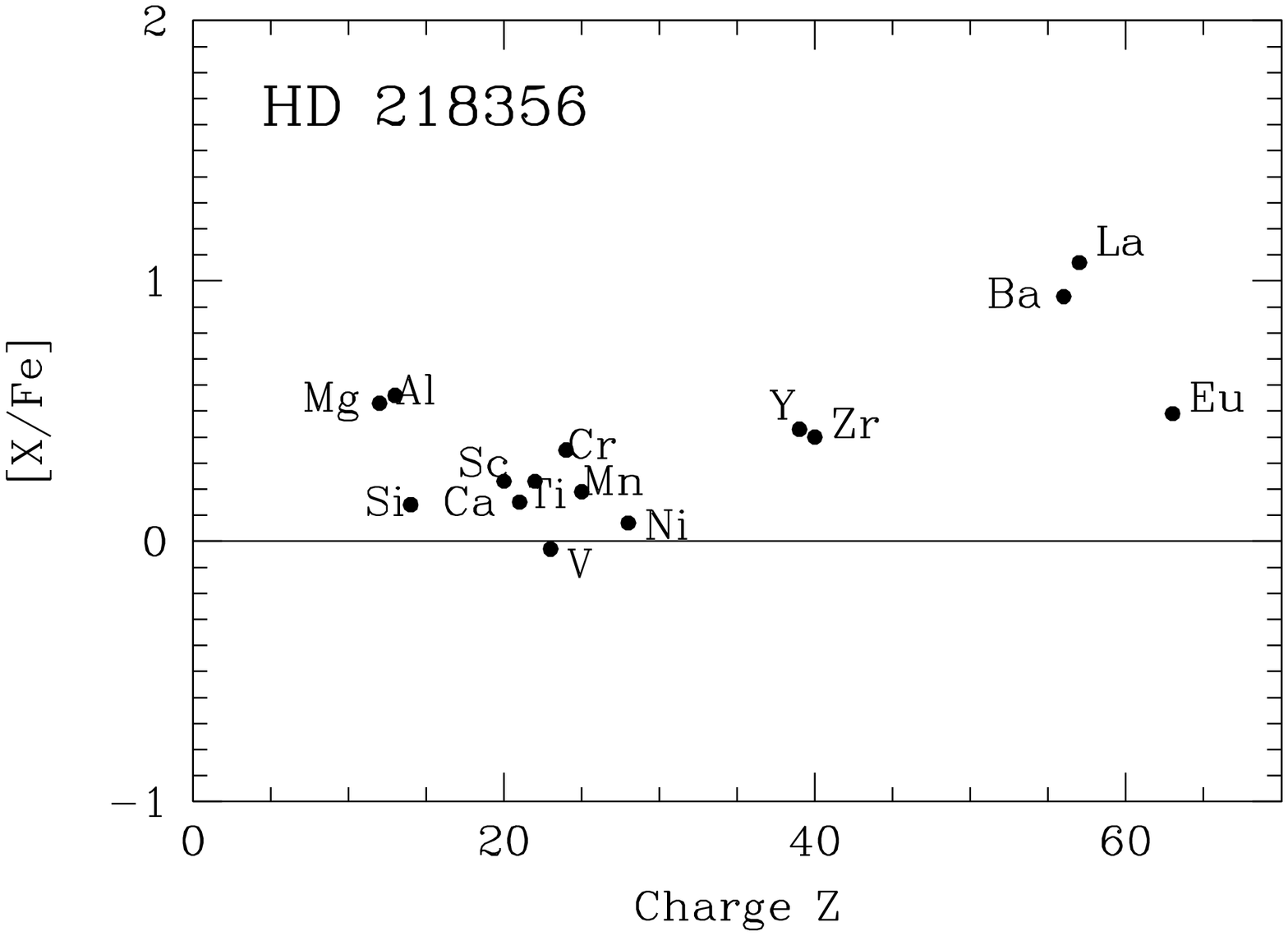}
\includegraphics[width=4.0cm]{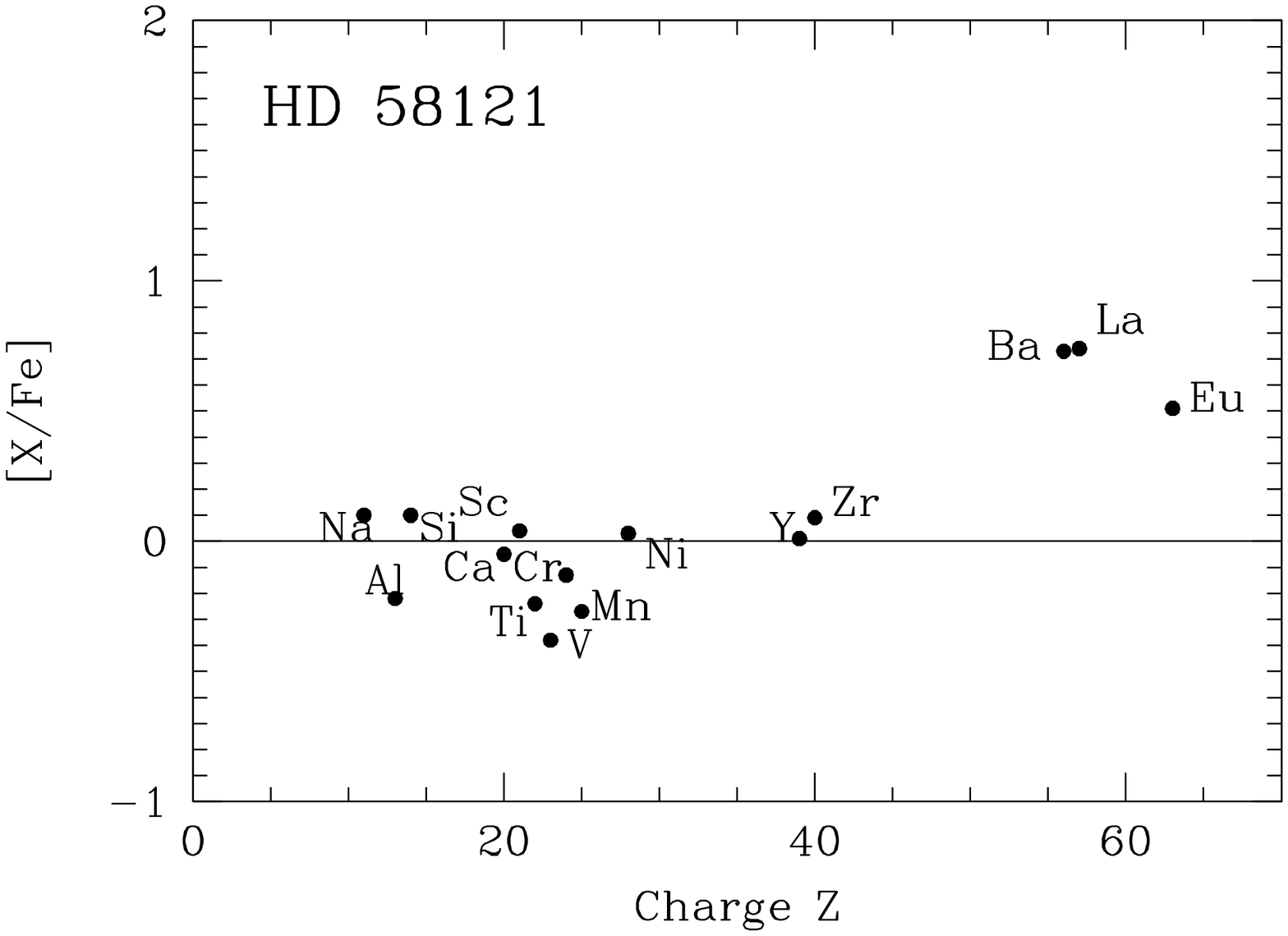}   \\   \vspace{-1.cm}
\includegraphics[width=4.0cm]{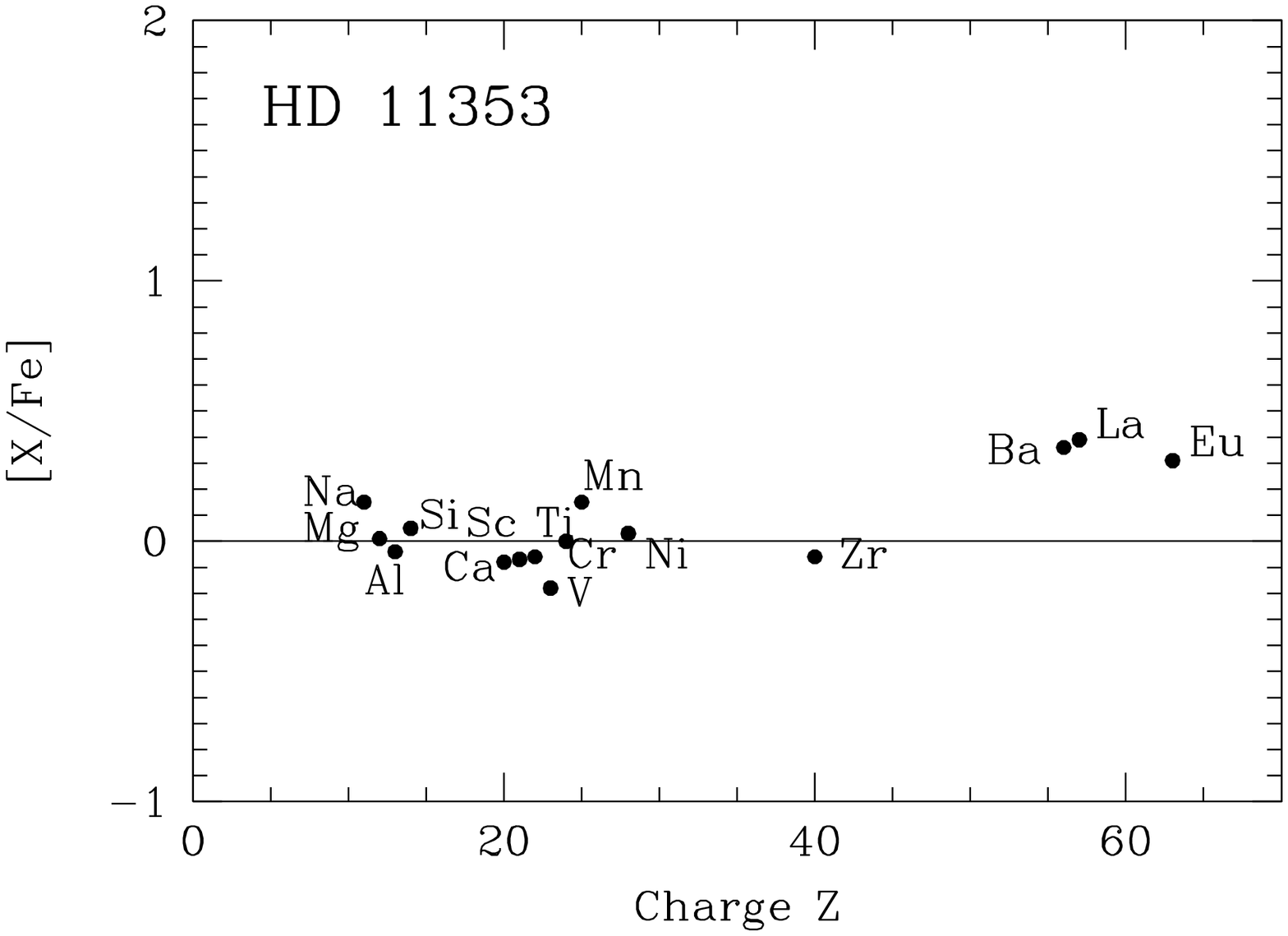}
\includegraphics[width=4.0cm]{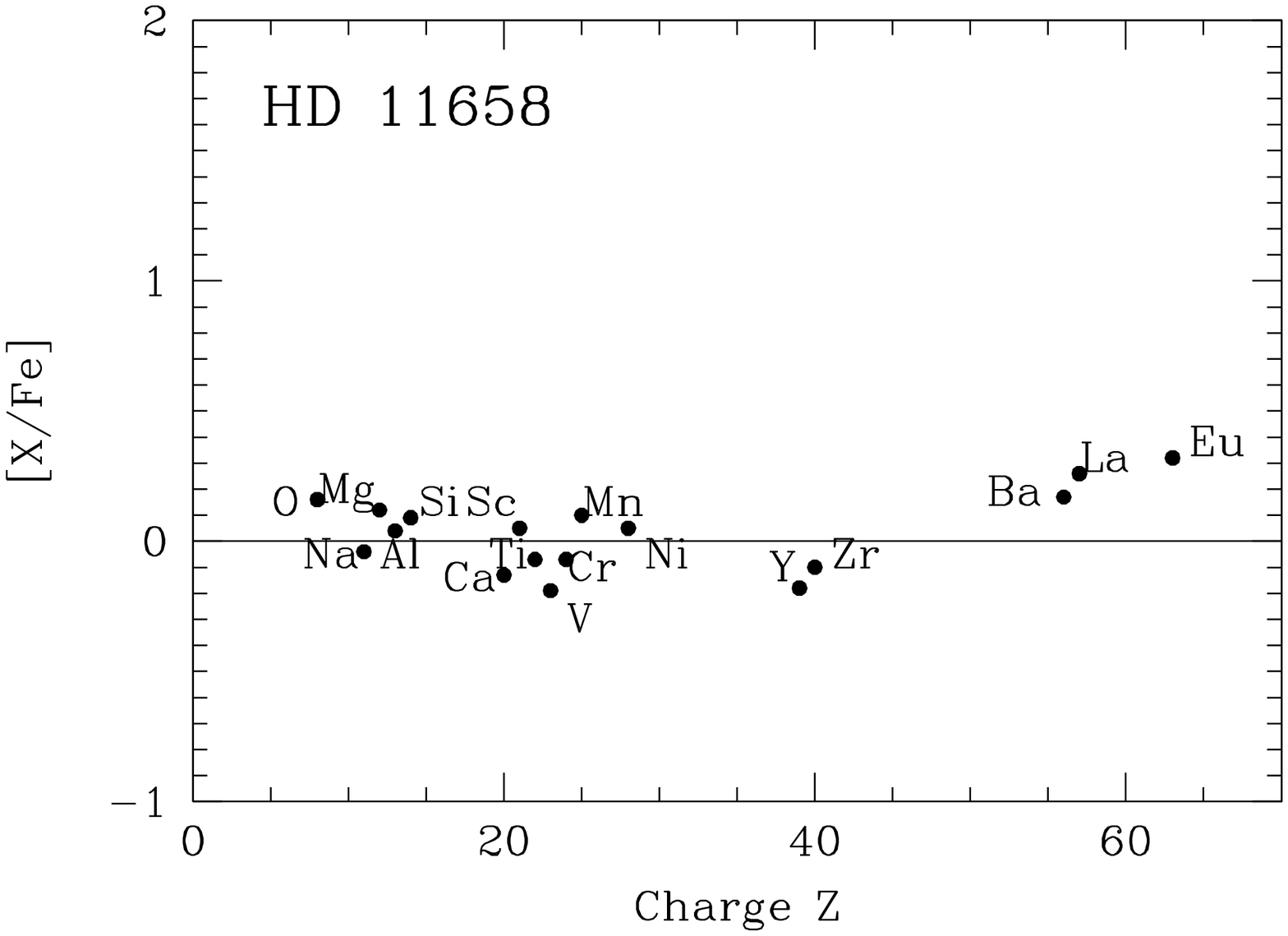}
\includegraphics[width=4.0cm]{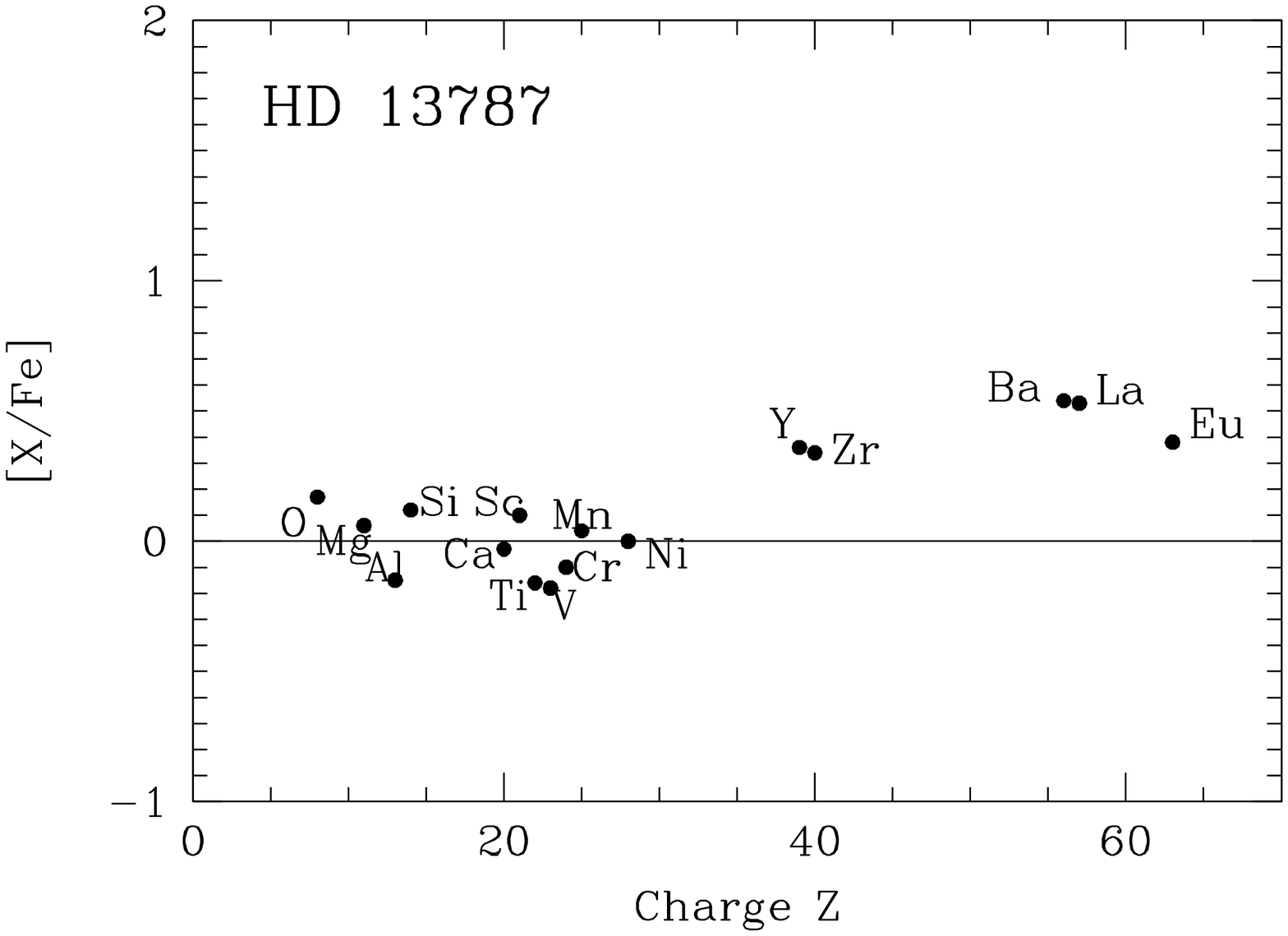}   \\   \vspace{-1.cm}
\includegraphics[width=4.0cm]{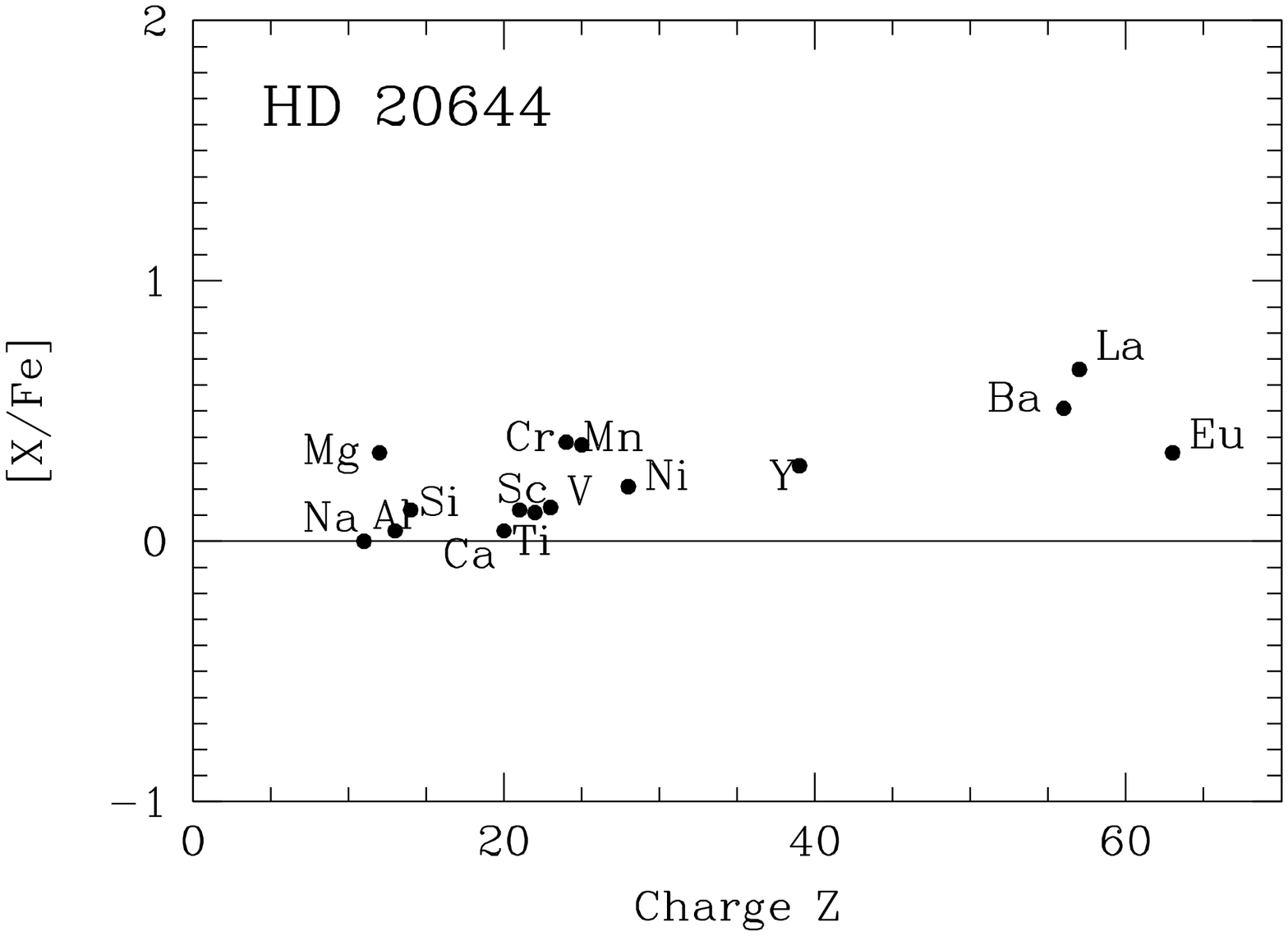}
\includegraphics[width=4.0cm]{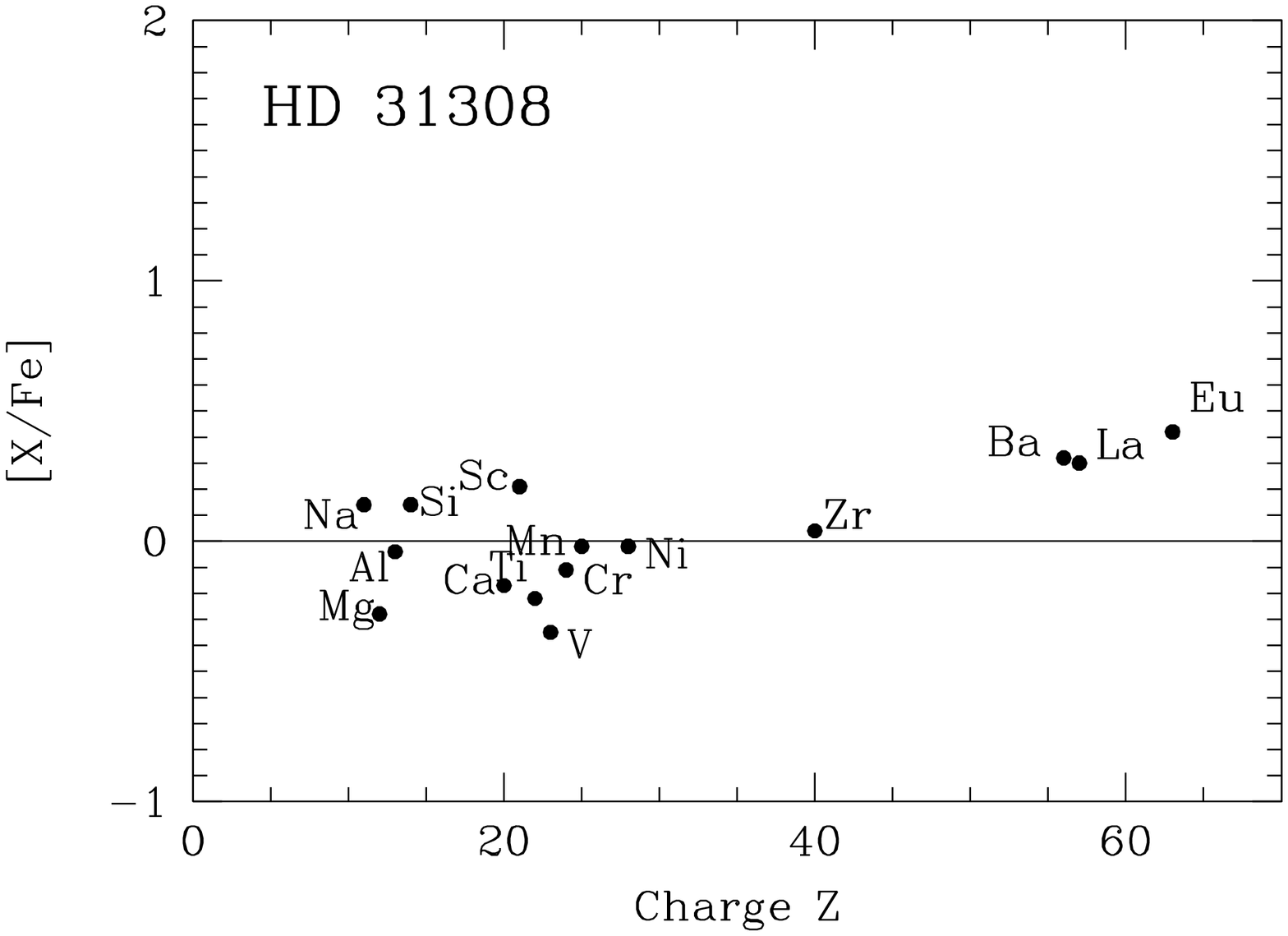}
\includegraphics[width=4.0cm]{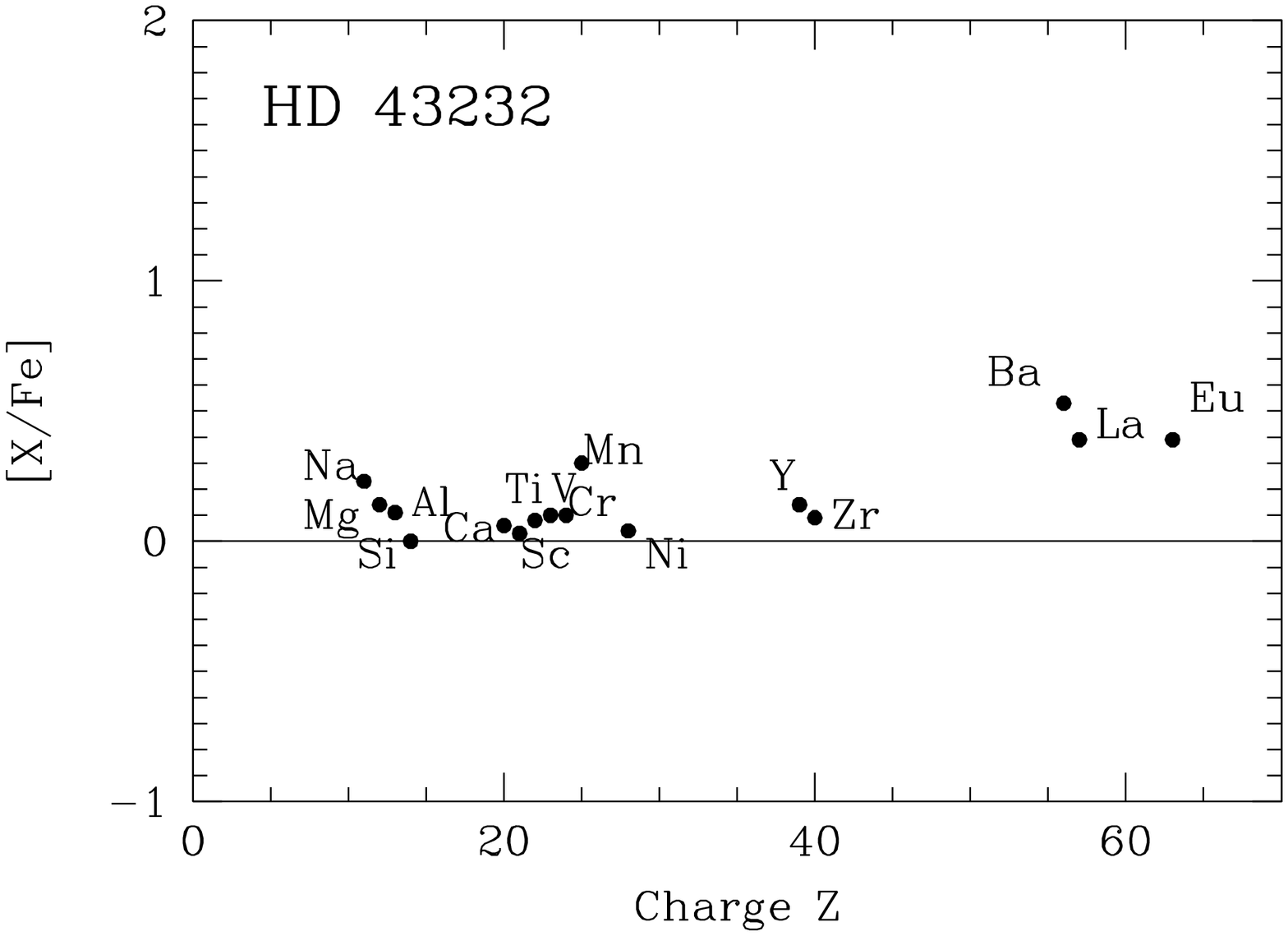}   \\   \vspace{-1.cm}
\includegraphics[width=4.0cm]{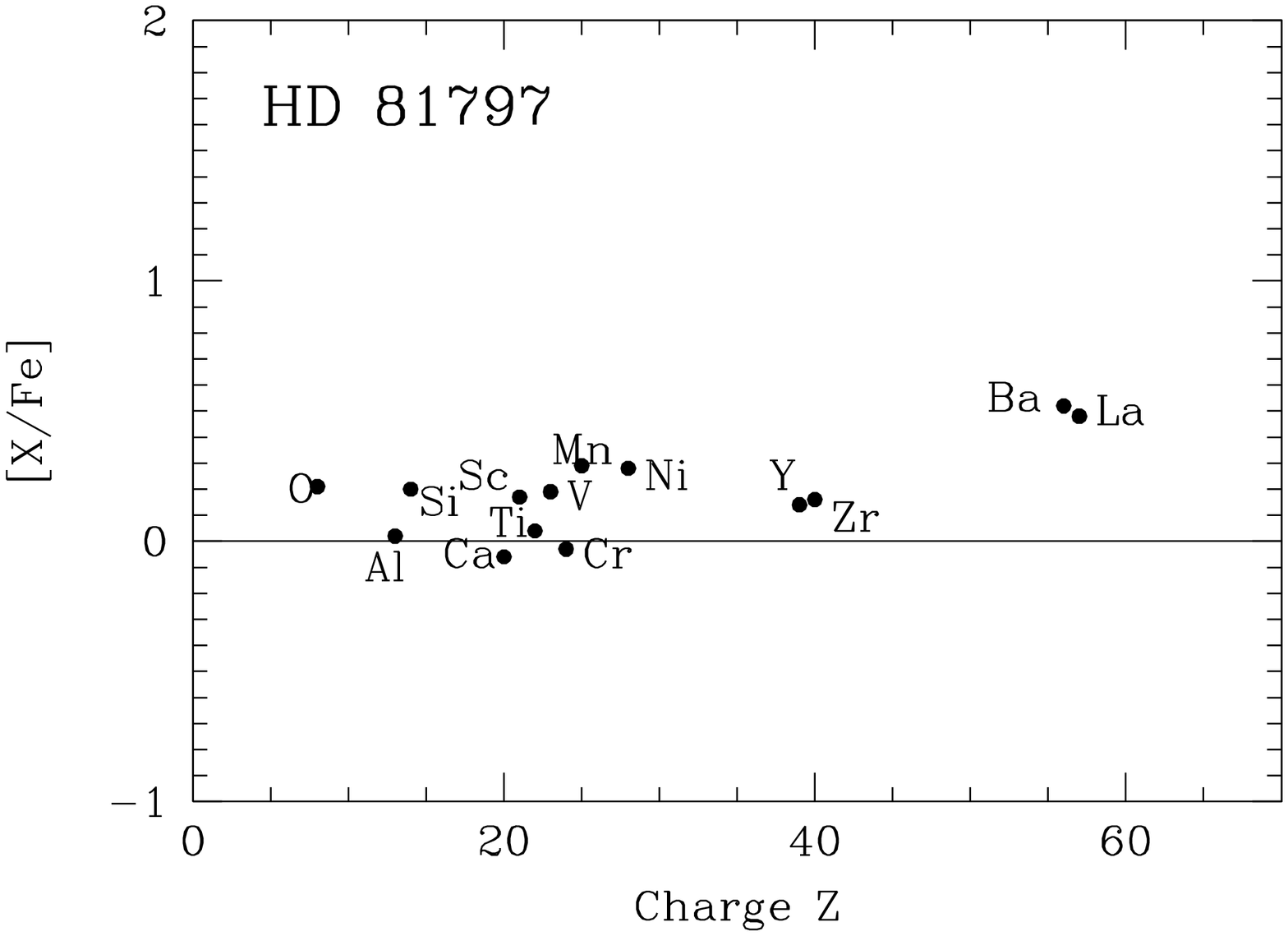}
\includegraphics[width=4.0cm]{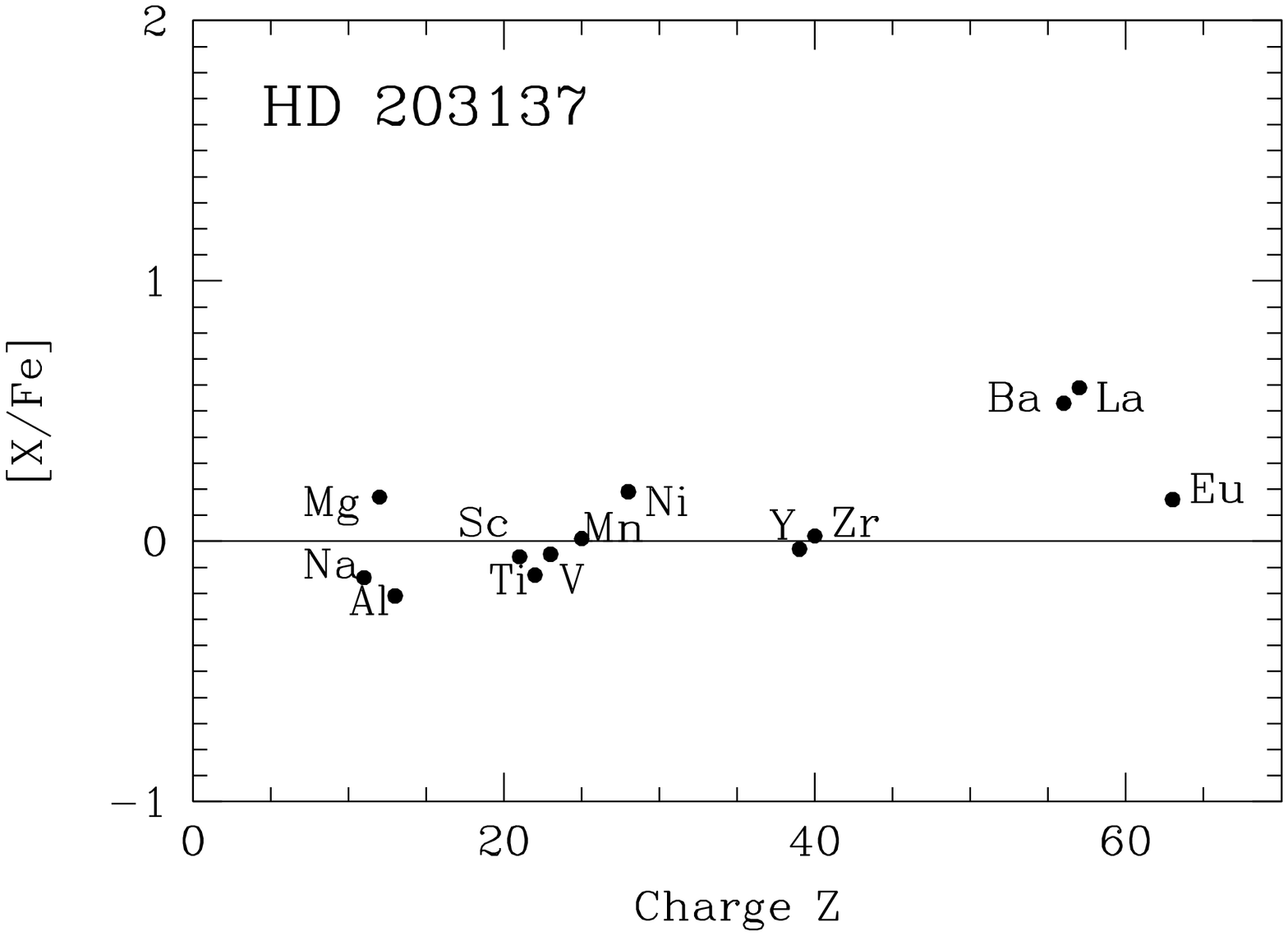}
\includegraphics[width=4.0cm]{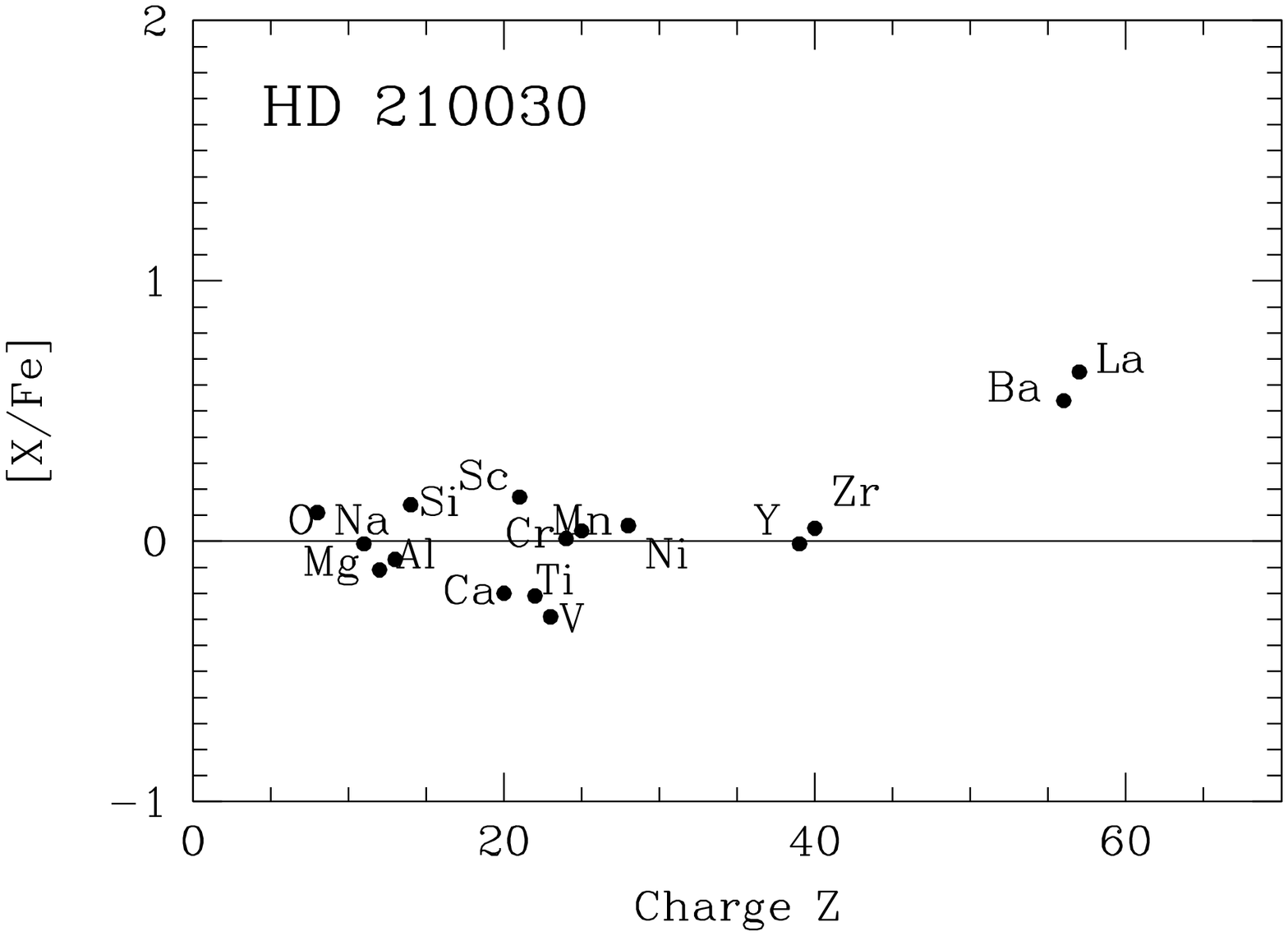}   \\   \vspace{-1.cm}
\includegraphics[width=4.0cm]{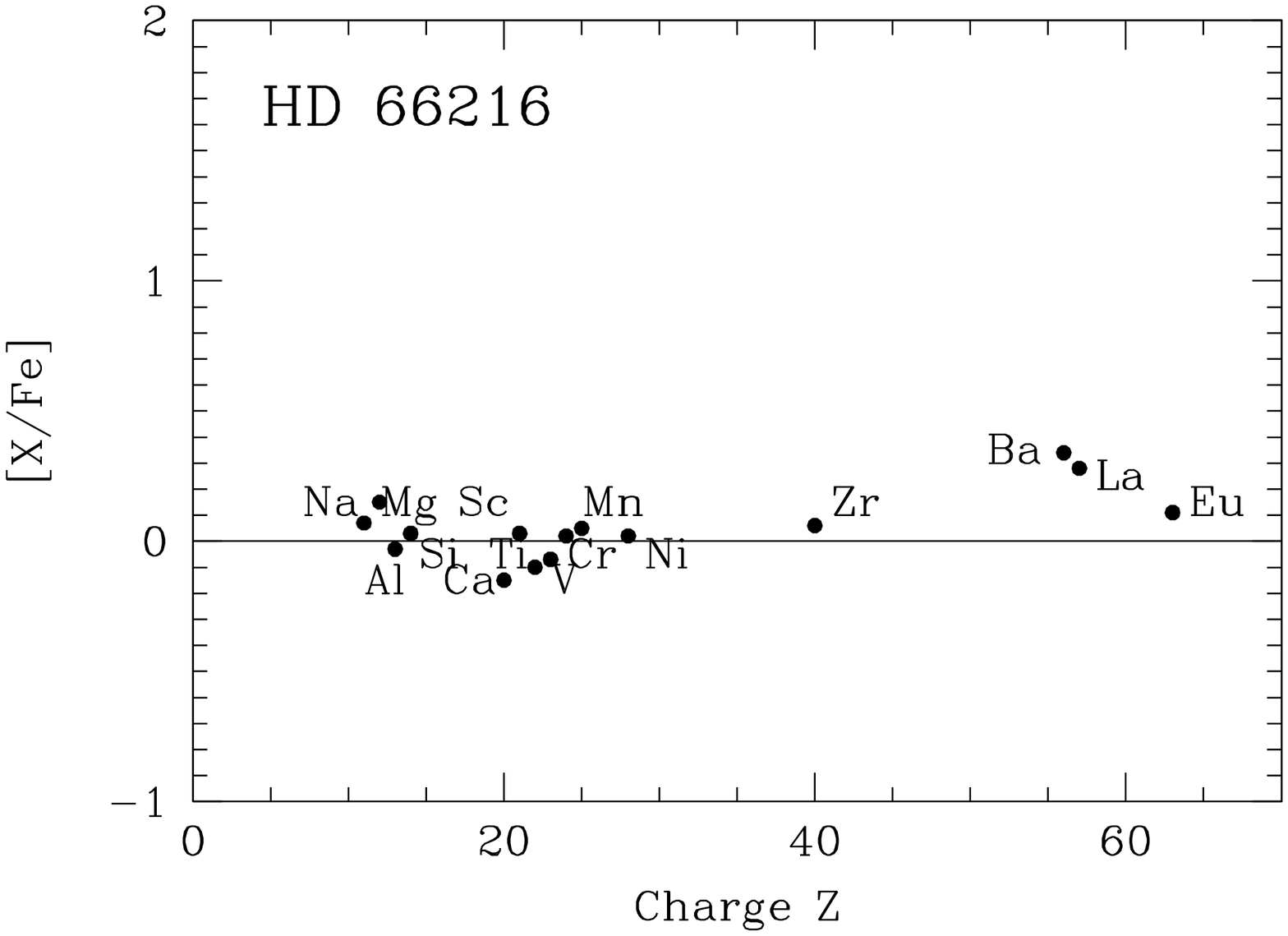}
\vspace{-1.cm}
\caption{Abundances of the 19 sample stars: (1) the top nine panels are for the strong Ba stars with
[Ba/Fe]$>$0.6, and the top six among them have Ba intensity Ba$>$2, but HD\,218356 has Ba=1.5 and
HD\,58121 has Ba=1.0; (2) the remaining ten panels are for the 10 mild Ba stars with Ba intensity of Ba$\leq$1.0. } \label{abun.eps}
\end{figure*}

Table~\ref{abundance1} presents the abundances
$\log \epsilon(X)$ (in the usual scale log$X$(H)=12.0) and the
corresponding [X/Fe] values for our sample of Ba stars.
The two full tables will be published online.
In Figure~\ref{abun.eps} we have plotted [X/Fe] vs. the atomic number Z for the 19 stars of our sample.
The abundance pattern of the elements from O to Ni is similar to the solar abundance pattern ([X/Fe] close to zero
although with some scatter), but the $n$-capture elements Y, Zr, Ba, La, Eu are overabundant relative to the Sun and
the overabundances are different in strong and mild Ba stars.
As commented in Lu (1991), the Ba intensity class was estimated visually according to Warner's procedure
(Warner 1965)  on a scale of 1 (weakest) to 5 (strongest).  Lu (1991, sect. 4.1) called Ba stars with a Ba
intensity between 2 and 5 ``strong'', and Ba stars with a Ba intensity between 0.3 and 1.5 ``weak''. More recently,
Jorissen et al. (1998), called Ba stars with a Ba intensity $\leq 2$ ``mild'', and Ba stars with a Ba intensity
of 3, 4 or 5 ``strong''.

In this work, we could roughly divided the sample stars as strong Ba stars with Ba intensity Ba=2-5 and mild
(weak) Ba stars with Ba$<$2. However, we will consider [Ba/Fe] abundance ratios as a stricter diagnostic for
strong or mild Ba stars.
The Ba intensities of the 19 stars in our sample are given in Table~\ref{simbad}(second column).
It shows that there are six stars with Ba=2-5
(HD\,18418, HD\,49641, HD\,58368, HD\,90127, HD\,95153, HD\,218356).
The Ba intensity of
HD\,224276 is not given by Lu (1991) but we identify it as a strong
Ba star since its heavy elements show a pattern similar to HD\,95193 and HD\,218356
(which are both strong Ba stars with Ba=2.0\,m), and its [Ba/Fe] is up to 0.78,
which is much higher than the solar value. Thus we suggest its Ba intensity could be assumed to be 2.0.
For HD\,212320 and HD\,58121, although their Ba intensities are 1.5 and 1.0, but their abundances are up
to [Ba/Fe] = 1.11 and 0.73, respectively. Thus we also classify them as ``strong Ba stars''.
Therefore, to diagnose strong Ba stars, we consider their [Ba/Fe] $>$ 0.60.
Then nine of the sample stars are strong Ba stars.

The top nine panels of Figure~\ref{abun.eps} show the abundance patterns of these nine strong Ba stars in our sample.
The $n$-capture elements in these nine strong Ba stars show obvious overabundance relative to those in the Sun.
The remaining ten panels of Figure~\ref{abun.eps} show the abundance patterns of ten mild Ba stars in our sample.
Among them, the
``Ba-like'' star HD\,66216 has
a pattern similar to the mild Ba stars HD\,11353 with Ba intensity of 0.1 and
HD\,31308 with Ba intensity of 0.2. Thus we classify
this Ba-like star as a ``mild Ba star'' and assume its Ba intensity is 0.2. These three stars have [Ba/Fe] of
0.34, 0.36 and 0.32, respectively.

Figure~\ref{abun.eps} shows that:

\noindent--In the strong Ba stars, the first peak $s$-process elements Y and Zr are
overabundant (unlike the $\alpha$- and iron peak elements).
The second peak $s$-process elements Ba and La, and the $r$-process element Eu are more overabundant
than the first peak ones.

\noindent--In the mild Ba stars the first peak elements are not overabundant, and have the same behavior
as the $\alpha$- and iron-peak elements. The second peak of $s$-process elements (Ba and La) is higher
than the first peak elements, but is less overabundant than that in the strong Ba stars.

Y and Zr belong to the first peak of the $s$-process (corresponding to the neutron magic number 50),
and the core He and shell C burning in massive stars mainly produce these elements
(Raiteri et al. 1991a, 1991b, 1992, 1993; Qian \& Wasserburg. 2007).
Ba, La and Eu belong to the second peak. Ba and La (corresponding to the neutron magic number 82)
are mainly generated during
shell He burning in AGB stars (Busso et al. 2001),
whereas the element Eu is dominantly produced through $r$-pocess during a supernova explosion (Sneden et al. 2008).
As mentioned in Busso et al. (2001, their Sect.2.1), those nuclei whose locations correspond to neutron magic numbers
(N = 50 and 82) are mainly of $s$-process origin, and act as
bottlenecks for the $s$-process path because of their low
$n$-capture cross sections. Consequently, for relatively low $s$-process efficiencies,
the neutron flux mainly feeds the
nuclei at the 1st peak, while for higher exposures, the
second peak species are favored.

The overabundaces of $n$-capture process elements are lower in mild Ba stars than in strong Ba stars,
which could be explained by the relatively weaker neutron exposure in their  AGB-companion stars.
Stronger neutron exposure will benefit stronger $s$-process neucleosynthesis and lead to higher
$s$-process element abundances. If the neutron exposure is not strong enough, the first peak of
Y and Zr could show weak overabundance and probably sometimes be close to the solar value.
A Ba star having a longer orbital period, meaning there is a larger distance of the two member stars
in the binary system, will have less accretion efficiency from its AGB-companion star with
$s$-process overabundances. In addition, we only have  \ion {Y}{i} and  \ion{Zr}{i} lines to
derive Y and Zr abundances in this work. \ion {Y}{i} and  \ion{Zr}{i} will probably result in lower
Y and Zr abundances than the \ion {Y}{ii} line (Gratton \& Sneden 1994; Yushchenko et al. 2004) and
\ion{Zr}{ii} line (Allen \& Barbuy (2006a, sect.4.7).

\subsection{Ba intensity versus [Ba/Fe]}

\begin{figure}
\centering
\includegraphics[width=6.0cm]{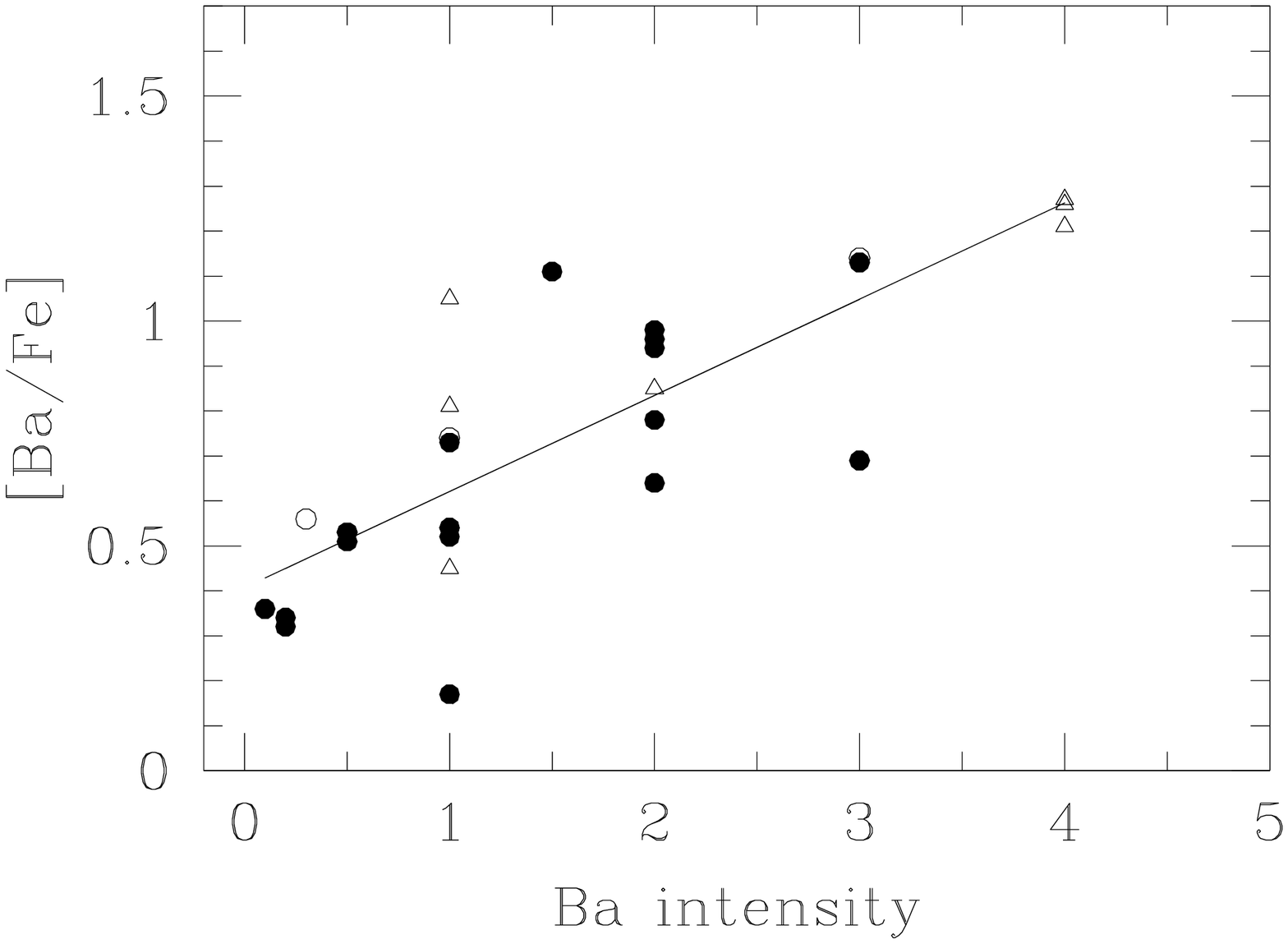}
\vspace{-1.5cm}
\caption{The [Ba/Fe] vs. Ba intensity relations for the 19 sample stars (the filled circles),
and the data points from Liang et al. (2003, the open circles) and
Liu et al. (2009, the open triangles). The solid line is linear least-square
fit for these data points.
}
\label{babafe.eps}
\end{figure}

It is interesting to show the relation between [Ba/Fe] and Ba intensity, which is given in Figure \ref{babafe.eps}
for our 19 sample stars (the filled circles) and some Ba stars
from Liang et al. (2003, the open circles) and
Liu et al. (2009, the open triangles). A linear least-square fit is obtained for
these 29 Ba stars as shown by the solid line.
This shows a correlation or consistency between Ba intensity and [Ba/Fe] (also see Lu \& Upgren 1985),
but here we include many objects with lower Ba intensity down to 0.1. Thus our work extends to Ba stars
with much weaker overabundances of $n$-capture process elements, which will be useful for understanding
the abundance patterns of Ba stars, from mild to strong Ba cases.
The general trend of the increasing [Ba/Fe] following Ba intensity is clearly shown in Figure \ref{babafe.eps}.
However, from the [Ba/Fe] value and the abundance patterns, we suggest the two stars HD\,212320 and HD\,58121
are strong Ba stars with [Ba/Fe] larger than 0.6, although their Ba intensities are 1.5 and 1.0, respectively.

\subsubsection{[X/Fe] vs. [Fe/H]}

It is interesting to discuss the relation of  [X/Fe] vs. [Fe/H]
for the 17 elements analyzed in our sample of stars.
Our sample stars are distributed in a narrow
range of [Fe/H]: $-0.32$$<$[Fe/H]$<$+0.23 and thus most of them should be disk stars.

We found that for Na and the iron-peak elements (Sc, Ti, V, Cr, Mn and
Ni), [X/Fe] is independent of [Fe/H]. For other elements (Al, $\alpha$-elements, $n$-capture elements), [X/Fe]
shows slight anti-correlations with [Fe/H].

\begin{figure*}[b]
\centering
\includegraphics[width=4.0cm]{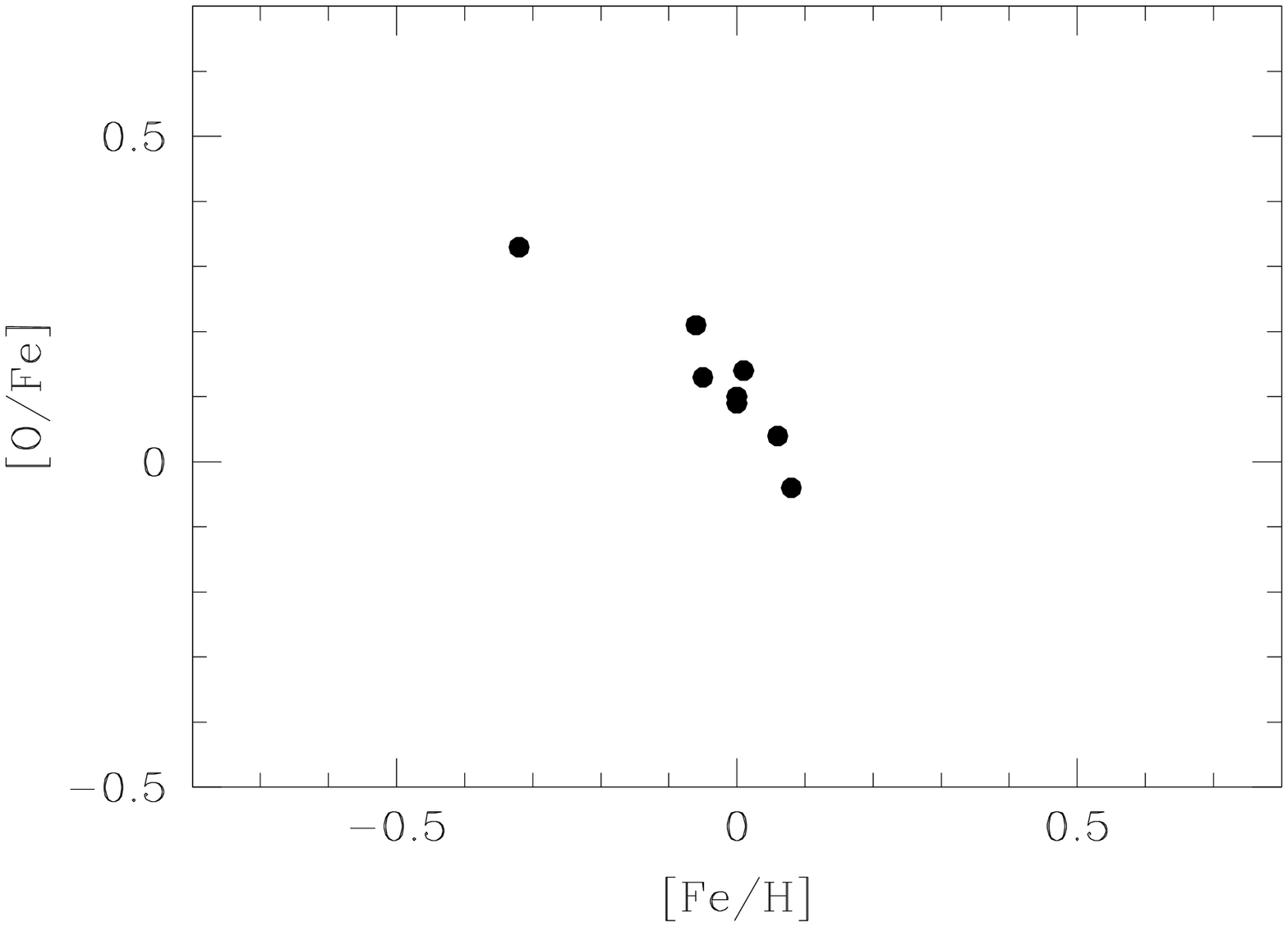}
\includegraphics[width=4.0cm]{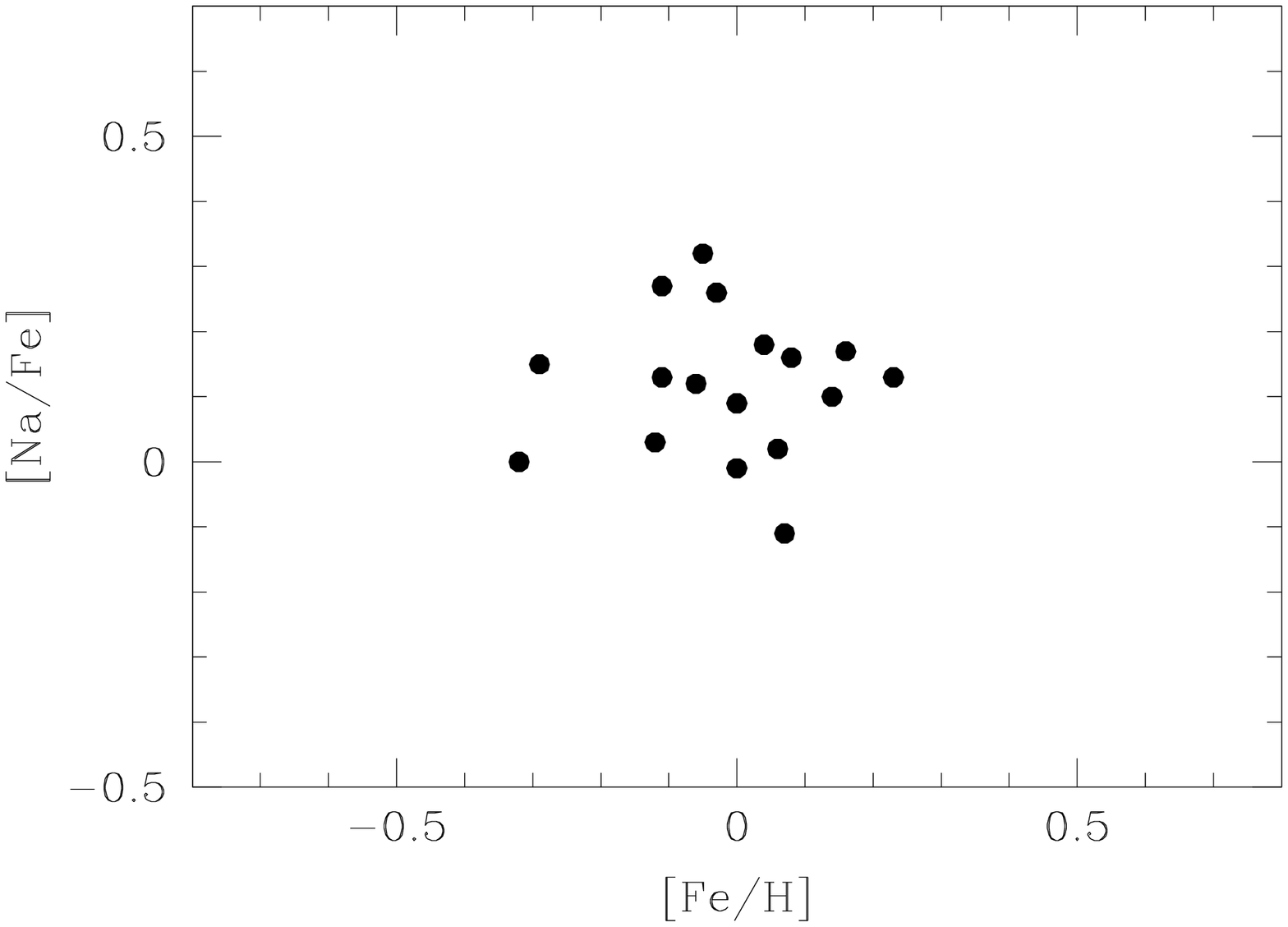}
\includegraphics[width=4.0cm]{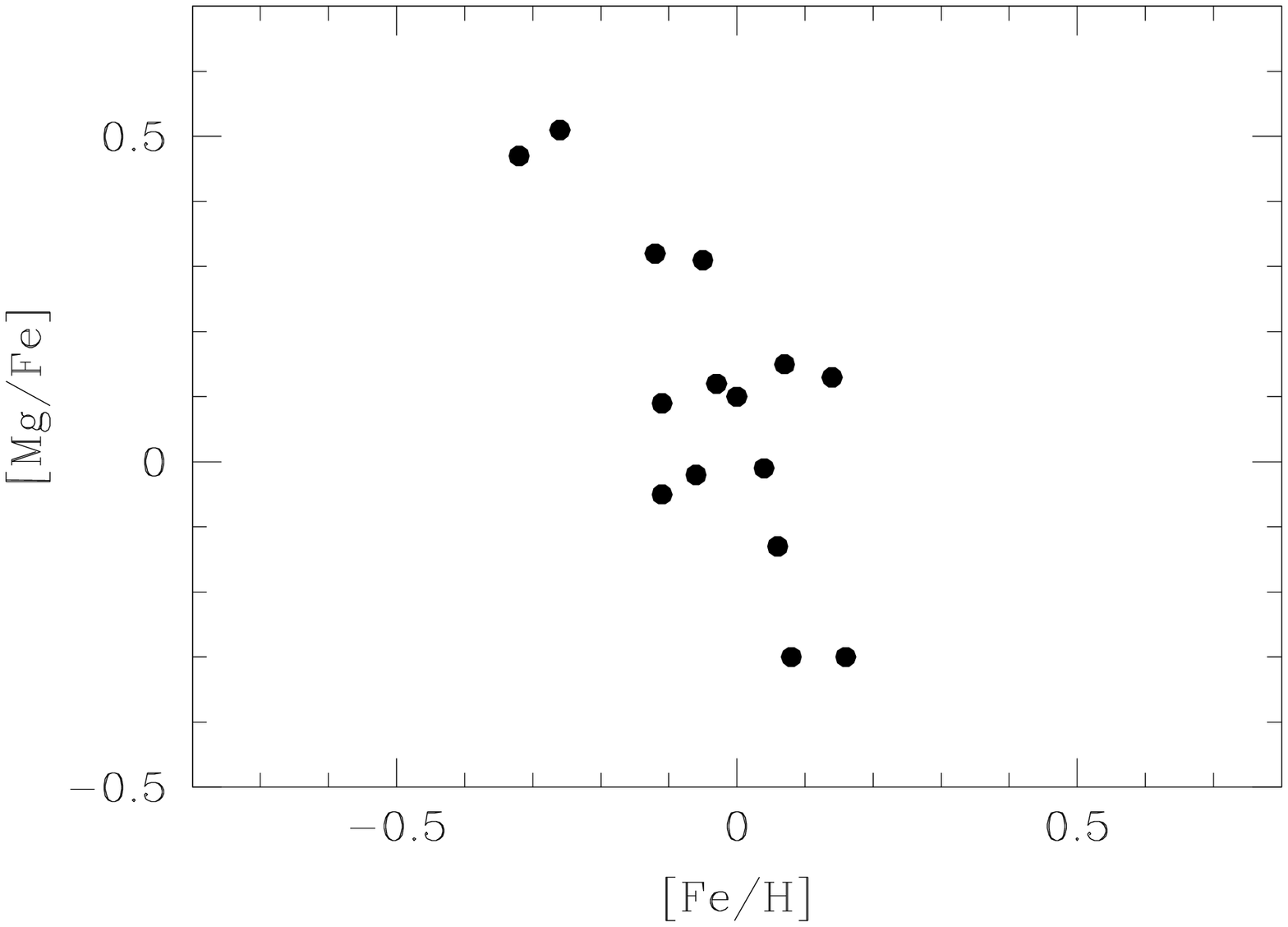} \\  \vspace{-1.cm}
\includegraphics[width=4.0cm]{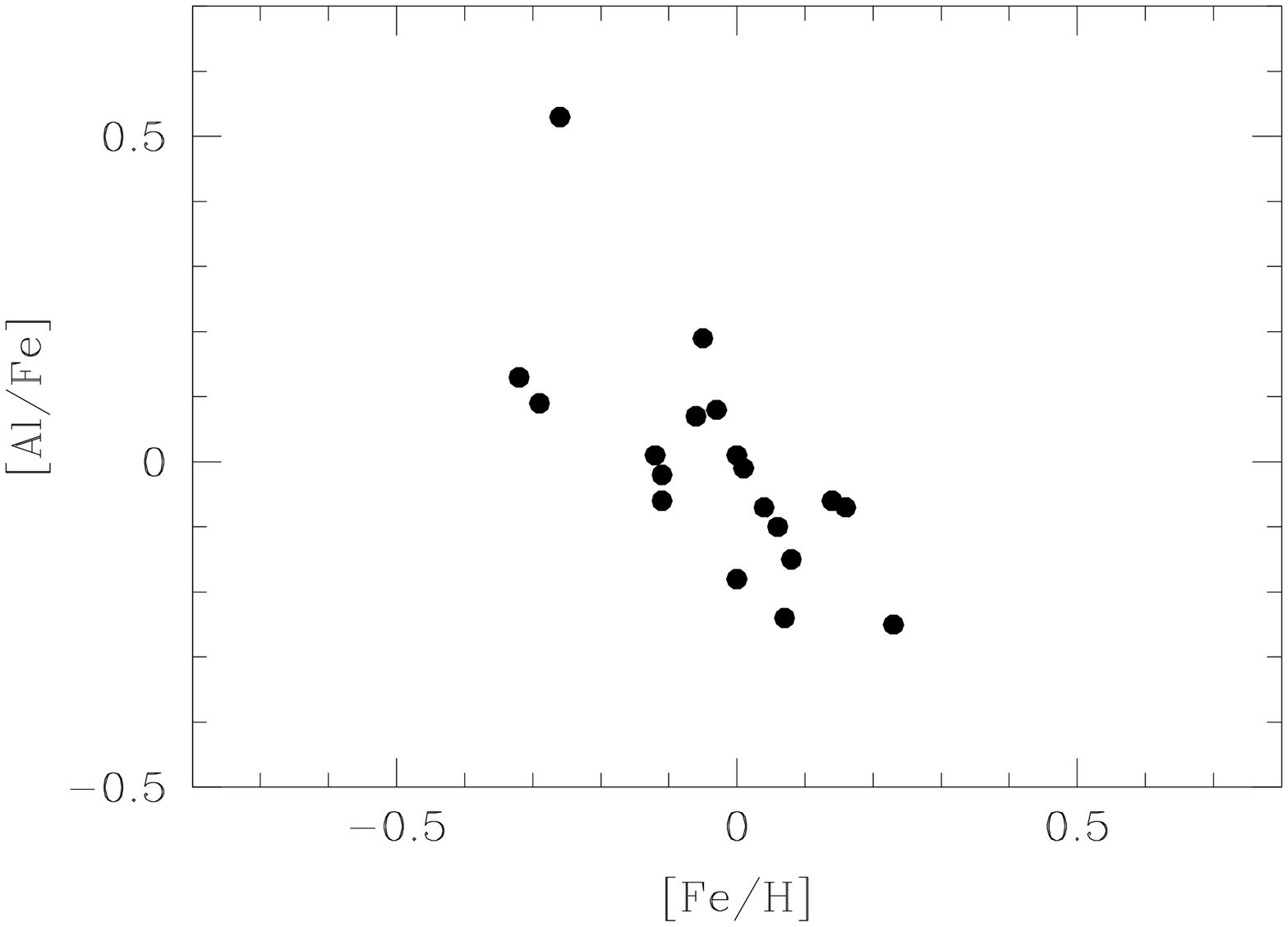}
\includegraphics[width=4.0cm]{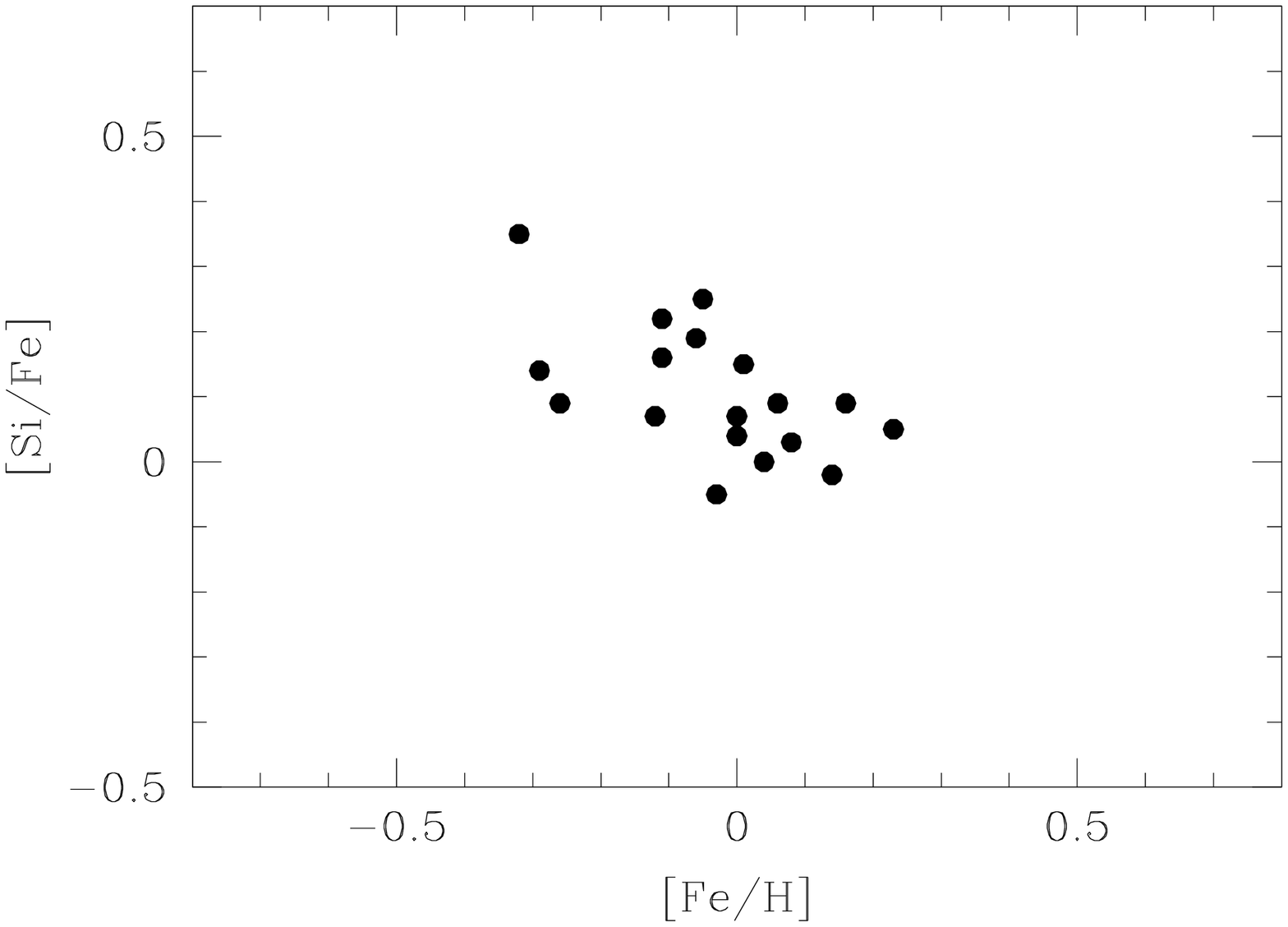}
\includegraphics[width=4.0cm]{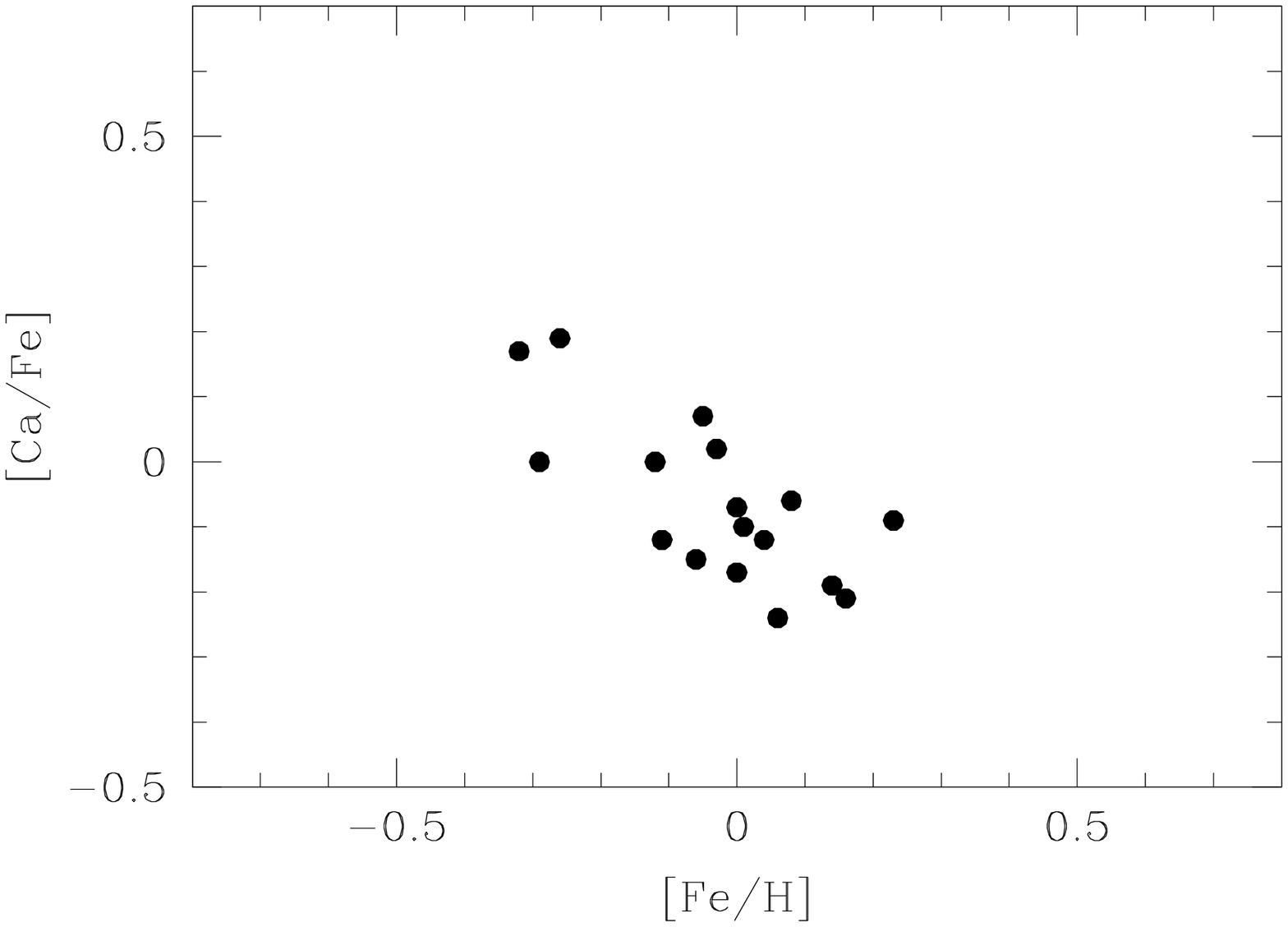}  \\ \vspace{-1.cm}
\includegraphics[width=4.0cm]{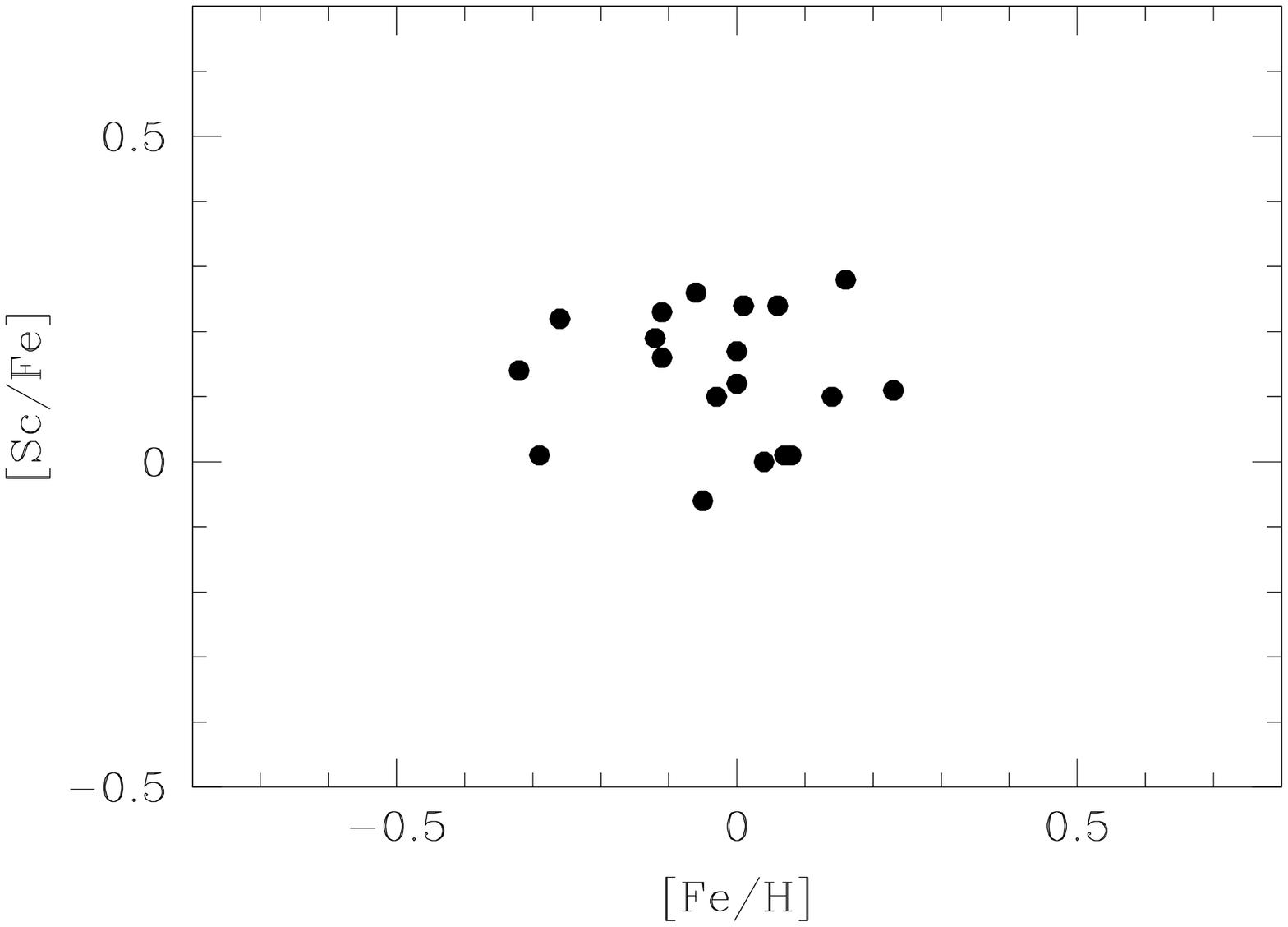}
\includegraphics[width=4.0cm]{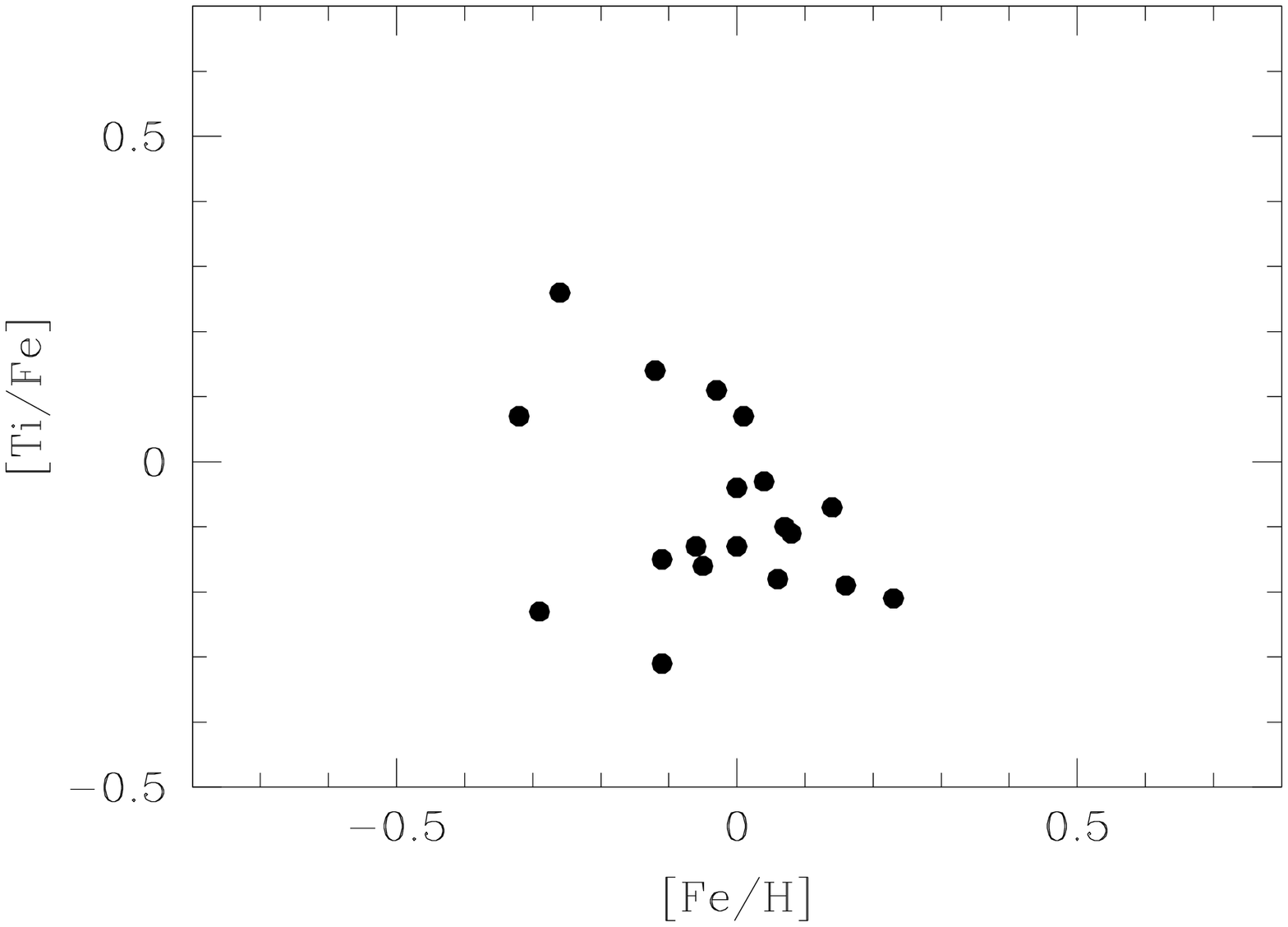}
\includegraphics[width=4.0cm]{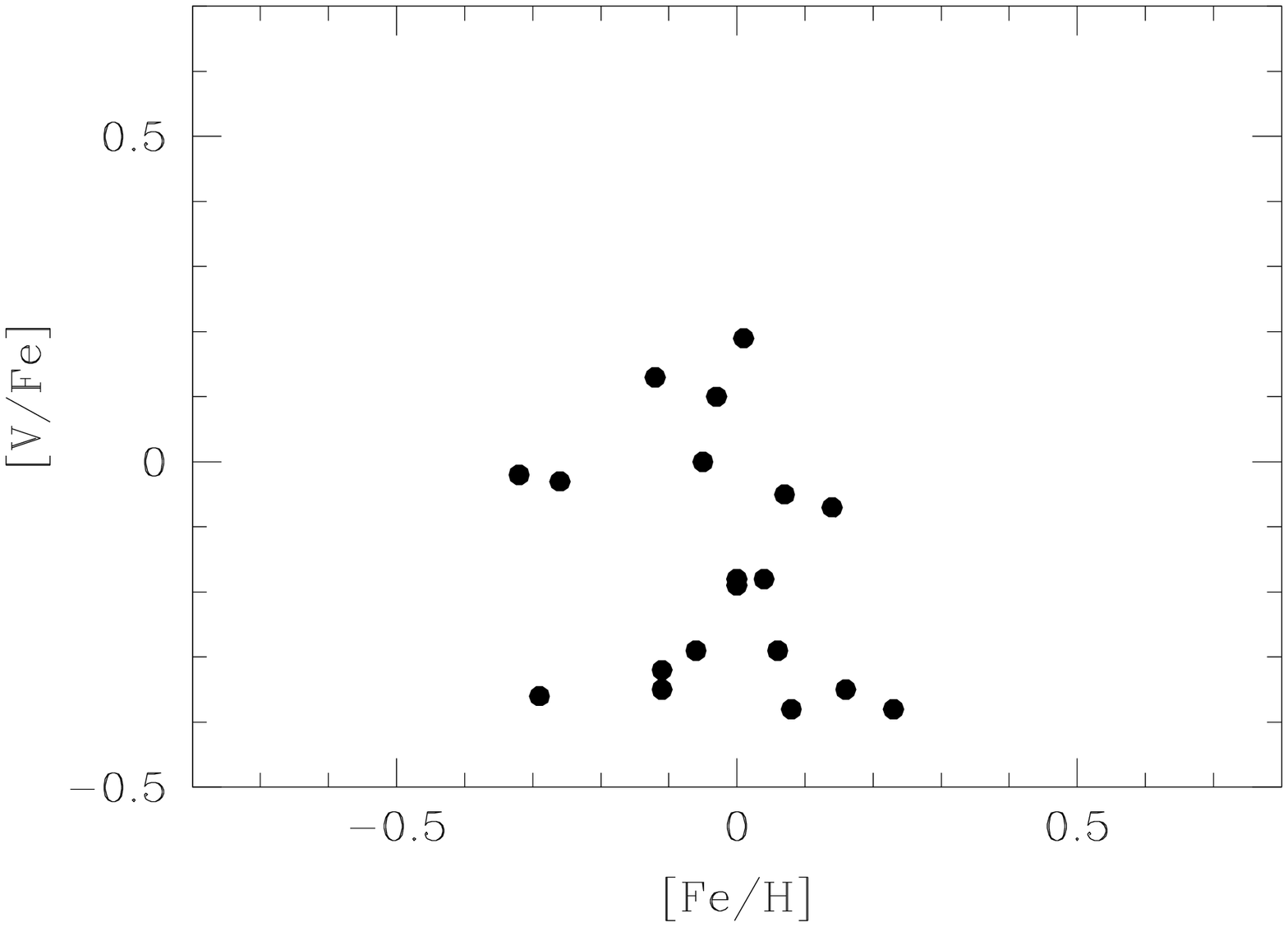}   \\   \vspace{-1.cm}
\includegraphics[width=4.0cm]{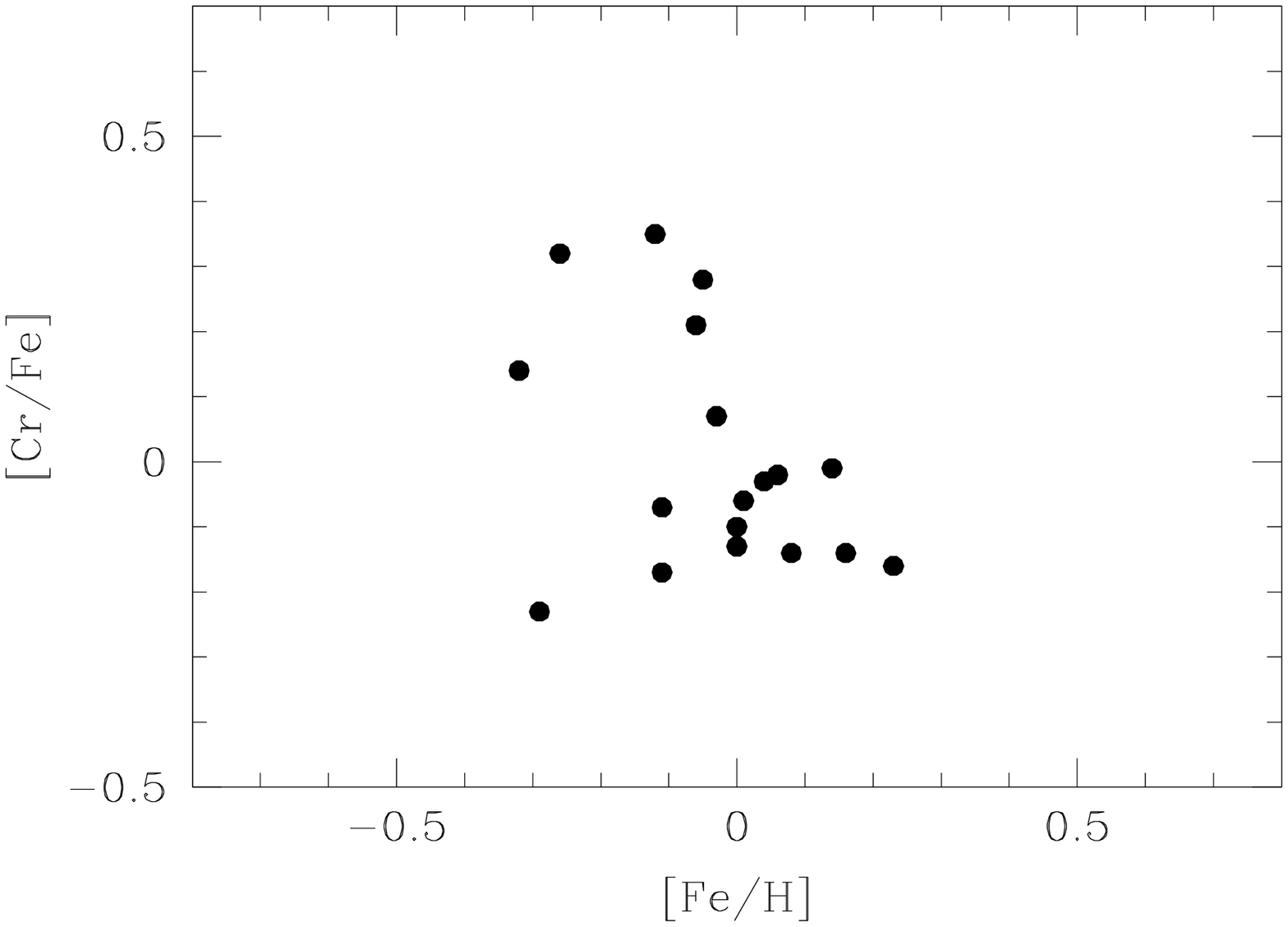}
\includegraphics[width=4.0cm]{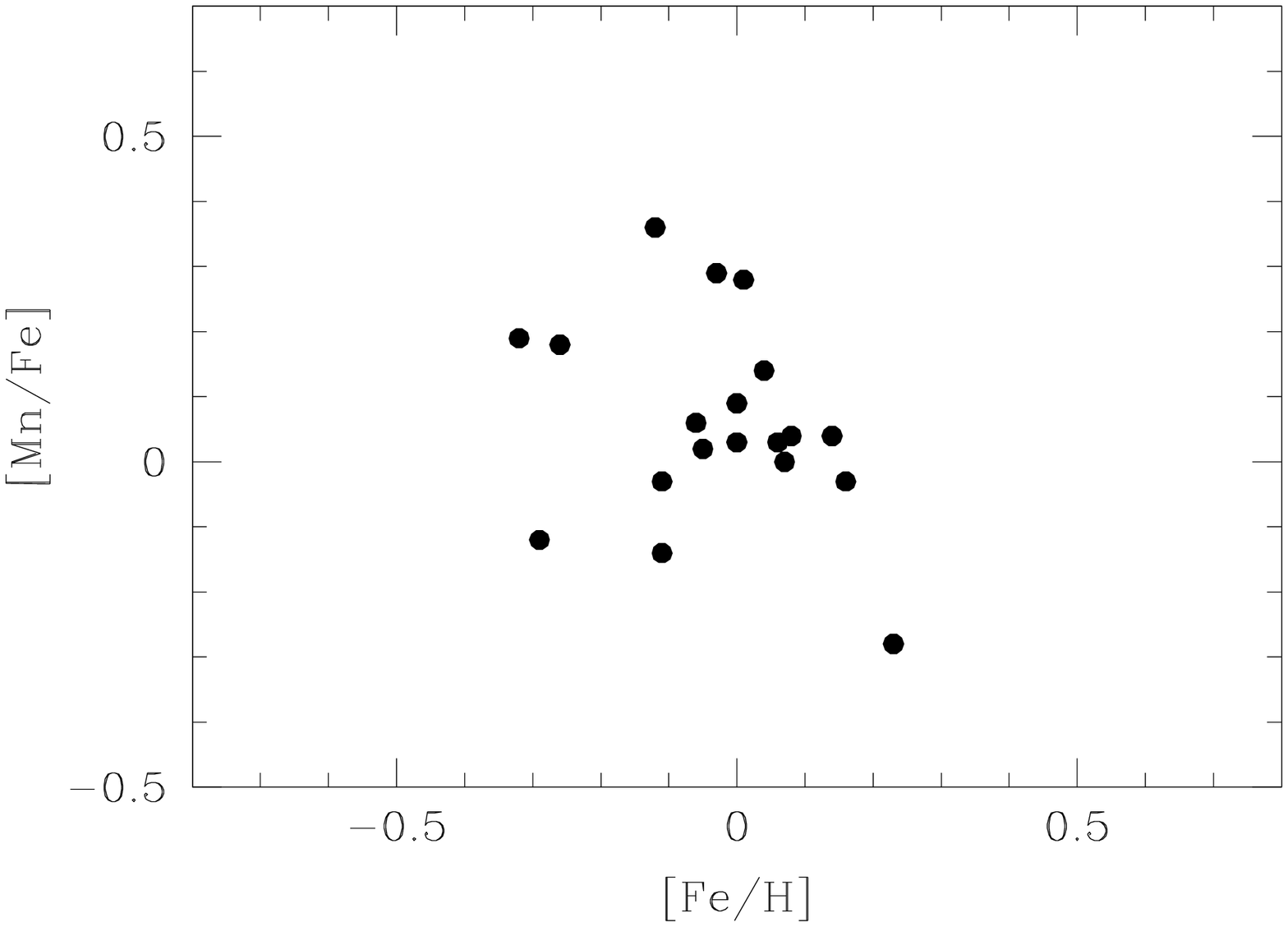}
\includegraphics[width=4.0cm]{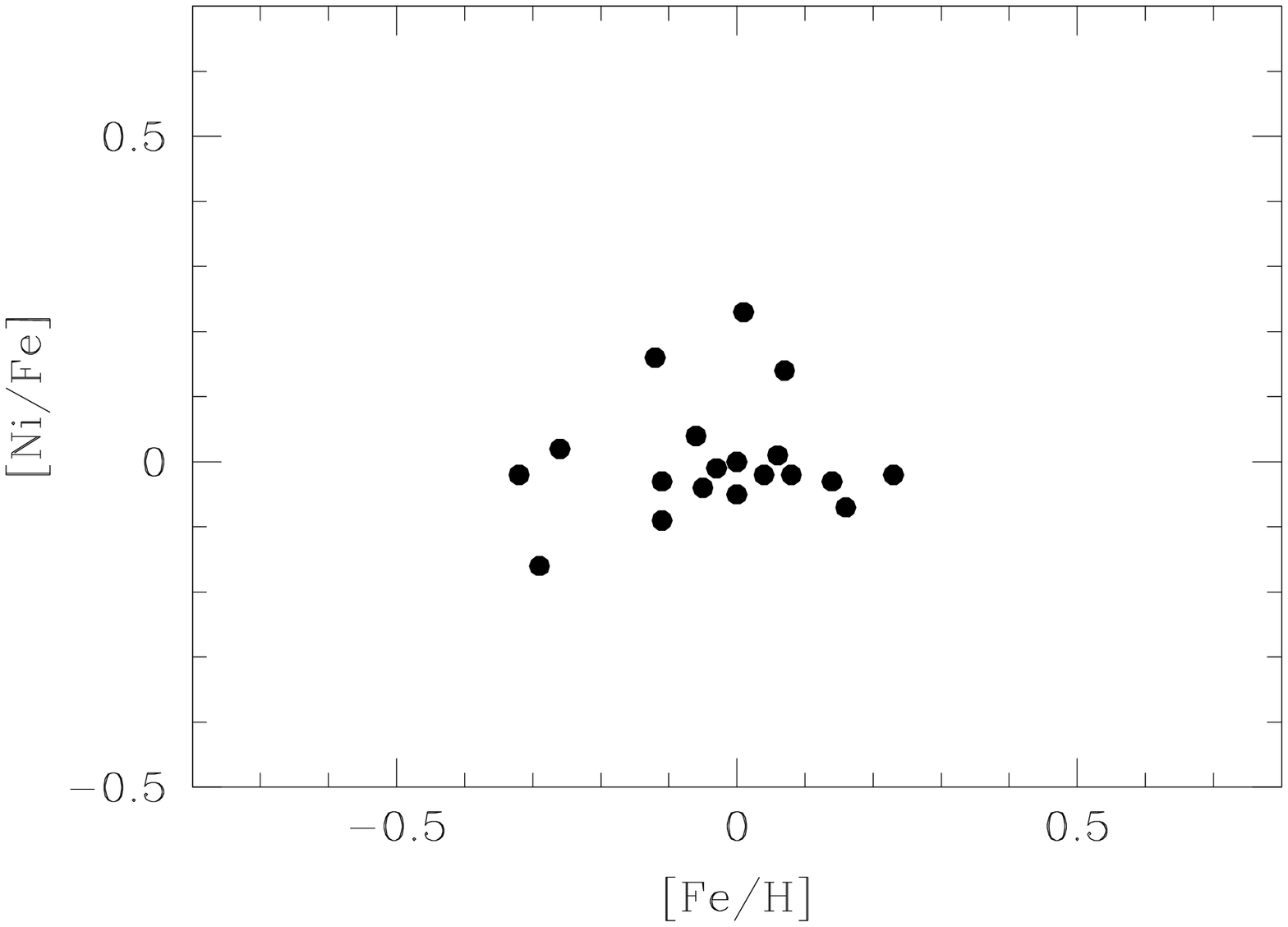}   \\   \vspace{-1.cm}
\includegraphics[width=4.0cm]{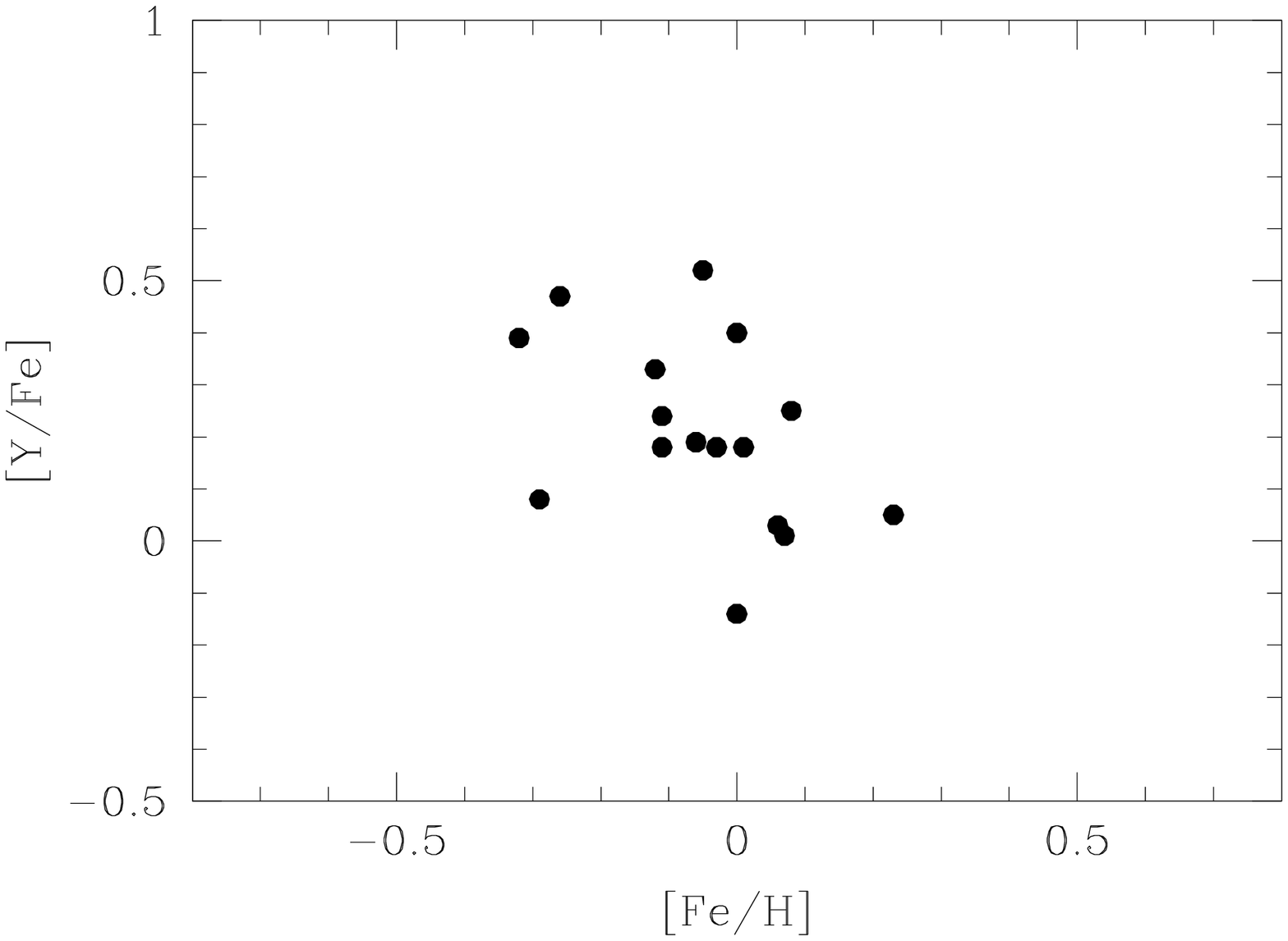}
\includegraphics[width=4.0cm]{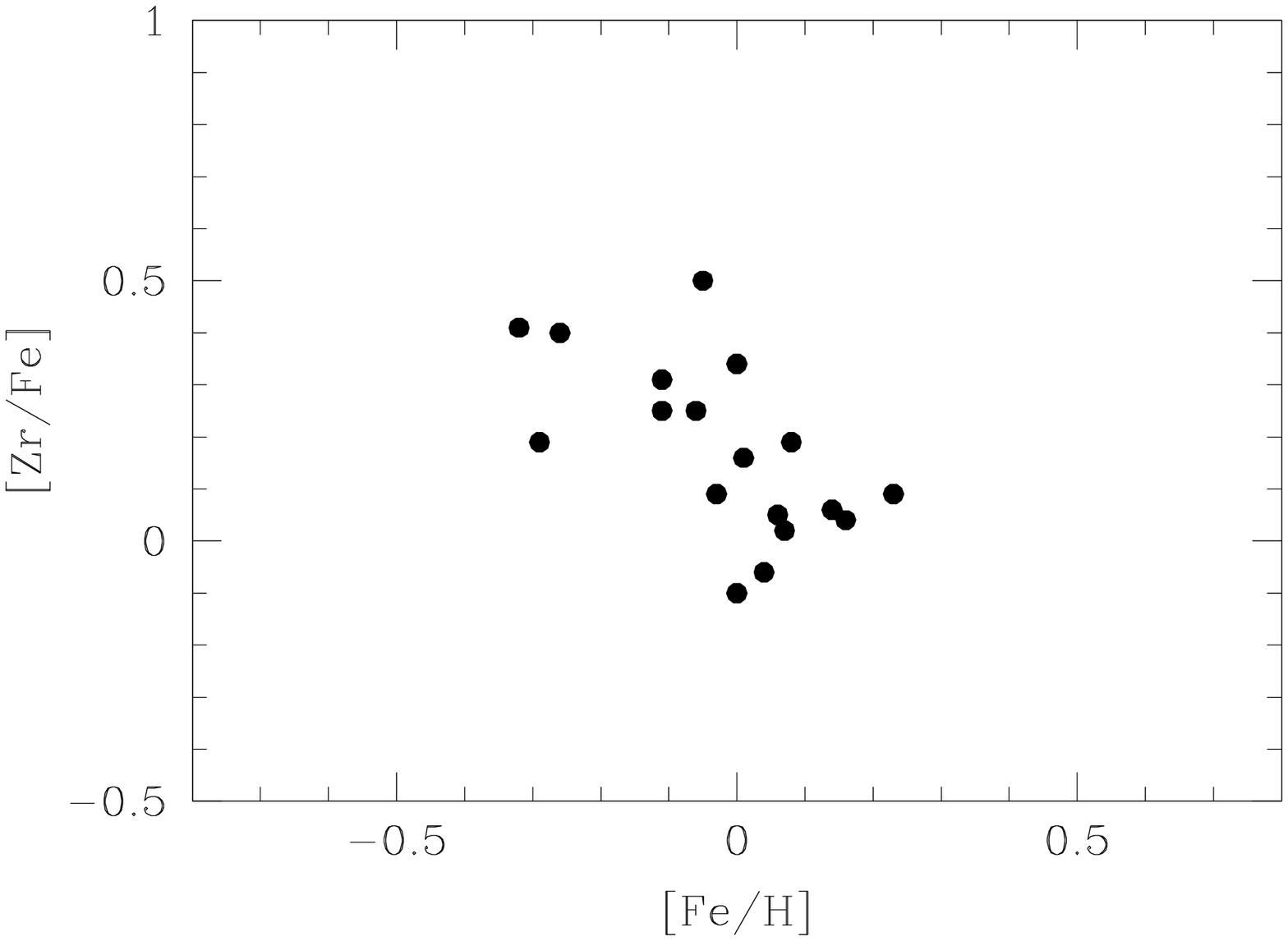}
\includegraphics[width=4.0cm]{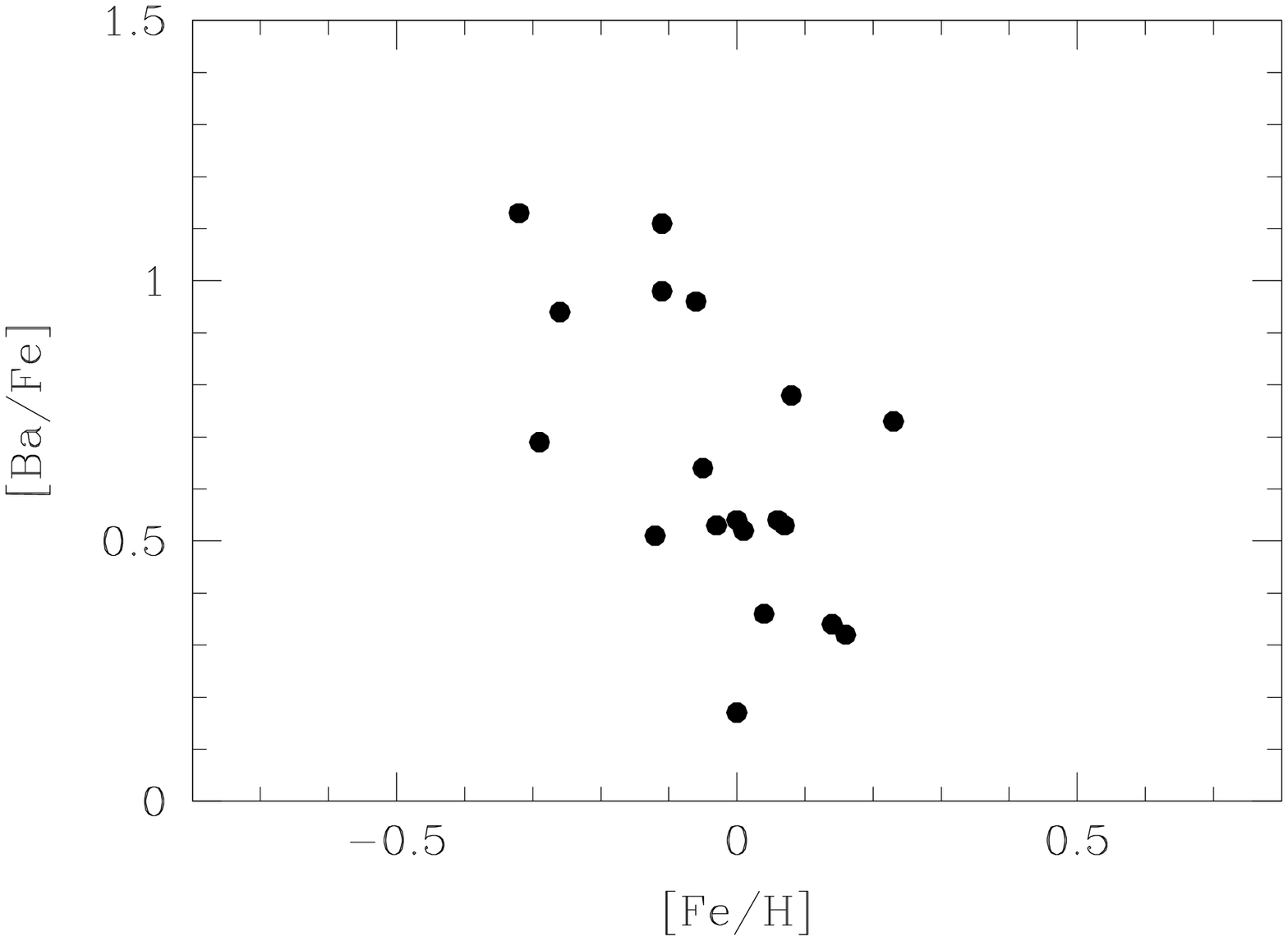}  \\   \vspace{-1.cm}
\includegraphics[width=4.0cm]{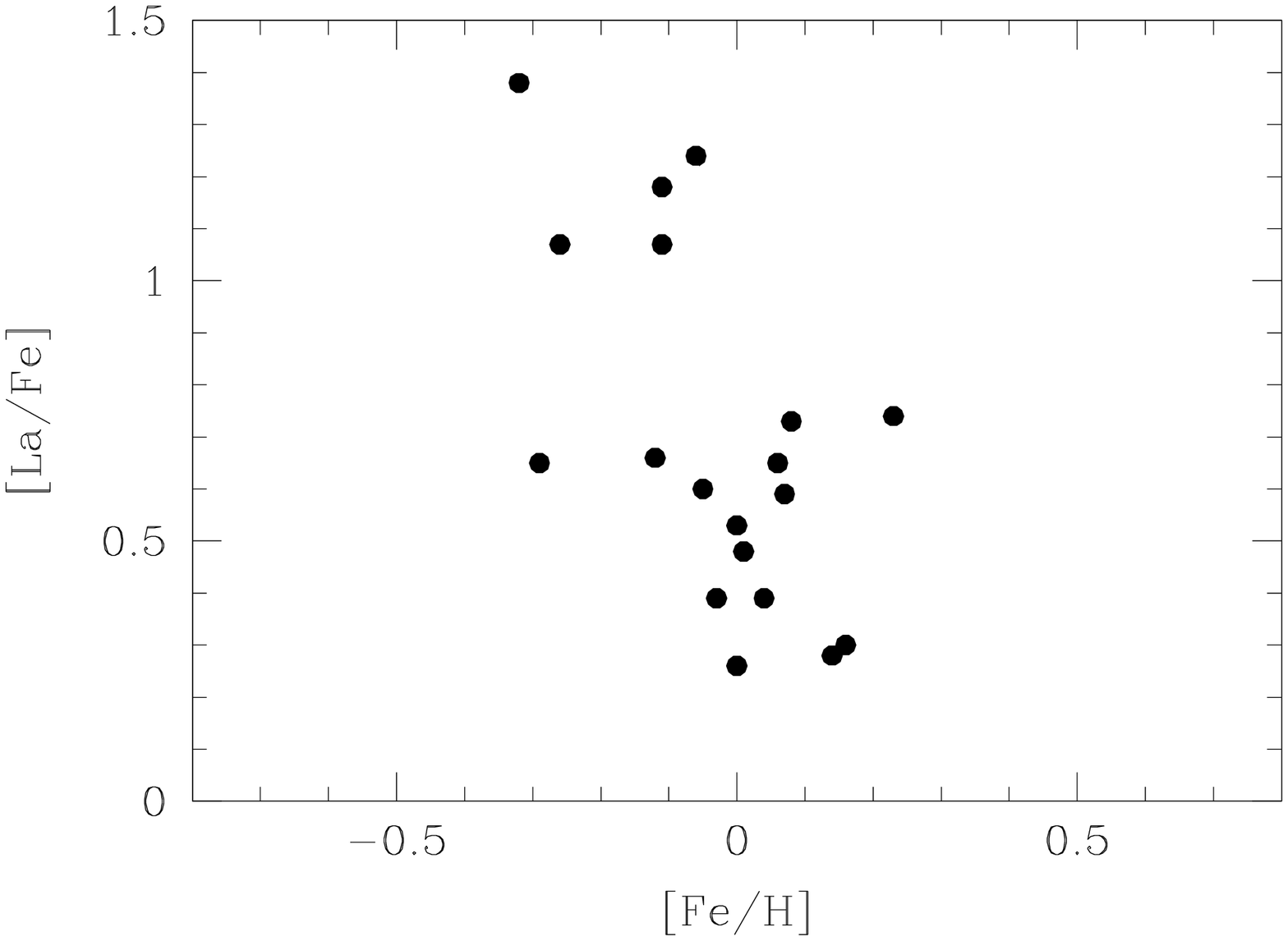}
\includegraphics[width=4.0cm]{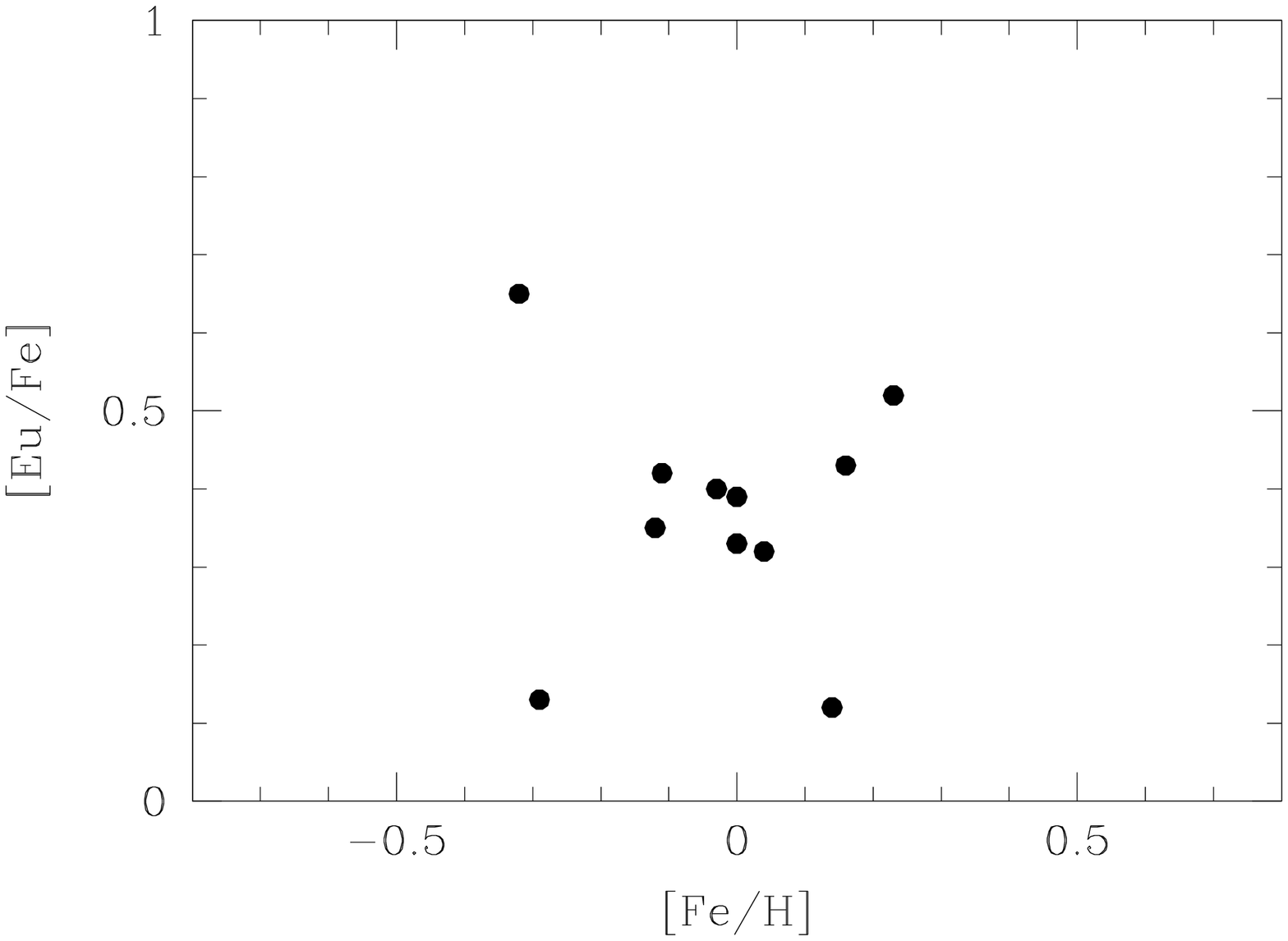}   \vspace{-1.cm}
\caption{[X/Fe] vs. [Fe/H] relations for the 19 sample stars, from iron-peak elements to the $n$-capture elements. } \label{xfefeh.eps}
\end{figure*}

Figure~\ref{xfefeh.eps} displays these trends.
For $n$-capture elements, our samples of tars shows increasing [X/Fe] with decreasing [Fe/H].
This anti-correlation is less obvious for Eu.
[Ba/Fe] and [La/Fe] show the most overabundant values in our sample stars, up to +1.13 and +1.38,
respectively. Eu shows relatively weak overabundances (up to 0.65) since Eu has more contributions from
the r-process. However, [Y/Fe] and [Zr/Fe] are not exceptionally overabundant with
median values of 0.15 and 0.18, respectively. The ranges of [X/Fe],
the median values and the line numbers used in the abundance analyses
are given in Table~\ref{valueXFe} for these 17 elements that were analyzed in our sample
stars.

\subsubsection{Comparison with the common stars in literature}

\begin{figure}
\centering
\includegraphics[width=4.0cm]{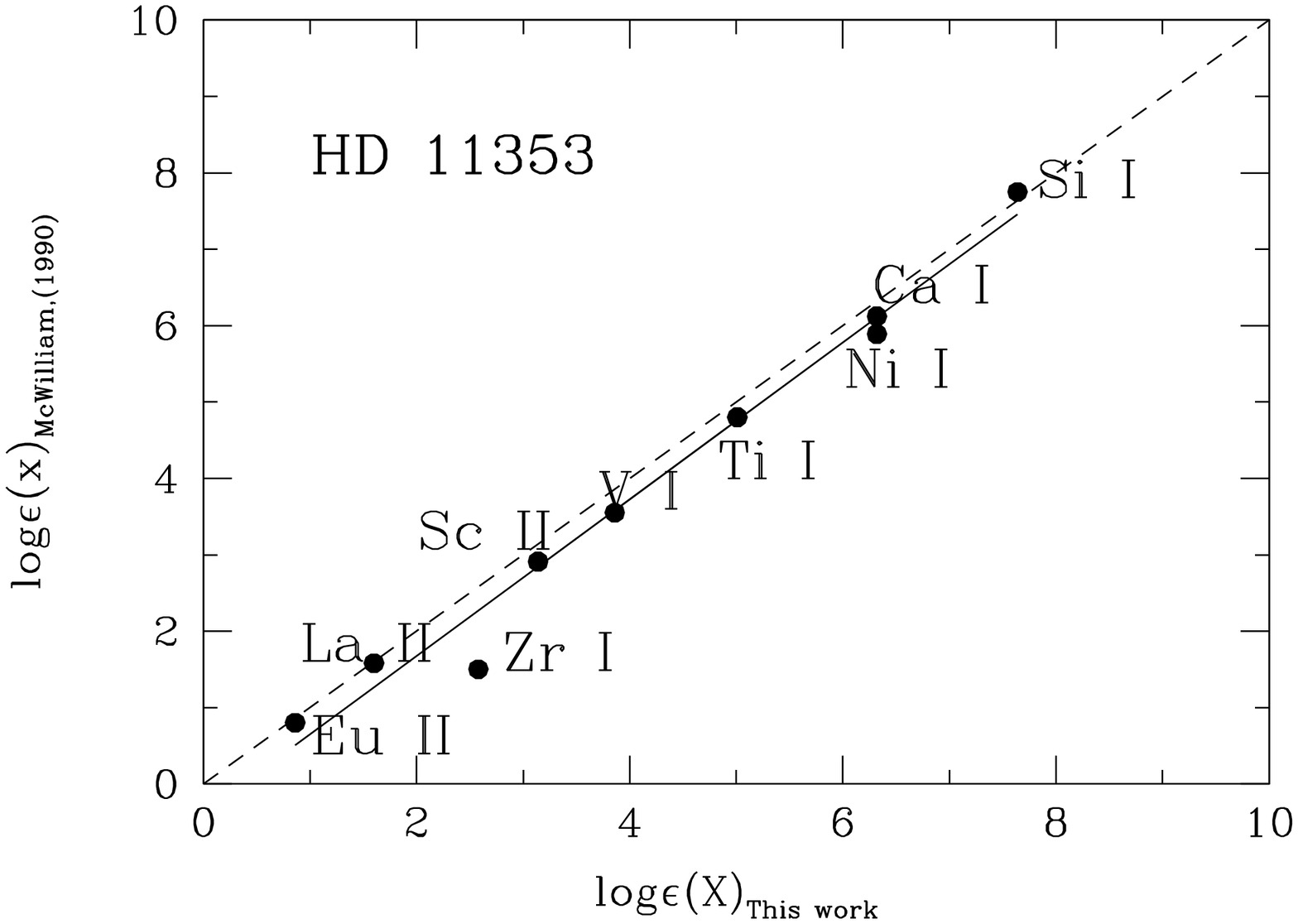}
\includegraphics[width=4.0cm]{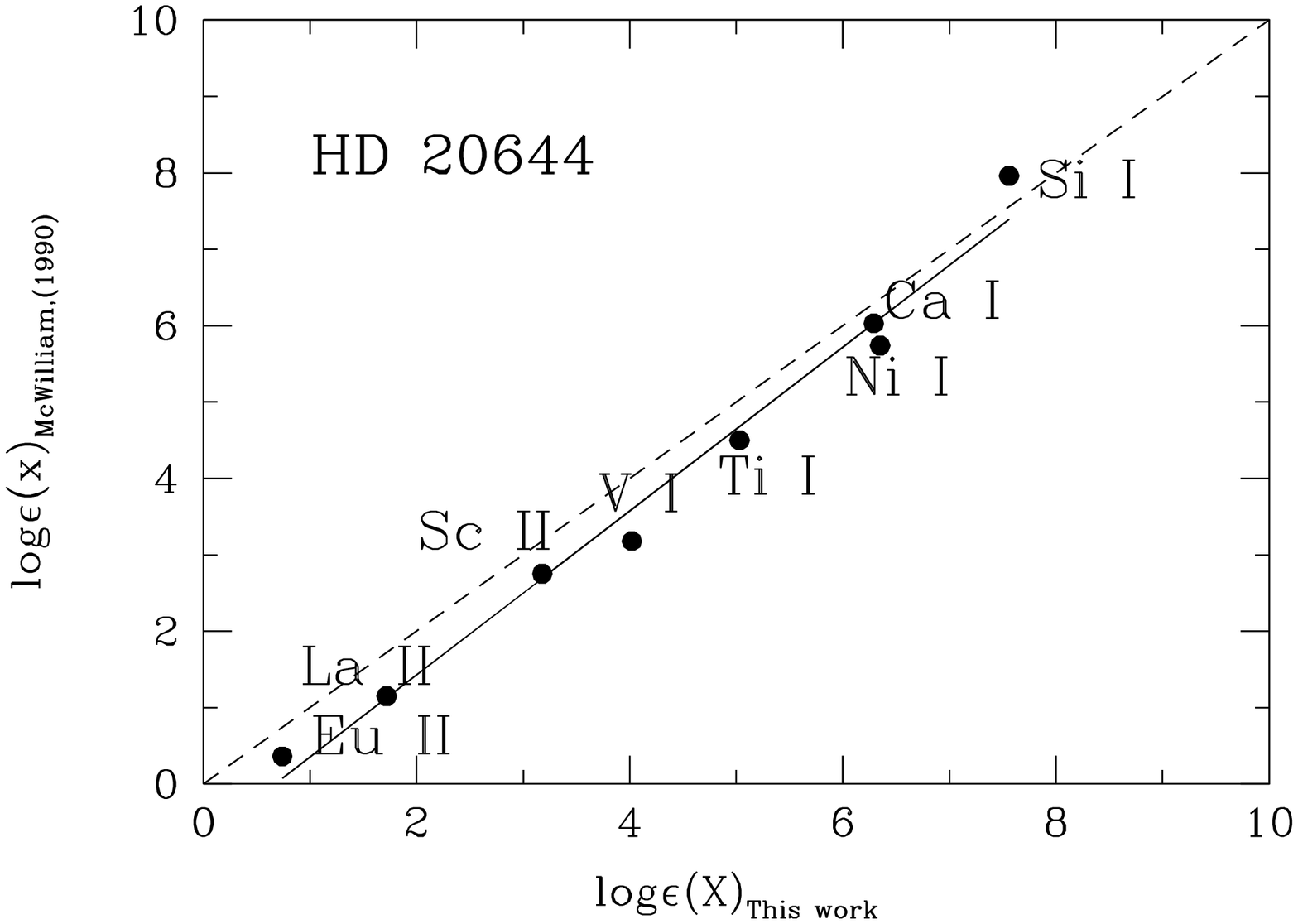}
\includegraphics[width=4.0cm]{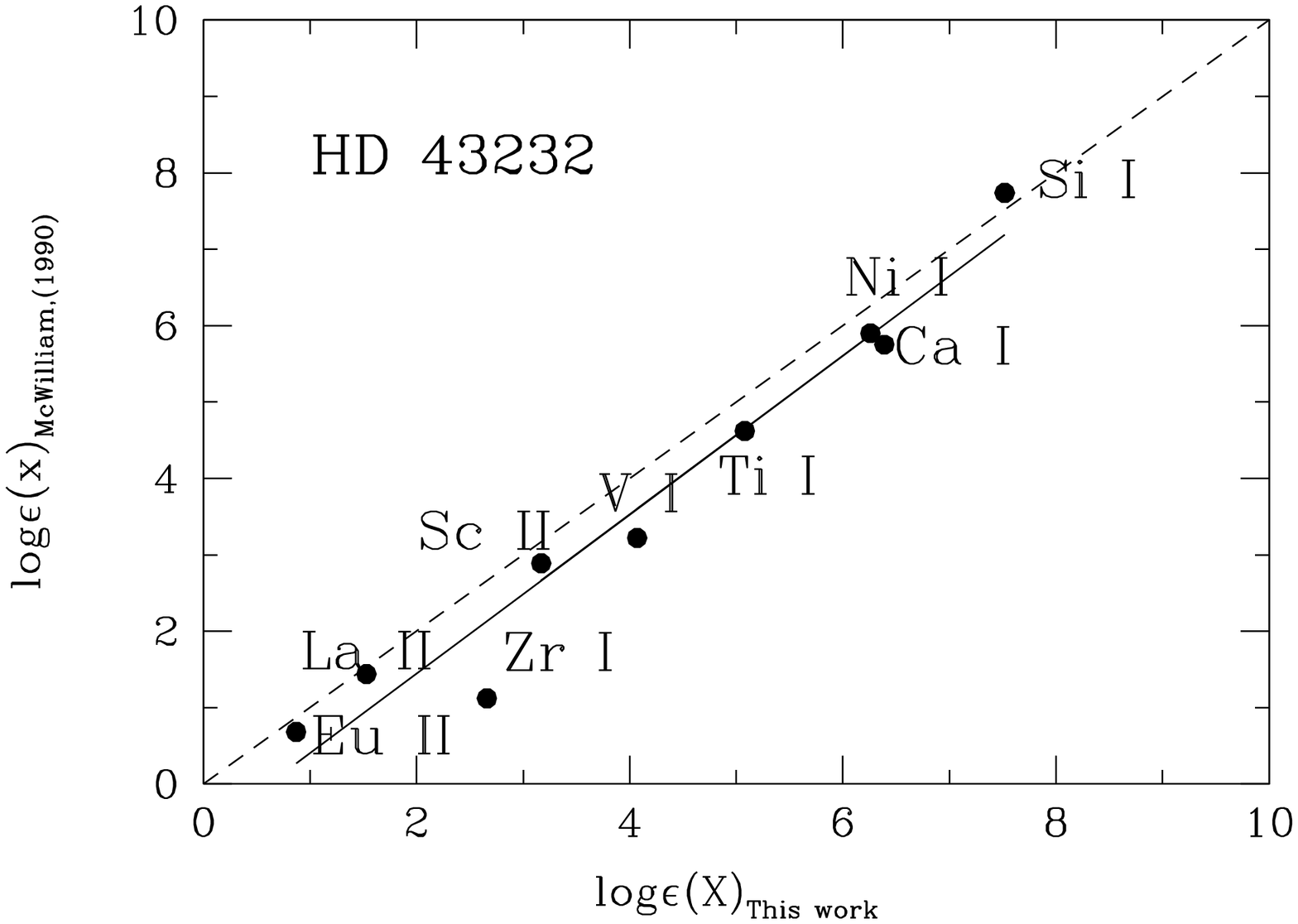}  \\   \vspace{-1.cm}
\includegraphics[width=4.0cm]{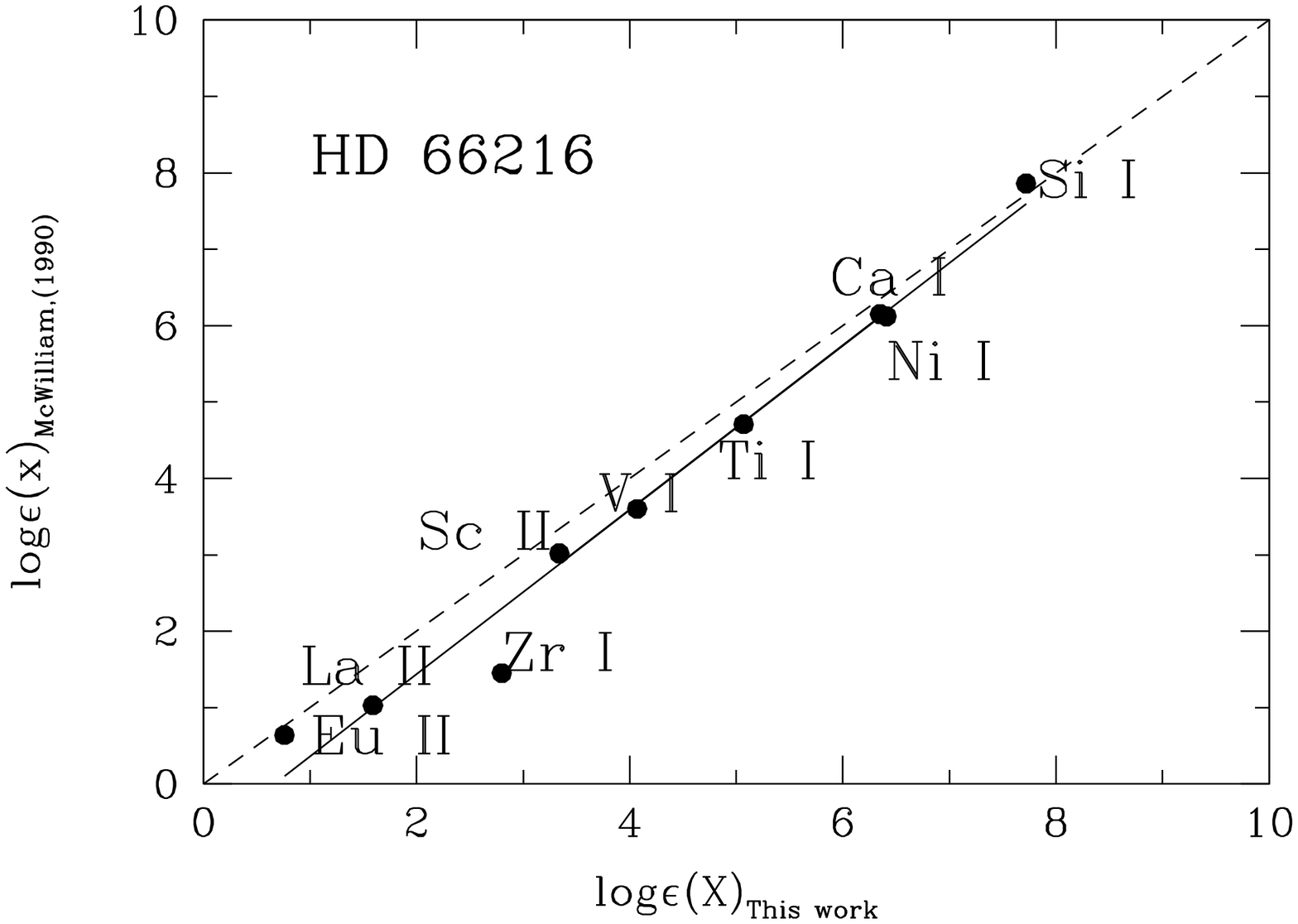}
\includegraphics[width=4.0cm]{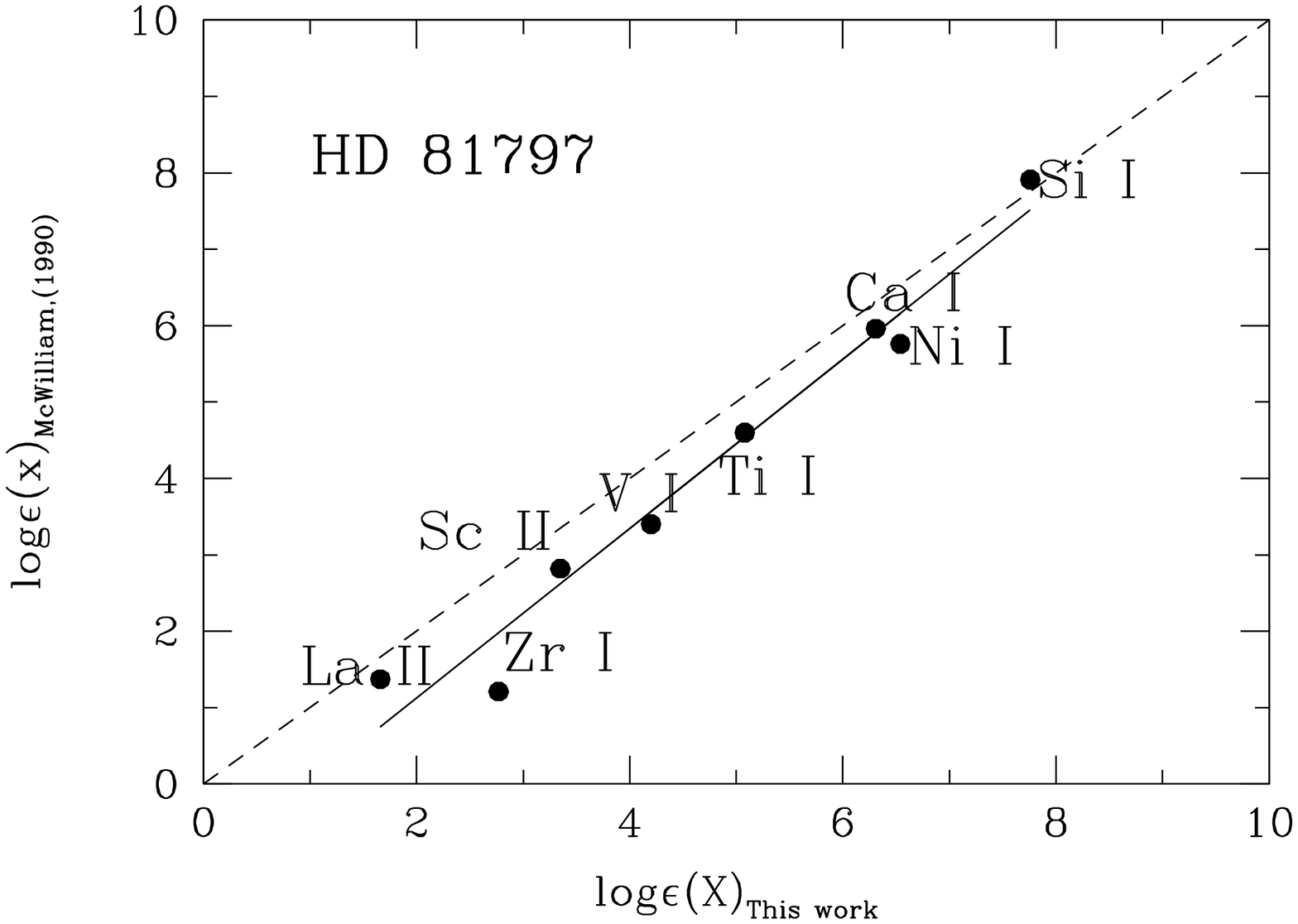}
\includegraphics[width=4.0cm]{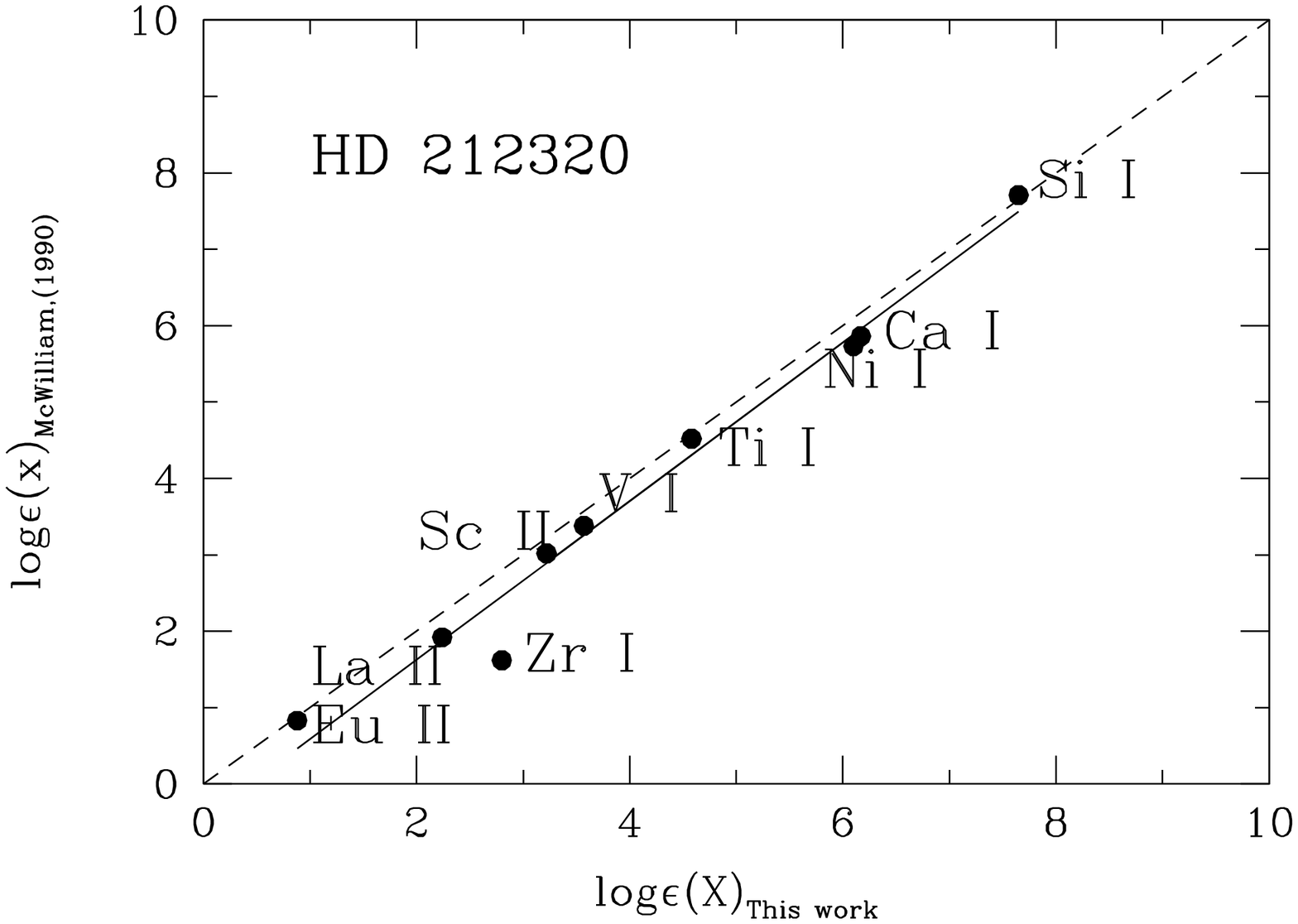}     \vspace{-1.cm}
\caption {Comparison of $\log\epsilon(X)$ for ten elements of the six common stars in this work and McWilliam (1990). The solid
lines are the least-square fits to the data and the dashed diagonal
lines are the one-to-one relation. } \label{abun_compare.eps}
\end{figure}

To compare the abundances of our sample Ba stars with those of the
common stars in  McWilliam (1990), we compare the $\log\epsilon(X)$ abundances for the six common stars in our
work with those of McWilliam (1990) as shown in Figure~\ref{abun_compare.eps}.
They indicate that our abundances are generally
lower than theirs. The reason could be the result of
different atmospheric parameters (Table~\ref{parameter}), the
different atmospheric models and atomic data, etc. in the two works.

\section{Kinematic analysis and orbital period}

\subsection{Kinematic analysis}

The kinematic studies of the stars in the solar neighborhood are based on
the velocities U, V and W measured along axes pointing respectively
towards the Galactic center, the direction of Galactic rotation and
the North Galactic Pole. The accurate distances (parallax) and the
proper motions available in the Hipparcos Catalogue (ESA 1997),
combined with stellar radial velocity, make it possible to derive
reliable space velocities.  For our sample stars, the proper motion,
parallax and radial velocity values are taken from SIMBAD. For some of
them without radial velocity measurements, we calculate these radial
velocities from the observed spectra and correct them by using the
{\sl observatory/astutil/rvcorrect} program in
IRAF\footnote{IRAF is distributed by the National Optical
Astronomical Observatory, which is operated by the Association of
Universities for Research in Astronomy, Inc., under cooperative
agreement with the National Science Foundation.}.

\begin{figure*}
\sidecaption
\includegraphics[width=3.5cm]{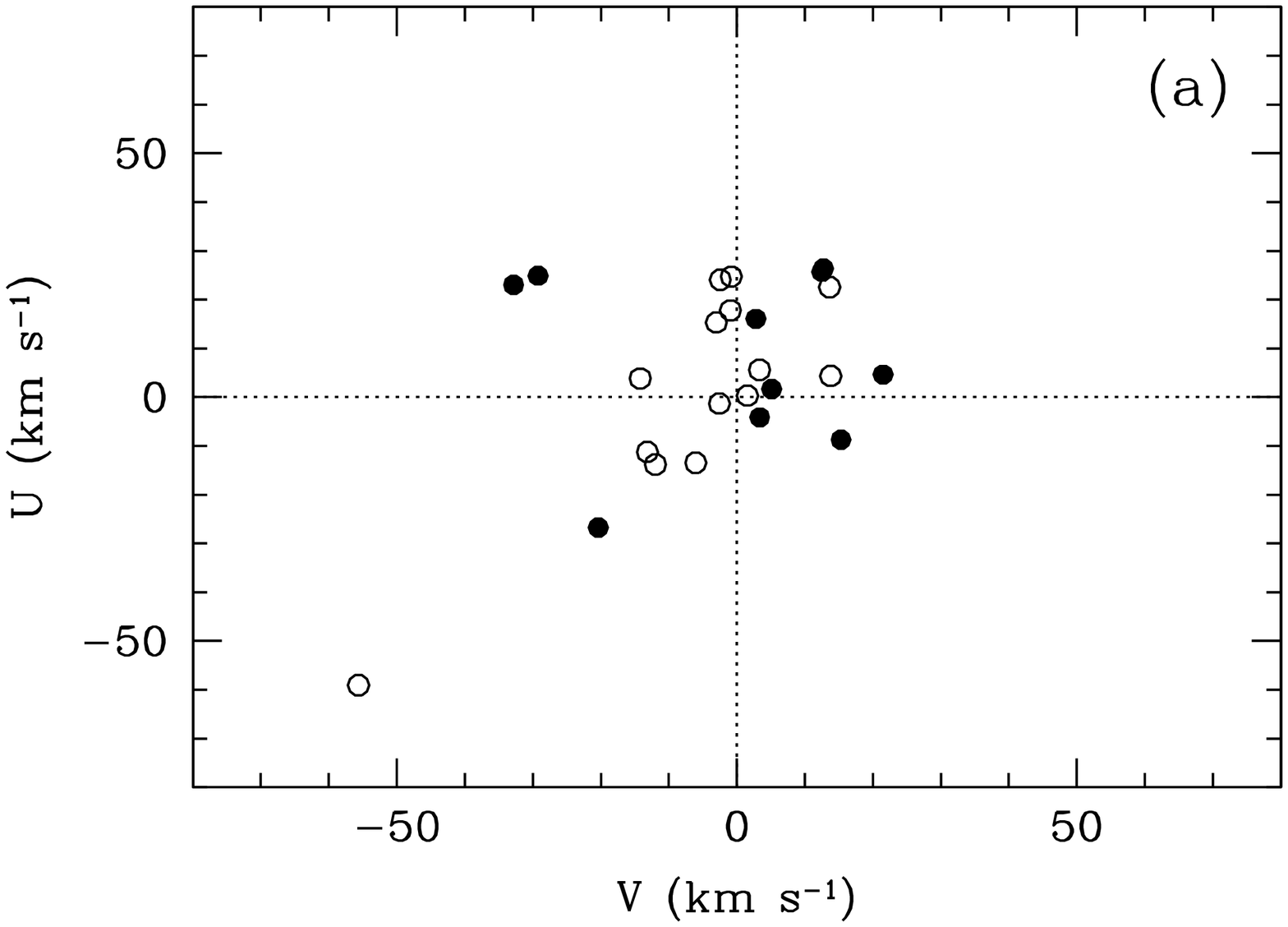}
\includegraphics[width=3.5cm]{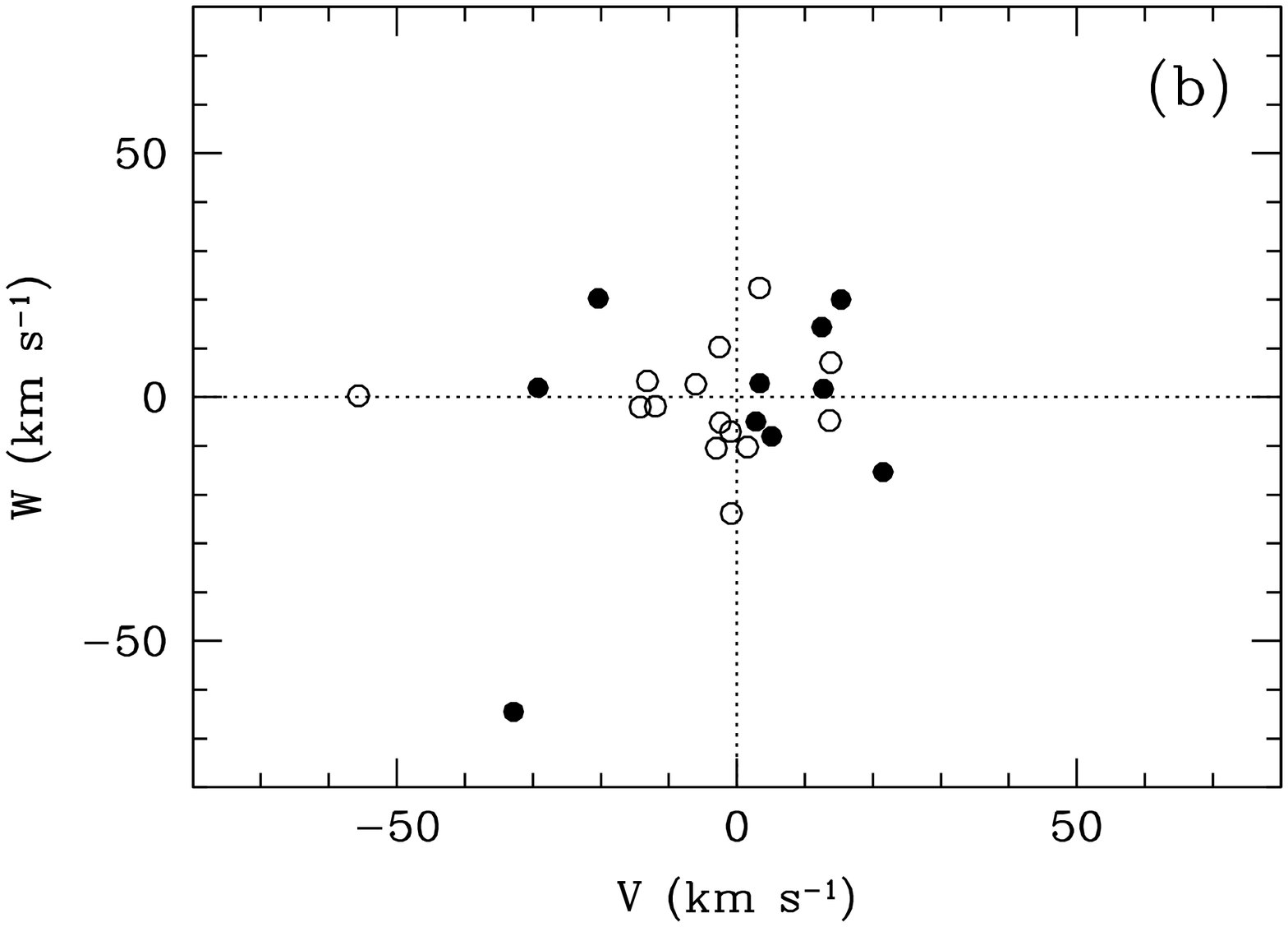}
\includegraphics[width=3.5cm]{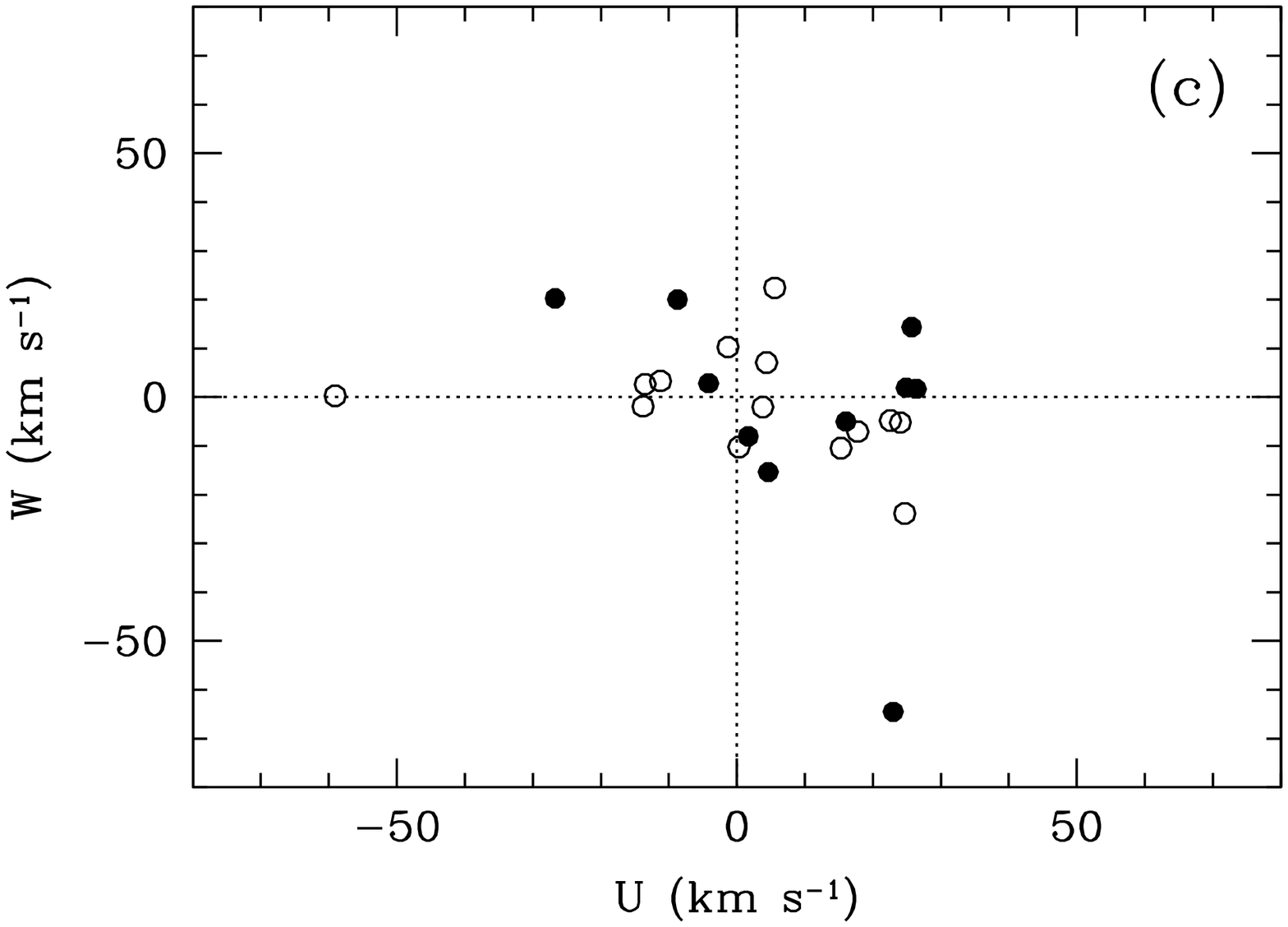}
\includegraphics[width=3.5cm]{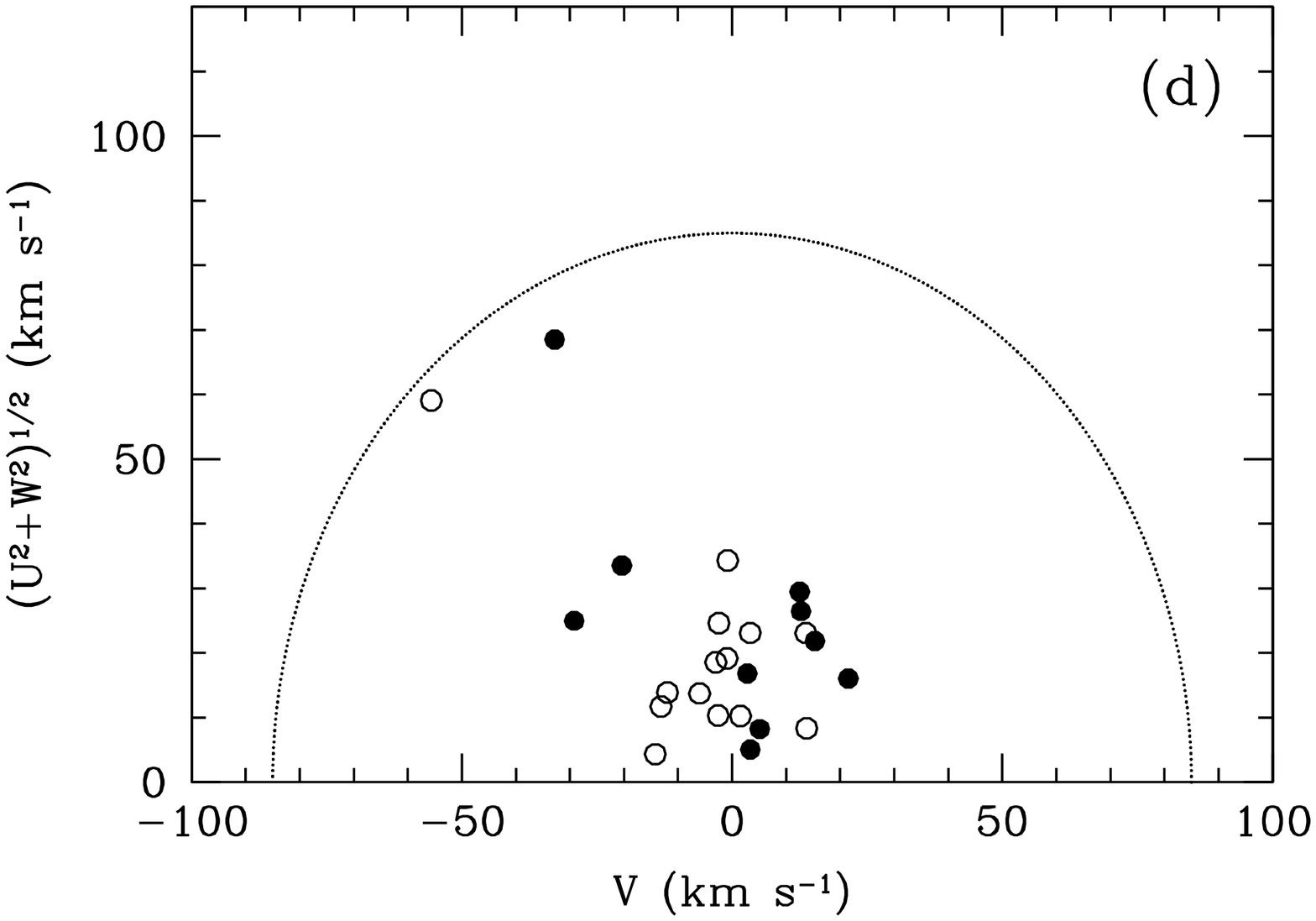}
\vspace{-1.cm}
\caption {The U, V and W velocities of our sample stars.
The filled circles refer to strong Ba stars (with
Ba$\geq$2) and the open circles refer to mild stars (with Ba$<$2).}
\label{uvw.eps}
\end{figure*}

The calculation of the space velocity with respect to the Sun is
based on the method presented by Johnson \& Soderblom (1987).  The
correction of space velocity to the Local Standard of Rest is based
on a solar motion of $\rm(U,V,W)_{\odot}$=(10.1, 4.0, 6.7) km s$^{-1}$
(see Hogg et al. 2005, which is slightly different from the values
given in Binney \& Merrifield 1998). The mean U, V and W velocities and
their distributions are shown in Figure~\ref{uvw.eps}.

The mild and strong Ba stars do not present obvious
differences in the UV, WV and WU planes (Fig.~\ref{uvw.eps}a,b,c),
and most of the stars are
in the disk. The disk properties of most stars are further
confirmed by Figure~\ref{uvw.eps}d, their Toomre
diagrams ($\rm (U^2+W^2)^{1/2}$ vs. V velocity). There the
semi-circular line refers to the limit of thin disk stars, $V_{\rm
tot}$=$(U^2+V^2+W^2)^{1/2}<85$\,km\,s$^{-1}$ suggested in Chen et al.
(2004).

In Lu (1991), the histogram of U, V, W velocities and their
distributions in the UV, UW, VW planes (their figs.\,8, 9, 10) indicate that the strong
Ba stars have a higher velocity dispersion than the weak Ba stars and are,
therefore, probably members of the old disk population, whereas the
weak Ba stars are members of the young disk population.
But in our work, such a difference is not clear, as shown in
Figure~\ref{uvw.eps}. Maybe the reason is that we have fewer sample stars.

As Jorissen et al. (1998) mentioned, there are several pieces of evidence
that mild Ba stars belong to a somewhat younger population
than strong Ba stars. Except for the velocity dispersion found in Lu (1991),
Mennessier et al. (1997) suggested that mild Ba stars are mostly clump
giants with a mass in the range 2.5-4.5 M$_{\odot}$, whereas strong
Ba stars populate the giant branch and have masses in the
range 1-3 M$_{\odot}$. These mass estimates are consistent with those
derived from the mass-function distributions in Jorissen et al. (1998) (their fig.9).
They also found that
although mild Ba stars are dominated by high-mass objects, there is a small tail of
less massive objects which suggests that mild Ba stars are
indeed a mixture of populations H and L (their fig.11, where H and L mean high and low mass
respectively).
Some mild Ba stars having low orbital period should be in old population.
Metallicity is also a factor that affects the overabundances of $s$-process elements.
Being old objects, they should have rather low mass, but the third dredge-up becomes
more efficient for low metallicities (Busso et al. 2001).

\subsection{Orbital period}

Because Ba stars are binaries, the overabundance of $n$-capture process elements is due
to the transfer of s-enriched material from a previously-AGB-companion star
(now a white dwarf) via mass transfer or wind accretion. Thus orbital period is an
important parameter that characterizes a Ba star in a binary system.
As shown theoretically by
figure 4 in Han et al. (1995) and observationally by figure 3 in
Za\v{c}s (2000), strong Ba stars have orbital periods of $\sim$1000 days,
but the orbital periods of mild Ba stars are longer or shorter.
Ba stars with long orbital periods arise from wind accretion. The
longer the orbital periods are, the lower the accretion efficiency is.
Ba stars with short orbital periods are from stable Roche lobe overflow or a common envelope.
The shorter the
orbital periods are, the earlier during TP-AGB phase the onset of mass
transfer occurs.

There are five sample stars in the present work whose information on orbital periods
can be obtained from Jorissen et al. (1998):
HD\,49641, HD\,58121, HD\,58368 and HD\,95193 are classified as strong Ba stars in this work,
and their orbital periods are 1768\,d, 1214.3\,d, 672.7\,d and 1653.7\,d, respectively.
HD\,66216 is classified as a mild Ba star in this work,
and its orbital period is 2438\,d. The orbital periods of these Ba stars are generally consistent
with the suggestions of Han et al. (1995) and Za\v{c}s (2000).
We can understand that this longer orbital period of HD\,66216 may result in lower accretion
efficiency of $n$-capture process elements from the AGB-companion star through wind accretion
or wind exposure (Han et al. 1995), and thus there is less overabundance of $n$-capture process
elements on it compared to the other four strong Ba stars.
In addition, Jorissen et al. (1998) presented the compiled observations of orbital periods
of some strong Ba stars in their table\,2a, of some mild Ba stars in their table\,1a.
As shown in figures 2 and 4 of Jorissen et al. (1998), the orbital periods of Ba stars are in
a wide range from $\sim$200 to 11000 days (with one mild Ba star as 80.53 days), but
mild Ba stars generally have much higher orbital periods than the strong Ba stars (as shown by the
horizontal axes of their figs.11 and 6). Also, the short-period mode ($P<1500$ d) comprised
of nearly-circular systems is lacking among mild Ba stars (their Sect.10). It is a pity we
did not have the orbital periods for the rest of our 14 sample stars in this study.
We hope in the future orbital period information about these remaining stars can be obtained.

\section{Discussion and conclusions}

The chemical compositions of 19 Ba stars were obtained on the basis of high resolution and high $S/N$ ratio
Echelle spectra. We carried out a classical LTE analysis of these stars. From their effective temperature,
gravity and metallicity, the 19 Ba stars are giants and belong to the Galactic disk.

Among the 19 stars of our sample, six stars have Ba index larger than 2 and are classified as ``strong Ba stars''.
For these stars we found [Ba/Fe]$>$0.6. In HD\,224276 (which has no known Ba index), we found its [Ba/Fe]=0.78
and its abundance pattern is similar to those of other strong Ba stars (HD\,95193, HD\,218356). Thus this star is
also a strong Ba star and we assume its Ba index as 2.0. The Ba index of HD\,212320 is equal to 1.5, but we found that
it has [Ba/Fe]=1.11 thus we suggest it should be classified as a strong Ba star. For HD\,58121, this star has
[Ba/Fe]=0.73. Although its Ba index is 1.0, we assume this star has been misclassified and it should also be
a strong Ba star. Finally in our sample there are nine strong Ba stars with [Ba/Fe]$>$0.6.
The other ten stars can be considered to be mild Ba stars with +0.17$<$[Ba/Fe]$<$+0.54.
In particular the star HD\,66216,
classified as ``Ba-like star'', should be a mild Ba star with [Ba/Fe]=0.34.

The abundance distributions and the abundance scatter are quite consistent with what is observed for other Ba stars
(Allen \& Barbuy 2006b; Smiljanic et al. 2007).
In strong Ba stars, the $n$-capture process elements are all overabundant, but this overabundance is higher
for the second peak elements (about +0.9\,dex for [Ba/Fe] and [La/Fe], and +0.5\,dex for [Eu/Fe]) than for the
first peak elements (about 0.3\,dex for Y and Zr).
In mild Ba stars, [Y/Fe] and [Zr/Fe] are close to zero (not much overabundance of the first
peak elements), but the elements of the second peak: Ba, La and Eu are overabundant (about +0.4\,dex for [Ba/Fe]
and [La/Fe], and +0.3\,dex for [Eu/Fe]), this overabundance is lower than in the strong Ba stars.

The abundances of the $n$-capture elements (Y, Zr, Ba and La but not Eu) show a slight anti-correlation
with metallicity [Fe/H], which could mean that the lower metallicity benefits the nucleosynthesis of
$n$-capture process elements.

The kinematic parameters (UVW velocities) of stars in the sample confirm that they are disk stars.
Contrary to the findings of Lu (1991) and Jorissen et al. (1998), we did not find an obvious significant
difference between the velocities of the strong and mild Ba stars.
The reason could be that we have fewer stars than them.

The orbital periods are obtained for five of our sample stars, which show that the mild Ba star
HD\,66216 (2438\,d) has longer orbital period than the other four strong Ba stars HD\,49641, HD\,58121, HD\,58368 and
HD\,95193 (1768, 1214.3, 672.7 and 1653.7\,d, respectively). The orbital periods are consistent with
Han et al. (1995) and Za\v{c}s (2000), and are in the range of Ba stars given by Jorissen et al. (1998).
Further work to obtain the orbital periods of the remaining 14 sample stars will be useful to better
understand their evolution.

\begin{acknowledgements}
We thank the referees for an extensive and helpful review containing
very relevant scientific suggestions, which greatly improve this
paper. We are grateful to Prof. Jianrong Shi, Kefeng Tan, Shu Liu
and Wenyuan Cui for helping with data reduction and analysis.
Guochao Yang thanks Guohu Zhong, Yi Hu, Pin L$\ddot{\rm u}$, Qiang Liu,
Yueyang Zhang, He Gao, Dong Gao and Anbing Ren for their helpful
advices and encouragements. This work was supported by the Natural Science
Foundation of China (NSFC) Foundation under Nos.11273011, U1231119, 10973006,
11003002, 11273026, 10933001, 10973015; and the National Basic Research
Program of China (973 Program) Nos.2007CB815404, 2007CB815403, 2007CB815406.
\end{acknowledgements}



\begin{table*}[p]
\scriptsize \caption{Basic Data on the Sample of Stars.} \label{simbad}
\setlength\tabcolsep{2.5pt} \centering
\begin{tabular}{lllccrcrlrcrrrrrrr}
\hline \noalign{\smallskip} Star name & Ba & Sp. & $l$ & $b$ & ra. &
$\delta_{\rm ra.}$ & dec. & $\delta_{\rm dec.}$ & $\pi$ \,(mas) &
$\delta\pi$ & B & V & K\\
\noalign{\smallskip} \hline \noalign{\smallskip}
HD\,11353 &0.1 &K0  &165.8823&$-$68.0481& 38.78  &0.94&$-$38.04&0.53&12.59&0.85&4.873&3.738&  1.016 [0.250]\\
HD\,11658 &1.0c&K0  &132.9747&$-$09.9246& 90.98  &0.59&$-$30.35&0.50& 4.93&0.84&8.350&7.250&  4.675 [0.017]\\
HD\,13787 &1.0 &K0  &140.2071&$-$20.8720&$-$19.87&0.81&$-$7.72 &0.69& 5.23&1.01&8.250&7.300&  5.113 [0.023]\\
HD\,18418 &3.0 &K0  &148.7200&$-$18.0025& $-$4.84&1.27&$-$7.10 &0.83& 2.99&1.18&8.530&7.540&  5.232 [0.020]\\
HD\,20644 &0.5 &K4  &158.1375&$-$23.4606& $-$8.21&0.88&$-$16.42&0.64& 5.09&0.90&6.076&4.472&  0.877 [0.172]\\
HD\,31308 &0.2:&G5  &203.9237&$-$28.4207&  8.38  &0.95&$-$2.84 &0.69& 5.05&1.02&8.340&7.330&  5.057 [0.026]\\
HD\,43232 &0.5 &K1.5&214.4047&$-$10.9789& $-$6.18&0.78&$-$20.06&0.65& 5.06&0.90&5.332&3.987&  1.087 [0.246]\\
HD\,49641 &3.0c&K0  &209.4241&$+$01.2500& $-$7.08&0.82&$-$10.46&0.49& 0.73&0.88&8.500&7.130&  4.253 [0.015]\\
HD\,58121 &1.0m&G7  &211.1724&$+$10.1346&  0.54  &0.89&$-$0.26 &0.57& 2.82&0.95&9.090&7.920&  5.377 [0.021]\\
HD\,58368 &2.0c&K0  &210.0191&$+$11.0233&  2.74  &0.94&$-$0.74 &0.61& 2.36&0.97&8.990&7.970&  5.773 [0.023]\\
HD\,66216 &bl  &K2  &193.8981&$+$27.1407&$-$28.53&0.83&$-$35.25&0.64&12.66&0.78&6.087&4.944&  ----   ----- \\
HD\,81797 &1.0:&K3  &241.4881&$+$29.0455&$-$14.49&0.95&33.25   &0.53&18.40&0.78&3.486&2.004&$-$1.127[0.208]\\
HD\,90127 &2.0 &K0  &249.3141&$+$42.1085& 11.61  &1.01&$-$5.73 &0.57& 4.55&0.87&8.440&7.270&  4.758 [0.027]\\
HD\,95193 &2.0m&G8  &266.1248&$+$40.5624& $-$6.56&1.19&$-$2.65 &0.79& 2.30&1.03&9.280&8.300&  6.050 [0.021]\\
HD\,203137&0.5 &K4.5&092.0166&$+$00.4142&  7.77  &0.66&$-$2.11 &0.54& 1.91&0.62&8.870&7.000&  2.263 [0.278]\\
HD\,210030&1.0 &K0  &048.8171&$-$47.8983&  9.39  &1.07&$-$27.46&0.56& 6.07&0.95&8.570&7.470&  5.014 [0.016]\\
HD\,212320&1.5:&G3  &055.6592&$-$49.6862& $-$6.21&1.19&8.95    &0.87& 7.10&0.93&6.912&5.922&  3.891 [0.270]\\
HD\,218356&2.0m&K0  &095.1201&$-$31.7089&  0.26  &0.56&$-$33.31&0.41& 6.07&0.67&6.067&4.767&  1.765 [0.220]\\
HD\,224276&:   &K2  &113.0484&$-$15.8091&  6.21  &0.70&$-$0.01 &0.62& 2.09&0.96&9.48 &8.45 &  6.125 [0.020]\\
\noalign{\smallskip} \hline
\end{tabular}

Notes: Ba intensity in column 2. c: certain Ba stars, m: marginal Ba
stars, bl: Ba-likely stars, and ``:" refers to
the case of barium intensity uncertain.
\end{table*}

\begin{table*}
\centering
\scriptsize
\caption{ The Reddening Values of the Color Indices.}
\label{E.table}
\setlength\tabcolsep{2.5pt}
\begin{tabular}{lccccccccccccccccc}
\hline
\noalign{\smallskip}
 Star name & $(B-V)_{o}$ & $(V-K)_{o}$ & $(b-y)_{o}$ & $E(B-V)$ & $D$\,(kpc) & $A_{\rm B}$ &  $A_{\rm V}$ &
 $A_{\rm K}$ & $A_{\rm b}$ & $A_{\rm y}$ & $(B-V)$ & $(V-K)$ & $(b-y)$ \\
\noalign{\smallskip}
\hline
\noalign{\smallskip}
HD\,11353  & 1.135 & 2.722 & ---   & 0.016     & 0.079 & 0.071 & 0.055 & 0.006 & 0.067 & 0.054 & 1.119 & 2.673 & ---   \\
HD\,11658  & 1.100 & 2.575 & ---   & 0.045     & 0.203 & 0.194 & 0.149 & 0.017 & 0.182 & 0.147 & 1.055 & 2.442 & ---   \\
HD\,13787  & 0.950 & 2.187 & ---   & 0.030     & 0.191 & 0.128 & 0.098 & 0.011 & 0.120 & 0.097 & 0.920 & 2.100 & ---   \\
HD\,18418  & 0.990 & 2.308 & ---   & 0.051     & 0.334 & 0.220 & 0.169 & 0.019 & 0.206 & 0.167 & 0.939 & 2.158 & ---   \\
HD\,20644  & 1.604 & 3.595 & ---   & 0.202$^a$ & 0.196 & 0.870 & 0.668 & 0.074 & 0.816 & 0.660 & 1.402 & 3.001 & ---   \\
HD\,31308  & 1.010 & 2.273 & ---   & 0.017     & 0.198 & 0.074 & 0.057 & 0.006 & 0.069 & 0.056 & 0.993 & 2.223 & ---   \\
HD\,43232  & 1.345 & 2.900 & ---   & 0.114$^a$ & 0.198 & 0.491 & 0.377 & 0.042 & 0.460 & 0.373 & 1.231 & 2.565 & ---   \\
HD\,49641  & 1.370 & 2.877 & ---   & 0.147$^a$ & 1.370 & 0.636 & 0.489 & 0.054 & 0.597 & 0.483 & 1.223 & 2.442 & ---   \\
HD\,58121  & 1.170 & 2.543 & 0.667 & 0.024     & 0.355 & 0.103 & 0.079 & 0.009 & 0.097 & 0.078 & 1.146 & 2.473 & 0.649 \\
HD\,58368  & 1.020 & 2.197 & ---   & 0.021     & 0.424 & 0.090 & 0.069 & 0.008 & 0.085 & 0.069 & 0.999 & 2.135 & ---   \\
HD\,66216  & 1.143 & ---   & ---   & 0.009     & 0.079 & 0.040 & 0.031 & 0.003 & 0.038 & 0.030 & 1.134 & ---   & ---   \\
HD\,81797  & 1.482 & 3.131 & ---   & 0.010     & 0.054 & 0.042 & 0.032 & 0.004 & 0.039 & 0.032 & 1.472 & 3.102 & ---   \\
HD\,90127  & 1.170 & 2.512 & ---   & 0.032     & 0.220 & 0.138 & 0.106 & 0.012 & 0.130 & 0.105 & 1.138 & 2.418 & ---   \\
HD\,95193  & 0.980 & 2.250 & ---   & 0.047     & 0.435 & 0.202 & 0.155 & 0.017 & 0.190 & 0.153 & 0.933 & 2.112 & ---   \\
HD\,203137 & 1.870 & 4.737 & ---   & 0.048     & 0.524 & 0.208 & 0.160 & 0.018 & 0.195 & 0.158 & 1.822 & 4.595 & ---   \\
HD\,210030 & 1.100 & 2.456 & ---   & 0.026     & 0.165 & 0.112 & 0.086 & 0.009 & 0.105 & 0.085 & 1.074 & 2.380 & ---   \\
HD\,212320 & 0.990 & 2.031 & ---   & 0.044     & 0.141 & 0.188 & 0.144 & 0.016 & 0.176 & 0.143 & 0.946 & 1.903 & ---   \\
HD\,218356 & 1.300 & 3.002 & 0.835 & 0.086     & 0.165 & 0.372 & 0.286 & 0.032 & 0.349 & 0.283 & 1.214 & 2.748 & 0.768 \\
HD\,224276 & 1.030 & 2.325 & ---   & 0.071     & 0.478 & 0.307 & 0.236 & 0.026 & 0.288 & 0.233 & 0.959 & 2.115 &  ---  \\
\noalign{\smallskip}
\hline
\end{tabular}

Notes: $(B-V)_{o}$, $(V-K)_{o}$, $(b-y)_{o}$ and $(B-V)$, $(V-K)$,
$(b-y)$ are the values observed and after extinction correction respectively. $D$
is obtained directly from the parallaxes ($D$ = 1/$\pi$). The
superscript $a$ indicates that these stars are heavily reddened,
thus their colors (and temperatures) are uncertain, and the resulting
masses and ages may have larger uncertainties.
\end{table*}

\begin{table*}
\centering
\scriptsize
 \caption{The atmospheric parameters of the sample stars and the comparison of six common stars with
 that of McWilliam (1990, Mc).}
\label{parameter}
\setlength\tabcolsep{2.5pt}
\begin{tabular}{lccccrrrrcc}
\hline
\noalign{\smallskip}
 Star name & $T_{\rm eff}(V-K)$ & $T_{\rm eff}(b-y)$ & $T_{\rm eff}(B-V)$ & log\,$g$ & [Fe/H] &
 $\xi_{\rm t}\, (\rm km\, s^{-1})$ & $T_{\rm eff \rm (Mc)}$ & log $g_{(\rm Mc)}$ & $\rm [Fe/H]_{(Mc)}$
 & $\xi_{\rm t \rm (Mc)}$ \\
\noalign{\smallskip}
\hline
\noalign{\smallskip}
 HD\,11353   & 4467 &  ---  &  4592  &  2.13 &    0.04  & 1.45 & 4600&2.70&$-$0.13&2.2 \\
 HD\,11658   & 4660 &  ---  &  4699  &  2.61 &    0.00  & 1.50 & --- &--- &---    &--- \\
 HD\,13787   & 5007 &  ---  &  4966  &  2.90 &    0.00  & 1.40 & --- &--- &---    &--- \\
 HD\,18418   & 4934 &  ---  &  4845  &  2.36 & $-$0.29  & 1.60 & --- &--- &---    &--- \\
 HD\,20644   & 4231 &  ---  &  4115  &  1.18 & $-$0.12  & 1.60 & 3980&1.56&$-$0.31&2.6 \\
 HD\,31308   & 4884 &  ---  &  4865  &  2.84 &    0.16  & 1.60 & --- &--- &---    &--- \\
 HD\,43232   & 4551 &  ---  &  4386  &  1.40 & $-$0.03  & 1.55 & 4270&2.22&$-$0.18&2.6 \\
 HD\,49641   & 4645 &  ---  &  4351  &  1.05 & $-$0.32  & 1.50 & --- &--- &---    &--- \\
 HD\,58121   & 4642 &  4743 &  4588  &  2.41 &    0.23  & 1.35 & --- &--- &---    &--- \\
 HD\,58368   & 4965 &  ---  &  4777  &  2.47 & $-$0.11  & 1.50 & --- &--- &---    &--- \\
 HD\,66216   & ---  &  ---  &  4588  &  2.43 &    0.14  & 1.38 & 4540&2.73&0.03   &1.7 \\
 HD\,81797   & 4175 &  ---  &  4034  &  1.23 &    0.01  & 1.50 & 4120&1.77&$-$0.12&2.6 \\
 HD\,90127   & 4680 &  ---  &  4536  &  2.36 & $-$0.06  & 1.43 & --- &--- &---    &--- \\
 HD\,95193   & 4992 &  ---  &  4924  &  2.62 & $-$0.05  & 1.57 & --- &--- &---    &--- \\
 HD\,203137  & 3630 &  ---  &  3609  &  0.60 &    0.07  & 1.45 & --- &--- &---    &--- \\
 HD\,210030  & 4722 &  ---  &  4679  &  2.73 &    0.06  & 1.50 & --- &--- &---    &--- \\
 HD\,212320  & 5027 &  ---  &  4880  &  2.63 & $-$0.11  & 1.50 & 4790&2.87&$-$0.25&2.2 \\
 HD\,218356  & 4398 &  4311 &  4374  &  1.54 & $-$0.26  & 1.45 & --- &--- &---    &--- \\
 HD\,224276  & 4994 &  ---  &  4910  &  2.56 &    0.08  & 1.45 & --- &--- &---    &--- \\
\noalign {\smallskip}
\hline
\end{tabular}
\end{table*}


\begin{table*}
\centering \scriptsize \caption{Elemental Abundances of the Sample
Stars.} \label{abundance1} \setlength\tabcolsep{2pt}
\begin{tabular}{lrrrrrrrrrrrrrrrrrrrrrrrrr}
\hline \noalign{\smallskip} &&
\multicolumn{2}{c}{HD\,11353 } && \multicolumn{2}{c}{HD\,11658 } && \multicolumn{2}{c}{HD\,13787 } && \multicolumn{2}{c}{HD\,18418 }
&& \multicolumn{2}{c}{HD\,20644 } && \multicolumn{2}{c}{HD\,31308 } && \multicolumn{2}{c}{HD\,43232 } \\
\cline{3-4} \cline{6-7} \cline{9-10} \cline{12-13} \cline{15-16} \cline{18-19} \cline{21-22} \\
el && $\log\epsilon$(X) & [X/Fe] && $\log\epsilon$(X) & [X/Fe] &&
$\log\epsilon$(X) & [X/Fe] && $\log\epsilon$(X) & [X/Fe] &&
$\log\epsilon$(X) & [X/Fe] && $\log\epsilon$(X) & [X/Fe] &&
$\log\epsilon$(X) & [X/Fe]  \\
\noalign{\smallskip} \hline \noalign{\smallskip}
\ion{Fe}{i} && 7.55& ---    && 7.51& ---    && 7.51& ---    && 7.21& ---    && 7.39& --- && 7.67& ---    && 7.48& ---    \\
\ion{Fe}{ii}&& 7.61& ---    && 7.53& ---    && 7.53& ---    && 7.21& ---    && 7.51& --- && 7.78& ---    && 7.38& ---    \\
\ion{O}{i}  && --- & ---    && 8.99& 0.16   && 9.00& 0.17   &&  ---&  ---   && --- & --- && --- & ---    && --- & ---    \\
\ion{Na}{i} && 6.52& 0.15   && 6.29& $-$0.04&& 6.39& 0.06   && 6.16& 0.12   && 6.22& 0.00&& 6.63& 0.14   && 6.53& 0.23   \\
\ion{Mg}{i} && 7.63& 0.01   && 7.70& 0.12   && --- & ---    && --- & ---    && 7.81& 0.34&& 7.46& $-$0.28&& 7.69& 0.14   \\
\ion{Al}{i} && 6.47& $-$0.04&& 6.51& 0.04   && 6.32& $-$0.15&& 6.30& 0.12   && 6.40& 0.04&& 6.59& $-$0.04&& 6.55& 0.11  \\
\ion{Si}{i} && 7.64& 0.05   && 7.64& 0.09   && 7.67& 0.12   && 7.45& 0.19   && 7.56& 0.12&& 7.85& 0.14   && 7.52& 0.00   \\
\ion{Ca}{i} && 6.32& $-$0.08&& 6.23& $-$0.13&& 6.33& $-$0.03&& 6.11& 0.04   && 6.29& 0.04&& 6.35& $-$0.17&& 6.39& 0.06   \\
\ion{Sc}{ii}&& 3.14& $-$0.07&& 3.22& 0.05   && 3.27& 0.10   && 2.82& $-$0.06&& 3.18& 0.12&& 3.54&  0.21  && 3.17& 0.03   \\
\ion{Ti}{i} && 5.01& $-$0.06&& 4.96& $-$0.07&& 4.87& $-$0.16&& 4.48& $-$0.26&& 5.03& 0.11&& 4.97& $-$0.22&& 5.08& 0.08   \\
\ion{V}{i}  && 3.86& $-$0.18&& 3.81& $-$0.19&& 3.82& $-$0.18&& 3.35& $-$0.36&& 4.02& 0.13&& 3.81& $-$0.35&& 4.07& 0.10   \\
\ion{Cr}{i} && 5.71&    0.00&& 5.60& $-$0.07&& 5.57& $-$0.10&& 5.18& $-$0.20&& 5.94& 0.38&& 5.72& $-$0.11&& 5.74& 0.10   \\
\ion{Mn}{i} && 5.58& 0.15   && 5.49& 0.10   && 5.43& 0.04   && 4.99& $-$0.11&& 5.65& 0.37&& 5.53& $-$0.02&& 5.66& 0.30   \\
\ion{Ni}{i} && 6.32&    0.03&& 6.30& 0.05   && 6.25&    0.00&& 5.85& $-$0.11&& 6.35& 0.21&& 6.39&$-$0.02 && 6.26& 0.04   \\
\ion{Y}{i}  && --- & ---    && 2.06& $-$0.18&& 2.60&  0.36  && 1.99& 0.04   && 2.42& 0.29&& --- & ---    && 2.35& 0.14   \\
\ion{Zr}{i} && 2.58& $-$0.06&& 2.50& $-$0.10&& 2.94& 0.34   && 2.50& 0.19   && --- & --- && 2.80& 0.04   && 2.66& 0.09   \\
\ion{Ba}{ii}&& 2.53& 0.36   && 2.30& 0.17   && 2.67& 0.54   && 2.53& 0.69   && 2.53& 0.51&& 2.61& 0.32   && 2.63& 0.53   \\
\ion{La}{ii}&& 1.60& 0.39   && 1.43& 0.26   && 1.70& 0.53   && 1.53& 0.65   && 1.72& 0.66&& 1.63& 0.30   && 1.53& 0.39   \\
\ion{Eu}{ii}&& 0.86& 0.31   && 0.83& 0.32   && 0.89& 0.38   && 0.34& 0.12   && 0.74& 0.34&& 1.09& 0.42   && 0.87& 0.39   \\
\noalign{\smallskip} \hline
\end{tabular}
\begin{tabular}{lrrrrrrrrrrrrrrrrrrrrrrrrr}
\hline \noalign{\smallskip} && \multicolumn{2}{c}{HD\,49641 } && \multicolumn{2}{c}{HD\,58121 }
&& \multicolumn{2}{c}{HD\,58368 } && \multicolumn{2}{c}{HD\,66216 } && \multicolumn{2}{c}{HD\,81797 } && \multicolumn{2}{c}{HD\,90127 }
&& \multicolumn{2}{c}{HD\,95193 }  \\
\cline{3-4} \cline{6-7} \cline{9-10} \cline{12-13} \cline{15-16} \cline{18-19} \cline{21-22}  \\
el && $\log\epsilon$(X) & [X/Fe] && $\log\epsilon$(X) & [X/Fe] &&
$\log\epsilon$(X) & [X/Fe] && $\log\epsilon$(X) & [X/Fe] &&
$\log\epsilon$(X) & [X/Fe] && $\log\epsilon$(X) & [X/Fe] &&
$\log\epsilon$(X) & [X/Fe] \\
\noalign{\smallskip} \hline \noalign{\smallskip}
\ion{Fe}{i} && 7.19& ---    && 7.74& ---     && 7.40& ---    &&  7.65& ---    && 7.52& ---    && 7.45& ---    && 7.46& ---      \\
\ion{Fe}{ii}&& 6.99& ---    && 7.84& ---     && 7.55& ---    &&  7.64& ---    && 7.52& ---    && 7.47& ---    && 7.51& ---      \\
\ion{O}{i}  && 8.91&  0.40  && --- & ---     && --- & ---    &&  --- & ---    && 9.05& 0.21   && 9.05& 0.28   && 8.98&  0.20    \\
\ion{Na}{i} && 5.98& $-$0.03&& 6.66& 0.10    && 6.46& 0.24   &&  6.54& 0.07   &&  ---&  ---   && 6.36& 0.09   && 6.57& 0.29     \\
\ion{Mg}{i} && 7.75& 0.49   && --- & ---     && 7.44& $-$0.03&&  7.87& 0.15   && --- & ---    && 7.52& 0.00   && 7.86& 0.33     \\
\ion{Al}{i} && 6.31& 0.16   && 6.48& $-$0.22 && 6.33& $-$0.03&&  6.58& $-$0.03&& 6.50& 0.02   && 6.51& 0.10   && 6.64& 0.22     \\
\ion{Si}{i} && 7.63& 0.40   && 7.88& 0.10    && 7.71& 0.27   &&  7.72& 0.03   && 7.76& 0.20   && 7.73& 0.24   && 7.80& 0.30     \\
\ion{Ca}{i} && 6.25& 0.21   && 6.54& $-$0.05 && 6.17& $-$0.08&&  6.35& $-$0.15&& 6.31& $-$0.06&& 6.19& $-$0.11&& 6.42& 0.11     \\
\ion{Sc}{ii}&& 2.92& 0.07   && 3.44&  0.04   && 3.15& 0.09   &&  3.34& 0.03   && 3.35& 0.17   && 3.30& 0.19   && 2.99& $-$0.13  \\
\ion{Ti}{i} && 4.75& 0.04   && 5.02& $-$0.24 && 4.74& $-$0.18&&  5.07& $-$0.10&& 5.08&  0.04  && 4.81& $-$0.16&& 4.79& $-$0.19  \\
\ion{V}{i}  && 3.66& $-$0.02&& 3.85& $-$0.38 && 3.54& $-$0.35&&  4.07& $-$0.07&& 4.20& 0.19   && 3.65& $-$0.29&& 3.95& 0.00     \\
\ion{Cr}{i} && 5.52& 0.17   && 5.77& $-$0.13 && 5.52& $-$0.04&&  5.83& 0.02   && 5.65& $-$0.03&& 5.85& 0.24   && 5.93& 0.31     \\
\ion{Mn}{i} && 5.27& 0.20   && 5.35& $-$0.27 && 5.26& $-$0.02&&  5.58& 0.05   && 5.69& 0.29   && 5.40& 0.07   && 5.37& 0.03     \\
\ion{Ni}{i} && 5.96& 0.03   && 6.51& 0.03    && 6.16& 0.02   &&  6.41& 0.02   && 6.54& 0.28   && 6.28&  0.09  && 6.21& 0.01     \\
\ion{Y}{i}  && 2.27& 0.35   && 2.48& 0.01    && 2.27& 0.14   &&   ---& ---    && 2.39&  0.14  && 2.33&  0.15  && 2.67& 0.48     \\
\ion{Zr}{i} && 2.69& 0.41   && 2.92& 0.09    && 2.74& 0.25   &&  2.80& 0.06   && 2.77& 0.16   && 2.79&  0.25  && 3.05& 0.50     \\
\ion{Ba}{ii}&& 2.94& 1.13   && 3.09& 0.73    && 3.00& 0.98   &&  2.61& 0.34   && 2.66& 0.52   && 3.03& 0.96   && 2.72& 0.64     \\
\ion{La}{ii}&& 2.23& 1.38   && 2.14&  0.74   && 2.13& 1.07   &&  1.59& 0.28   && 1.66& 0.48   && 2.35& 1.24   && 1.72& 0.60     \\
\ion{Eu}{ii}&& 0.83& 0.64   && 1.25& 0.51    && 0.81& 0.41   &&  0.76& 0.11   && --- & ---    && 0.93& 0.48   && 0.86& 0.40     \\
\noalign{\smallskip} \hline
\end{tabular}
\begin{tabular}{lrrrrrrrrrrrrrrrrrrrrrrrrrr}
\hline \noalign{\smallskip} && \multicolumn{2}{c}{HD\,203137 } && \multicolumn{2}{c}{HD\,210030 }&& \multicolumn{2}{c}{HD\,212320 }
&& \multicolumn{2}{c}{HD\,218356 } && \multicolumn{2}{c}{HD\,224276 } \\
\cline{3-4} \cline{6-7} \cline{9-10} \cline{12-13} \cline{15-16} \\
el && $\log\epsilon$(X) & [X/Fe] && $\log\epsilon$(X) & [X/Fe] &&
$\log\epsilon$(X) & [X/Fe] && $\log\epsilon$(X) & [X/Fe] && $\log\epsilon$(X) & [X/Fe] \\
\noalign{\smallskip} \hline \noalign{\smallskip}
\ion{Fe}{i} && 7.58& ---     && 7.57 & ---    && 7.40 & ---    && 7.25 & ---     && 7.59 & --- \\
\ion{Fe}{ii}&& 7.70& ---     && 7.71 & ---    && 7.56 & ---    && 7.29 & ---     && 7.65 & --- \\
\ion{O}{i}  && --- & ---     && 9.00 & 0.11   &&  --- &  ---   && ---  & ---     && 8.94 & 0.03    \\
\ion{Na}{i} && 6.26& $-$0.14 && 6.38 & $-$0.01&& 6.32 & 0.10   && ---  & ---     && 6.54 & 0.13 \\
\ion{Mg}{i} && 7.82& 0.17    && 7.53 & $-$0.11&& 7.58 & 0.11   && 7.87 & 0.53    && 7.38 & $-$0.28 \\
\ion{Al}{i} && 6.33& $-$0.21 && 6.46 & $-$0.07&& 6.37 & 0.01   && 6.79 & 0.56    && 6.43 & $-$0.12 \\
\ion{Si}{i} &&  ---&  ---    && 7.75 & 0.14   && 7.65 &  0.21  && 7.45 & 0.14    && 7.71 & 0.08 \\
\ion{Ca}{i} &&  ---&    ---  && 6.22 & $-$0.20&& 6.17 & $-$0.08&& 6.35 & 0.23    && 6.42 & $-$0.02 \\
\ion{Sc}{ii}&& 3.18& $-$0.06 && 3.40 & 0.17   && 3.22 & 0.16   && 3.08 & 0.15    && 3.19 & $-$0.06 \\
\ion{Ti}{i} && 4.97& $-$0.13 && 4.88 & $-$0.21&& 4.58 & $-$0.34&& 5.02 & 0.23    && 4.97 & $-$0.14 \\
\ion{V}{i}  && 4.02& $-$0.05 && 3.77 & $-$0.29&& 3.57 & $-$0.32&& 3.73 & $-$0.03 && 3.70 & $-$0.38 \\
\ion{Cr}{i} &&  ---&  ---    && 5.74 & 0.01   && 5.42 & $-$0.14&& 5.78 & 0.35    && 5.64 & $-$0.11 \\
\ion{Mn}{i} && 5.47& 0.01    && 5.49 & 0.04   && 5.15 & $-$0.13&& 5.34 & 0.19    && 5.52 & 0.05 \\
\ion{Ni}{i} && 6.51& 0.19    && 6.37 & 0.06   && 6.10 & $-$0.04&& 6.08 & 0.07    && 6.36 & 0.03    \\
\ion{Y}{i}  && 2.28& $-$0.03 && 2.29 & $-$0.01&& 2.33 & 0.20   && 2.43 & 0.43    && 2.53 & 0.21 \\
\ion{Zr}{i} && 2.69& 0.02    && 2.71 & 0.05   && 2.80 & 0.31   && 2.76 & 0.40    && 2.87 & 0.19 \\
\ion{Ba}{ii}&& 2.73& 0.53    && 2.73 & 0.54   && 3.13 & 1.11   && 2.83 & 0.94    && 2.99 & 0.78 \\
\ion{La}{ii}&& 1.83& 0.59    && 1.88 & 0.65   && 2.24 &  1.18  && 2.00 & 1.07    && 1.98 & 0.73 \\
\ion{Eu}{ii}&& 0.74& 0.16    && ---  & ---    && 0.88 & 0.48   && 0.76 & 0.49    && 1.10 & 0.51 \\
\noalign{\smallskip} \hline
\end{tabular}
\end{table*}

\begin{table*}
\centering \scriptsize \caption {Uncertainties in the Abundance
Analysis for a Sample Star HD\,66216. } \label{error.table1}
\setlength\tabcolsep{2.5pt}
\begin{tabular}{lcrrrrr}
\hline \noalign{\smallskip} \multicolumn{1}{c}{ } &
\multicolumn{1}{r}{$\sigma_{\rm EW}/\sqrt N$ } &\multicolumn{1}{r}
{$\Delta {T_{\rm eff}}$ } & \multicolumn{1}{r}{$\Delta {\log g}$} &
\multicolumn{1} {r}{$\Delta {\rm [Fe/H]}$ } &
\multicolumn{1}{r}{$\Delta {\xi_{\rm t}}$ }
& \multicolumn{1}{c}{$\sigma_{\rm total}$ }\\
\noalign{\smallskip} \multicolumn{1}{c}{ } & \multicolumn{1}{r}{} &
\multicolumn{1}{r}{$200$ } & \multicolumn{1}{r}{$0.2$ } &
\multicolumn{1}{r}{$0.15$ } & \multicolumn{1}{r}{$0.2$ }
& \multicolumn{1}{r}{LSM}\\
\noalign{\smallskip} \hline \noalign{\smallskip}
$\Delta{\rm [\ion {Fe}{i}/H]}$  &0.01 & 0.02 & 0.04 &0.02 &0.04 & 0.06 \\
$\Delta{\rm [\ion {Fe}{ii}/H]}$ &0.02 & 0.15 & 0.01 &0.03 &0.01 & 0.15 \\
$\Delta{\rm [\ion {Na}{i}/Fe]}$ &0.02 & 0.12 & 0.05 &0.05 &0.05 & 0.15 \\
$\Delta{\rm [\ion {Mg}{i}/Fe]}$ &0.05 & 0.02 & 0.01 &0.01 &0.01 & 0.06 \\
$\Delta{\rm [\ion {Al}{i}/Fe]}$ &0.04 & 0.08 & 0.03 &0.04 &0.03 & 0.11 \\
$\Delta{\rm [\ion {Si}{i}/Fe]}$ &0.02 & 0.12 & 0.01 &0.01 &0.02 & 0.12 \\
$\Delta{\rm [\ion {Ca}{i}/Fe]}$ &0.01 & 0.14 & 0.08 &0.08 &0.07 & 0.19 \\
$\Delta{\rm [\ion {Sc}{ii}/Fe]}$&0.03 & 0.05 & 0.03 &0.01 &0.04 & 0.08 \\
$\Delta{\rm [\ion {Ti}{i}/Fe]}$ &0.02 & 0.13 & 0.08 &0.09 &0.08 & 0.20 \\
$\Delta{\rm [\ion {V}{i} /Fe]}$ &0.07 & 0.1  & 0.11 &0.11 &0.11 & 0.23 \\
$\Delta{\rm [\ion {Cr}{i}/Fe]}$ &0.03 & 0.12 & 0.07 &0.07 &0.07 & 0.17 \\
$\Delta{\rm [\ion {Mn}{i}/Fe]}$ &0.02 & 0.12 & 0.06 &0.04 &0.06 & 0.15 \\
$\Delta{\rm [\ion {Ni}{i}/Fe]}$ &0.01 & 0.01 & 0.05 &0.03 &0.05 & 0.08 \\
$\Delta{\rm [\ion {Y}{i}/Fe]}$  &0.08 & 0.23 & 0.01 &0.01 &0.01 & 0.24 \\
$\Delta{\rm [\ion {Zr}{i}/Fe]}$ &0.04 & 0.22 & 0.01 &0.01 &0.01 & 0.22 \\
$\Delta{\rm [\ion {Ba}{ii}/Fe]}$&0.02 & 0.01 & 0.07 &0.01 &0.07 & 0.10 \\
$\Delta{\rm [\ion {La}{ii}/Fe]}$&0.03 & 0.04 & 0.01 &0.03 &0.01 & 0.06 \\
$\Delta{\rm [\ion {Eu}{ii}/Fe]}$&0.07 & 0.01 & 0.01 &0.04 &0.01 & 0.08 \\
\noalign {\smallskip} \hline
\end{tabular}
\end{table*}

\begin{table*}
\centering \scriptsize \caption {Uncertainties in the Abundance
Analysis for a Sample Star HD\,212320. } \label{error.table2}
\setlength\tabcolsep{2.5pt}
\begin{tabular}{lcrrrrr}
\hline \noalign{\smallskip} \multicolumn{1}{c}{ } &
\multicolumn{1}{r}{$\sigma_{\rm EW}/\sqrt N$ } &\multicolumn{1}{r}
{$\Delta {T_{\rm eff}}$ } & \multicolumn{1}{r}{$\Delta {\log g}$} &
\multicolumn{1} {r}{$\Delta {\rm [Fe/H]}$ } &
\multicolumn{1}{r}{$\Delta {\xi_{\rm t}}$ }
& \multicolumn{1}{c}{$\sigma_{\rm total}$ }\\
\noalign{\smallskip} \multicolumn{1}{c}{ } & \multicolumn{1}{r}{ } &
\multicolumn{1}{r}{$200$ } & \multicolumn{1}{r}{$0.2$ } &
\multicolumn{1}{r}{$0.15$ } & \multicolumn{1}{r}{$0.2$ }
& \multicolumn{1}{r}{LSM }\\
\noalign{\smallskip} \hline \noalign{\smallskip}
$\Delta{\rm [\ion {Fe}{i}/H]}$  & 0.01 & 0.13  & 0.01 & 0.01 & 0.07 & 0.15 \\
$\Delta{\rm [\ion {Fe}{ii}/H]}$ & 0.03 & 0.15  & 0.11 & 0.06 & 0.06 & 0.21 \\
$\Delta{\rm [\ion {O}{i}/Fe]} $ & 0.04 & 0.19  & 0.06 & 0.01 & 0.03 & 0.21 \\
$\Delta{\rm [\ion {Na}{i}/Fe]}$ & 0.06 & 0.16  & 0.01 & 0.01 & 0.03 & 0.17 \\
$\Delta{\rm [\ion {Mg}{i}/Fe]}$ & 0.03 & 0.06  & 0.01 & 0.01 & 0.02 & 0.07 \\
$\Delta{\rm [\ion {Al}{i}/Fe]}$ & 0.02 & 0.11  & 0.01 & 0.01 & 0.02 & 0.11 \\
$\Delta{\rm [\ion {Si}{i}/Fe]}$ & 0.03 & 0.02  & 0.03 & 0.02 & 0.04 & 0.06 \\
$\Delta{\rm [\ion {Ca}{i}/Fe]}$ & 0.02 & 0.21  & 0.03 & 0.02 & 0.09 & 0.23 \\
$\Delta{\rm [\ion {Sc}{ii}/Fe]}$& 0.08 & 0.02  & 0.08 & 0.04 & 0.08 & 0.15 \\
$\Delta{\rm [\ion {Ti}{i}/Fe]}$ & 0.01 & 0.17  & 0.01 & 0.03 & 0.07 & 0.19 \\
$\Delta{\rm [\ion {V}{i} /Fe]}$ & 0.04 & 0.19  & 0.01 & 0.01 & 0.06 & 0.20 \\
$\Delta{\rm [\ion {Cr}{i}/Fe]}$ & 0.03 & 0.18  & 0.01 & 0.01 & 0.06 & 0.19 \\
$\Delta{\rm [\ion {Mn}{i}/Fe]}$ & 0.03 & 0.22  & 0.02 & 0.01 & 0.12 & 0.25 \\
$\Delta{\rm [\ion {Ni}{i}/Fe]}$ & 0.01 & 0.13  & 0.02 & 0.03 & 0.08 & 0.15 \\
$\Delta{\rm [\ion {Y}{i}/Fe]}$  & 0.07 & 0.15  & 0.04 & 0.2  & 0.02 & 0.26 \\
$\Delta{\rm [\ion {Zr}{i}/Fe]}$ & 0.05 & 0.2   & 0.02 & 0.16 & 0.02 & 0.26 \\
$\Delta{\rm [\ion {Ba}{ii}/Fe]}$& 0.02 & 0.05  & 0.04 & 0.07 & 0.09 & 0.13 \\
$\Delta{\rm [\ion {La}{ii}/Fe]}$& 0.05 & 0.05  & 0.09 & 0.05 & 0.1  & 0.16 \\
$\Delta{\rm [\ion {Eu}{ii}/Fe]}$& 0.09 & 0.01  & 0.09 & 0.05 & 0.03 & 0.14 \\
\noalign {\smallskip} \hline
\end{tabular}
\end{table*}

\begin{table*}
\centering
\scriptsize
 \caption{The Range of 17 Elements in Our Sample Stars, and the Line Numbers used in the Abundance Analyses.}
\label{valueXFe}
\setlength\tabcolsep{2.5pt}
\begin{tabular}{lrrr|lrrrr}
\hline
\noalign{\smallskip}
 [X/Fe] & lines & range & median & [X/Fe] & lines & range & median \\
\noalign{\smallskip}
\hline
\noalign{\smallskip}
 \ion {O}{i}  & 3 & [0.03, 0.40]    & $ 0.18$  &   \ion {Cr}{i}     & 7 & [$-0.20$, 0.38] & $-0.02$  \\
 \ion {Na}{i} & 7 & [$-0.14$, 0.29] & $ 0.10$  &   \ion {Mn}{i}     & 3 & [$-0.27$, 0.37] & $ 0.05$  \\
 \ion {Mg}{i} & 9 & [$-0.28$, 0.53] & $ 0.12$  &   [\ion {Fe}{i}/H] &60 & [$-0.32$, 0.23] & $ 0.00$  \\
 \ion {Al}{i} & 6 & [$-0.22$, 0.56] & $ 0.01$  &   \ion {Ni}{i}     &44 & [$-0.11$, 0.28] & $ 0.03$  \\
 \ion {Si}{i} &38 & [0.00, 0.40]    & $ 0.14$  &   \ion {Y}{i}      &1  & [$-0.18$, 0.48] & $ 0.15$  \\
 \ion {Ca}{i} &27 & [$-0.20$, 0.23] & $-0.06$  &   \ion {Zr}{i}     &4  & [$-0.10$, 0.50] & $ 0.18$  \\
 \ion {Sc}{ii}& 5 & [$-0.13$, 0.21] & $ 0.07$  &   \ion {Ba}{ii}    & 3 & [$ 0.17$, 1.13] & $ 0.54$  \\
 \ion {Ti}{i} &12 & [$-0.34$, 0.23] & $-0.14$  &   \ion {La}{ii}    & 2 & [$ 0.26$, 1.38] & $ 0.65$  \\
 \ion {V}{i}  & 3 & [$-0.38$, 0.19] & $-0.18$  &   \ion {Eu}{ii}    & 1 & [$ 0.11$, 0.64] & $ 0.40$  \\
\noalign {\smallskip}
\hline
\end{tabular}
\end{table*}

\end{document}